\documentclass[prd,superscriptaddress,floatfix,notitlepage,nofootinbib]{revtex4-1} 

\pdfoutput=1

\usepackage[american]{babel}
\usepackage[utf8x]{inputenc}
\usepackage[T1]{fontenc}
\usepackage{lmodern}

\usepackage{amsfonts}
\usepackage{mathtools}
\usepackage{amssymb}
\usepackage{empheq}
\usepackage{natbib}
\usepackage{ifthen}
\usepackage{blindtext}
\usepackage{graphicx}
\usepackage{xspace}
\usepackage{epsfig}
\usepackage{adjustbox}
\usepackage{booktabs}
\usepackage{multirow}
\usepackage{dcolumn}
\usepackage{bm}
\usepackage{bbm}
\usepackage{dsfont}
\usepackage{mathrsfs}
\usepackage{enumerate}
\usepackage{braket}
\usepackage{enumitem}
\usepackage{pifont}
\usepackage{cancel}
\usepackage{tabularx}
\usepackage{microtype}
\usepackage{siunitx}
\usepackage{float}
\usepackage{simplewick}
\usepackage{marginnote}
\usepackage{relsize}
\usepackage[
dvipsnames,usenames]{xcolor}
\usepackage{tikz}
\usepackage{hyperref}
\definecolor{navyblue}{rgb}{0.0, 0.0, 0.5}
\definecolor{coralred}{rgb}{1.0, 0.25, 0.25}
\definecolor{green(munsell)}{rgb}{0.0, 0.66, 0.47}
\hypersetup{colorlinks=true,linkcolor=navyblue,citecolor=coralred,urlcolor=green(munsell),linktocpage=true,pdfencoding=auto}
\usepackage[all]{hypcap}

\newcommand{\hp}{\hphantom}
\DeclarePairedDelimiter{\abs}{\lvert}{\rvert}

\makeatletter
\newlength{\apb@width}
\newcommand{\autoparbox}[2][c]{\settowidth{\apb@width}{#2}\parbox[#1]{\apb@width}{#2}}

\makeatother
\newcommand\EQ[1]{Eq.~\eqref{eq:#1}}
\newcommand\eqsI[1]{Eqs.~\eqref{eq:#1}}
\newcommand{\eqsII}[2]{Eqs.~\eqref{eq:#1}, \eqref{eq:#2}}

\newcommand{\be}{\begin{equation}}
\newcommand{\ee}{\end{equation}}
\newcommand{\ba}{\[\begin{aligned}}
\newcommand{\ea}{\end{aligned}\]}

\newcommand\hlight[1]{\tikz[overlay,remember picture,baseline=-\the\dimexpr\fontdimen22\textfont2\relax]\node[rectangle,fill=blue!50,rounded corners,fill opacity = 0.2,draw,thick,text opacity =1] {$#1$};}

\def\k{{\boldsymbol{k}}}

\def\q{{\boldsymbol{q}}}
\def\p{{\boldsymbol{p}}}

\def\param{{\boldsymbol{\theta}}}

\renewcommand\[{\left[}
\renewcommand\]{\right]}
\renewcommand{\vec}{\bm}

\newcommand{\deltaNL}{\delta_g}

\newcommand{\dirac}[1]{(2\pi)^3\delta^{(3)}_{\mathrm{D}}(#1)}

\newcommand{\eu}{\mathrm{e}}
\newcommand{\dif}{\mathrm{d}}
\newcommand\del{\partial}

\newcommand\vers[1]{\hat{\vec{#1}}}

\newcommand{\D}[2]{{\cal D}^{#1}_{\hp{#1}#2}}
\newcommand{\iD}[2]{{\cal I}^{#1}_{\hp{#1}#2}}
\newcommand{\I}[2]{{\cal I}^{#1}_{\hp{#1}#2}}

\newcommand{\M}[3]{#1^{#2}_{\hp{#2}#3}}

\begin{document}

\begin{flushright}
CERN-TH-2023-130
\end{flushright}

\title{Cosmological Information in Perturbative Forward Modeling}

\author{Giovanni Cabass}
\affiliation{School of Natural Sciences, Institute for Advanced Study, Princeton, NJ 08540, United States}
\author{Marko Simonovi{\'c}}
\affiliation{Theoretical Physics Department, CERN, 1 Esplanade des Particules, Geneva 23, CH-1211, Switzerland}
\author{Matias Zaldarriaga}
\affiliation{School of Natural Sciences, Institute for Advanced Study, Princeton, NJ 08540, United States}

\date{\today}
\begin{abstract}
\noindent 
We study how well field-level inference with a perturbative forward model can constrain cosmological parameters compared to conventional analyses. We exploit the fact that in perturbation theory the field-level posterior can be computed analytically in the limit of small noise. In the idealized case where the only relevant parameter for the nonlinear evolution is the nonlinear scale, we argue that information content in this posterior is the same as in the $n$-point correlation functions computed at the same perturbative order. In the real universe other parameters can be important, and there are possibly enhanced effects due to nonlinear interactions of long and short wavelength fluctuations that can either degrade the signal or increase covariance matrices. We identify several different parameters that control these enhancements and show that for some shapes of the linear power spectrum they can be large. This leads to degradation of constraints in the standard analyses, even though the effects are not dramatic for a $\Lambda$CDM-like cosmology. The aforementioned long-short couplings do not affect the field-level inference which remains optimal. Finally, we show how in these examples calculation of the perturbative posterior motivates new estimators that are easier to implement in practice than the full forward modeling but lead to nearly optimal constraints on cosmological parameters. These results generalize to any perturbative forward model, including galaxies in redshift space.
\end{abstract}

\maketitle


\tableofcontents


\section{Introduction}
\label{sec:introduction}

\noindent Traditional methods to infer cosmological parameters from large-scale structure galaxy surveys are based on measuring and analyzing the $n$-point correlation functions. The $2$-point function or its Fourier transform, the power spectrum, is the most commonly used statistics in the data analysis. Higher-order statistics, such as the $3$-point function or bispectrum, are employed in the data analysis less often. There are several reasons why tackling the higher-order $n$-point functions is difficult. They are harder to estimate from the data and predict theoretically, particularly taking into account geometry of the survey, selection functions and other observational effects. Furthermore, the length of the data vector grows very fast with each new $n$-point function added, making the estimates of the covariance matrices computationally unfeasible. While some progress has been made in order to circumvent some of these issues (see e.g. \cite{Scoccimarro:2000sn,Sefusatti:2006pa,Baldauf:2014qfa,Angulo:2014tfa,Gil-Marin:2014sta,Slepian:2015hca,Gil-Marin:2016wya,Slepian:2016kfz,Pearson:2017wtw,Eggemeier:2018qae,Oddo:2019run,MoradinezhadDizgah:2020whw,Eggemeier:2021cam,Alkhanishvili:2021pvy,Oddo:2021iwq,Baldauf:2021zlt,Ivanov:2021kcd,Philcox:2021kcw,Philcox:2020zyp,Philcox:2020vbm,Philcox:2021ukg,Pardede:2022udo,Rizzo:2022lmh,Ivanov:2023qzb,DAmico:2022ukl,DAmico:2022osl,Pardede:2023ddq}), more work has to be done before the higher order statistics can be routinely used in the data analysis. 

One alternative to using $n$-point functions is forward modeling~\cite{Kitaura:2007pe,Jasche:2012kq,Wang:2013ep,Ata:2014ssa,Seljak:2017rmr,Modi:2018cfi}.\footnote{While the term forward modeling is usually intended as the procedure to obtain a dark matter or biased tracer field from the linear density field, in this paper we use it interchangeably with field-level inference.} In forward modeling one predicts the full nonlinear density field of matter and galaxies, given some cosmological parameters and a realization of the initial conditions. Such nonlinear field should be directly compared to the data, without using any summary statistics. This is nontrivial since the correct form of the likelihood is in general unknown, which motivates the use of the so-called likelihood-free inference (see for example~\cite{Alsing:2018eau,Alsing:2019xrx,Jeffrey:2020xve}). In order to constrain cosmological models, one has to vary not only all cosmological and nuisance parameters, but also all amplitudes and phases of the initial density field. This is clearly a formidable task, despite the recent progress in efficient sampling of large-dimensional parameter space~\cite{Jasche:2009hz,Jasche:2018oym,Dai:2022dso,Robnik:2022bzs,Modi:2022pzm}. However, setting these important technical challenges aside, forward modeling is argued to have a number of advantages. Most importantly, it includes all available information from the nonlinear density field, which corresponds to the information from all $n$-point correlation functions combined. It also allows for much easier combination of various types of data, as well as inclusion of all observational effects such as masks and selection functions, or various systematic effects, all of which can be forward modeled. In a nutshell, forward modeling is guaranteed to give the optimal constraints on cosmological parameters with optimal combination of different data sets, assuming technical difficulties related to sampling can be overcome. 

The full nonlinear forward modeling, which aims at describing distribution and properties of galaxies on very small scales, needs very advanced hydrodynamical cosmological simulations with realistic sub-grid models, that have to be run for each point in the immense parameter space. While this may be the ultimate way to interpret cosmological observations in the future, such data analysis is not feasible at present. A more modest but doable alternative is perturbative forward modeling. In this approach, given some realization of the initial conditions, the nonlinear density field is calculated in perturbation theory. Using the standard effective field theory methods applied to large-scale structure~\cite{McDonald:2009dh,Baumann:2010tm,Carrasco:2012cv,Porto:2013qua,Senatore:2014vja,2014JCAP...08..056A,Senatore:2014eva,Lewandowski:2014rca,Mirbabayi:2014zca,Angulo:2015eqa,Lewandowski:2015ziq,Abolhasani:2015mra,Vlah:2015sea,Blas:2015qsi,Vlah:2016bcl,Desjacques:2016bnm,Perko:2016puo,Chen:2020zjt,Cabass:2022avo}, this can be done for the dark matter field and biased tracers in real or redshift space~\cite{Baldauf:2015tla,Baldauf:2015zga,Schmittfull:2018yuk,Taruya:2018jtk,McQuinn:2018zwa,Schmittfull:2020trd,Modi:2019hnu,Schmidt:2020ovm,Kokron:2021xgh,Taruya:2021ftd,Obuljen:2022cjo}.\footnote{Note that even some hybrid forward models which rely on N-body simulations are essentially perturbative, as long as the expansion in density fluctuations is assumed in any one of the steps of producing the final galaxy density field. One typical example are models where the usual bias expansion is applied to the full nonlinear matter fields obtained from simulations, see for example~\cite{Kokron:2021xgh}. All the conclusions of this work apply to such models as well.} One important advantage of perturbative forward modeling is that it allows for rigorous definition of the field-level likelihood~\cite{Schmidt:2018bkr,Elsner:2019rql,Cabass:2019lqx,Schmidt:2020tao}, which can be made arbitrary precise on large enough scales. A lot of recent effort was dedicated to clarifying all technical aspects of this approach~\cite{Cabass:2020nwf,Schmidt:2020viy,Cabass:2020jqo} and making first applications to mock data from simulations~\cite{Barreira:2021ukk,Lazeyras:2021dar,Babic:2022dws,Andrews:2022nvv,Kostic:2022vok,Stadler:2023hea}, leading to encouraging proof-of-principle results. On the other hand, perturbative forward modeling has a major limitation---it is applicable only on large scales, where perturbation theory provides a good description of the nonlinear density field. Since the importance of nonlinearities in the perturbative regime is expected to be small by construction, it is interesting to ask whether perturbative forward modeling is different from analyses based on a few leading $n$-point correlation functions (for some earlier work on this, see Ref.~\cite{Schmidt:2018bkr}). 

A sharp answer can be given in a simple, Eulerian-like perturbative setup in which the only relevant parameter is the nonlinear scale $k_{\rm NL}$ and the variance of the density field approximated by a positive power of $k/k_{\rm NL}$ controls the nonlinear expansion. At a given order in perturbation theory the same nonlinear terms are used to make predictions both at the field level and for the $n$-point functions. It is then easy to argue that the exact same information is contained in both, provided that they are calculated at the same order. We will confirm this basic expectation using explicit calculations. However, the true nonlinear evolution in our universe is more complicated and there are effects that can change this simple picture. For example, loops in perturbation theory can lead to appearance of new, potentially large parameters, not controlled by the variance of the density field. Such large parameters usually appear due to the averaging over the couplings of long-wavelength and short-wavelength modes and it is in such cases that the field level inference can be superior to the analysis based on correlation functions. 

The most well-known example are large displacements generated by the long-wavelength density fluctuations, where the large parameter is the velocity dispersion. The long modes in this example displace the short-scale fluctuations by a large amount from their initial Lagrangian positions, significantly affecting the nonlinear density field. Averaging over these large displacements when computing the two-point function famously leads to the broadening of the BAO peak~\cite{Crocce:2007dt,Padmanabhan:2008dd,Sugiyama:2013gza}. This can substantially degrade the measurement of the BAO scale, one of the most important cosmological parameters inferred from galaxy surveys. On the other hand, assuming that the effects of large displacements are properly taken into account, the forward modeling can recover the full linear theory information about the position of the BAO peak, as demonstrated in~\cite{Babic:2022dws}. This is not a surprise, since the same knowledge of the long-short couplings at the map level is exploited in the BAO reconstruction algorithms, used in practice to sharpen the BAO peak and improve measurement of its position~\cite{Eisenstein:2006nk,Padmanabhan:2008dd,Schmittfull:2017uhh}.

In this paper we explore less-known examples of different type where averaging over the long-short interactions leads to degradation of errors rather than dilution of signal. Also in such cases the field-level inference can be more optimal than the conventional analysis based on the $n$-point functions. We identify new large parameters associated to each of these examples and show that they are related to the variance of powers of the density contrast $\delta$. For instance, depending on the shape of the linear power spectrum, the variance of~$\delta^2$ on large scales can be very large, even if the variance of~$\delta$ is small. In the analyses using $n$-point functions, these large parameters can lead to sizable contributions to the covariance matrices, impacting the inference of all cosmological parameters. For example, long modes can modulate the short-scale fluctuations, which can lead to a large scatter in the power spectrum on small scales once the average over the long modes is taken~\cite{Rimes:2005xs,Hamilton:2005dx,Sefusatti:2006pa,Mohammed:2016sre,Barreira:2017kxd}. As we will see, this contribution to the small scale covariance matrix is exactly controlled by the variance of~$\delta^2$. The long-short couplings can impact the covariance on large scales as well. The short modes can couple (through the nonlinear bias) to produce the long-wavelength field of biased tracers\footnote{Note that this is not true for dark matter, since the mass and momentum conservation imply that the long modes produced by interactions of the short modes are suppressed by $k^2/q^2$, where $\k$ is a wavenumber of the long mode and $\q$ a typical wavenumber of the short modes. This can be checked explicitly in perturbation theory. For instance, in the limit $\q_1\approx-\q_2$ the $F_2(\q_1,\q_2)$ kernels scale as $k^2/q_1^2$, where $\smash{\k=\q_1+\q_2}$. } with flat power spectrum. Once the short modes are averaged over, their contribution to the power spectrum and the covariance matrix (controlled again by the parameter related to the variance of~$\delta^2$) is indistinguishable from the shot noise. However, depending on the observed sample, the amplitude of this noise can be much larger than the Poisson prediction, producing larger error bars than expected. Unlike in the standard analyses, none of these issues impact cosmological inference at the field level, where all relevant long-shot couplings are automatically computed and taken into account.  

One important feature of these examples is that unexpectedly large contributions to the covariance do not come from the fully nonlinear regime, but rather from couplings of long and short modes. Such interactions are usually easier to compute in perturbation theory or measure in simulations. Given this, instead of doing the full forward modeling, one can look for simple modified estimators for cosmological parameters which have nearly optimal variance. This is similar to the case of local primordial non-Gaussianity in the cosmic microwave background (CMB) where the optimal estimators for~$\smash{f_{\rm NL}^{\rm local}}$ have to take into account the realization of the long modes in the survey~\cite{Creminelli:2006gc}. In many aspects we follow the logic of~\cite{Creminelli:2006gc} and adopt it to relevant cases in large-scale structure. One important result of our analysis is that the perturbatively-calculated posterior can be used to motivate the form of optimal estimators and we will show a couple of examples that illustrate this point. 

The paper is structured as follows: in Section~\ref{sec:from_likelihood_to_posterior-1} we derive perturbative expressions for errors on cosmological parameters at the field level (with and without marginalization over linear bias); in Section~\ref{sec:b1_marginalization} we compare power spectrum plus bispectrum to forward modeling; in Section~\ref{sec:beyond_simple_PT} we discuss three cases where additional parameters beside the variance of the density field are present in the game and consequently the field-level analysis can be different that standard ones; we conclude in Section~\ref{sec:conclusions}. Appendix~\ref{app:from_likelihood_to_posterior-2} collects some non-perturbative results beyond what we derive in Section~\ref{sec:from_likelihood_to_posterior-1}; Appendix~\ref{app:shot_noise_renormalization} shows how to perturbatively include a finite shot noise at the field level.

\section{From likelihood to posterior -- perturbative inversion}
\label{sec:from_likelihood_to_posterior-1}

\noindent In this section we show how to arrive at the expression for the full posterior and the Fisher matrix for cosmological parameters at the field level, in the limit of small noise. We apply this formula to the case where the only parameter in the theory is the nonlinear scale $k_{\rm NL}$ and derive explicit expressions for the posterior and the Fisher matrix at the one-loop order.

\subsection{Posterior in the limit of small noise}
\label{subsec:posterior_without_noise}

\noindent Let us imagine that the forward model for the nonlinear galaxy field $\deltaNL$ is known in terms of the linear density field $\delta$. We collect cosmological and bias parameters in $\param$. 
The forward model is then given by 
\begin{equation}
\label{eq:posterior_without_noise-1}
\deltaNL = \deltaNL[\delta,\param] + \epsilon_g 
\end{equation} 
for a given noise field $\epsilon_g$.
On large scales we can approximate the likelihood of the galaxy density field as a Gaussian 
\begin{equation}
\label{eq:posterior_without_noise-2}
{\cal L}[\hat{\delta}_g|\delta,\vec{\theta}] = {\rm normalization}\times\exp\bigg({-\frac{1}{2}}\int_{\k}\frac{|\hat{\delta}_g(\k) - \deltaNL[\delta,\param](\k)|^2}{P_{\epsilon}}\bigg)\,\,, 
\end{equation} 
where $P_{\epsilon}$ is the noise power spectrum. This form of the likelihood can be rigorously justified in the perturbative framework that we are going to use throughout the paper. The fiducial galaxy field $\hat{\delta}_g$ is given in terms of fiducial values of the initial field $\hat\delta$, fiducial values of parameters $\hat\param$ and a fiducial noise realization $\hat{\epsilon}_g$ by 
\begin{equation}
\label{eq:posterior_without_noise-3}
\hat{\delta}_g = \deltaNL[\hat{\delta},\vers{\theta}] + \hat{\epsilon}_g\,\,.
\end{equation} 
Note that $\delta$ is an independent variable that does not depend in $\param$, even though it is drawn from a Gaussian distribution with the variance $P(k,\param)$. The normalization in \EQ{posterior_without_noise-2} is such that 
\begin{equation}
\label{eq:posterior_without_noise-4}
\int{\cal D}\hat{\delta}_g\,{\cal L}[\hat{\delta}_g|\delta,\vec{\theta}] = 1\,\,. 
\end{equation}

Ultimately, we are interested in the posterior for cosmological parameters given some realization of the observed galaxy density field. This posterior is obtained by integrating the likelihood multiplied by the prior on initial conditions $\delta$. We will assume that $\delta$ has a Gaussian distribution with the variance given by the linear power spectrum $P(k)$. Defining $p(\param)$ to be the prior on cosmological parameters, the posterior is expressed as
\begin{equation}
\label{eq:posterior_without_noise-5}
{\cal P}[\param|\hat \delta_g] = {\rm normalization}\times\int{\cal D}\delta\,\exp\bigg({-\frac{1}{2}}\int_{\k}\frac{|\delta(\k)|^2}{P(k)} - \frac{1}{2}\int_{\k}\frac{|\hat{\delta}_g(\k) - \deltaNL[\delta,\param](\k)|^2}{P_{\epsilon}}\bigg) \times p(\param) \,\,, 
\end{equation} 
where we have now included the normalization of the prior in the overall factor. In what follows we will set $p(\param)=1$ for simplicity and having in mind situations in which constraints on cosmological parameters are dominated by the data. This choice does not change any of our conclusions and if needed the effect of the prior $p(\param)$ can be straightforwardly included in all our equations. Note that, contrary to our starting point, the power spectrum $P(k)$ does not depend on $\param$. It is easy to see why there is no loss of generality in making this assumption. Indeed, consider the case in which $\smash{P = P(k,\vec{\theta})}$, which is the case of interest in cosmology. It is then possible to perform the following change of variables $\smash{\delta(\vec{k})\to\delta(\vec{k})/P^{1/2}(k,\vec{\theta})}$. In terms of the new integration variables the prior is a normalized Gaussian with unit power spectrum, and all dependence of parameters is in the forward model. In the rest of the paper we will not need such a drastic change of variables. It is enough to use 
\begin{equation}
\label{eq:posterior_without_noise-6}
\text{$\delta(\k)\to\tau(k,\vec{\theta})\,\delta(\k)\,\,,$ \quad with \quad $\tau(k,\vec{\theta}) = \frac{{\cal M}(k,\vec{\theta})}{{\cal M}(k,\vers{\theta})}\,\,,$} 
\end{equation} 
where ${\cal M}(k,\vec{\theta})$ is the linear transfer function which relates the primordial potential to $\delta$ (for the amplitude of the linear field $A$, we simply have $\smash{\tau(k,\vec{\theta}) = A/\hat{A}}$). In this way the prior for the new $\delta$ field is a normalized Gaussian with power spectrum equal to the fiducial linear power spectrum, which exactly agrees with~\EQ{posterior_without_noise-5}. 

The main difficulty in calculating the posterior is to carry out the integral in \EQ{posterior_without_noise-5}. In general, this cannot be done analytically and one has to rely on numerical sampling of the likelihood.\footnote{A simple analytical solution exists only if the forward model is linear in the initial conditions. In this case the integrand is Gaussian in $\delta$ and the integral can be solved to obtain a well-known expression for the posterior in linear theory. We also refer to \cite{Seljak:2017rmr} for a discussion of how to carry out the path integral by expanding around a saddle point found numerically. Ref.~\cite{McQuinn:2020yes} instead discusses the saddle point in the high noise limit.} In most of the paper we will focus on the limit of small noise, relevant for dense spectroscopic samples on large scales. In this case one can expand the posterior in $P_\epsilon$ which simplifies calculations significantly. For the time being we focus only on the leading order in the $P_\epsilon\to0$ limit, in which the posterior becomes 
\begin{equation}
\label{eq:posterior_without_noise-7}
{\cal P}[\param|\hat \delta_g] = {\rm normalization}\times\int{\cal D}\delta\,\exp\bigg({-\frac{1}{2}}\int_{\k}\frac{|\delta(\k)|^2}{P(k)}\bigg)\,\delta^{(\infty)}_{\rm D}\Big(\hat{\delta}_g - \deltaNL[\delta,\param]\Big)\,\,, 
\end{equation} 
where the normalization is now only that of the prior and, in the same way as $P(k)$, it does not depend on $\param$ (hence we will drop it from now on to keep the notation as contained as possible). We leave the discussion of higher orders in $P_\epsilon$ for Section~\ref{sec:disconnected_diagrams}. In order to exploit the delta function in the integrand of the posterior, we can do the following change of variables $\delta\to\deltaNL$. The posterior can be then witten as
\be
\label{eq:posterior_without_noise-8}
\mathcal P[\param|\hat \delta_g] = \int \mathcal D \deltaNL \left|\frac{\partial \delta}{\partial \deltaNL} \right|\exp\bigg({-\frac{1}{2}}\int_{\k}\frac{|\delta[\deltaNL,\param](\k)|^2}{P(k)}\bigg)\,\delta^{(\infty)}_{\rm D}(\hat{\delta}_g - \deltaNL) \equiv\eu^{{\rm Tr}\ln J[\hat\delta_g,\param] - \frac 12 \chi^2_{\rm prior}[\hat\delta_g,\param]}\,\,,
\ee
where the two terms in the final result, one coming from the prior and the other one coming from the Jacobian, are denoted by $\chi^2_{\rm prior}$ and $J$ respectively. More explicitly
\be
\chi^2_{\rm prior}[\hat\delta_g,\param] \equiv \int_{\k}\frac{|\delta[\hat\deltaNL,\param](\k)|^2}{P(k)} \;, \qquad {\rm and } \qquad J[\hat\delta_g,\param] \equiv \left|\frac{\partial \delta[\deltaNL,\param]}{\partial \deltaNL} \right|_{\delta_g=\hat\delta_g} \,\,.
\ee

As expected, the final result depends on the realization of the galaxy density field $\hat\delta_g$ and parameters $\param$.
The key ingredient needed to find the posterior $\smash{\mathcal P[\param|\hat \delta_g]}$ is the inverse of the forward model $\smash{\delta[\deltaNL,\param]}$, which allows us to compute $\smash{\chi^2_{\rm prior}}$ and $J$. Finding $\smash{\delta[\deltaNL,\param]}$ is in general a very difficult task. However, in perturbative forward modeling the inverse model is also perturbative and can be calculated analytically. The full posterior can be then consistently computed up to a given power of the variance of the density field, which resembles the more familiar loop expansion for correlation functions. Such posterior can be used to do cosmological inference {\em without} the need to run MCMC, and on large scales it is guaranteed to lead to optimal constraints on cosmological parameters. We will show a few explicit perturbative examples throughout the paper. 

Before proceeding, let us add two more comments about~\EQ{posterior_without_noise-8}. First, we are assuming that the change of variables is one-to-one: in other words, we consider only the saddle point in the likelihood connected to linear theory. Other solutions will be present if the forward model is pushed to short scales (e.g.~due to shell crossing, see Ref.~\cite{Feng:2018for} for a discussion), but on large scales the assumption of a single solution is correct. The second observation regards the change of variables itself: we invert $\smash{\delta=\deltaNL^{-1}[\deltaNL,\param]}$ for varying $\param$. Indeed, the Dirac delta in \EQ{posterior_without_noise-7} is a Dirac delta in a space of dimension equal to the number of Fourier modes of the linear field (as is clear from the expression of the Gaussian likelihood, which involves an integral in ${\rm d}^3k$). Hence, even if $\smash{\delta_g=\hat{\delta}_g}$ we do \emph{not} obtain $\smash{\delta=\hat{\delta}}$: we do so only if $\smash{\param=\hat{\param}}$. The procedure is the same as what done in Ref.~\cite{Creminelli:2006gc}, in which similar calculations were carried out in the context of constraints on local primordial non-Gaussianity from higher-order statistics of the CMB.

\subsection{Posterior in the perturbative forward model}
\label{subsec:perturbative_forward_and_inverse-1}

\noindent In order to derive explicit expression for the posterior~$\mathcal P[\param|\hat \delta_g]$, we will assume that the perturbative forward model can be written as 
\be
\label{eq:perturbative_forward_model}
\deltaNL(\k) = \sum_{n=1}^{+\infty} \int_{\p_1,\dots,\p_n} \dirac{\k-\p_{1\cdots n}}\,X_{n}(\param; \p_1, \dots, \p_n)\,\delta(\p_1) \cdots \delta(\p_n) \equiv\sum_{n=1}^{+\infty}\deltaNL^{(n)}(\k) \,\,,
\ee
where $X_n$ are perturbation theory kernels. Note that the whole dependence on cosmological parameters is in the kernels~$X_n$. For example, this is the form of the nonlinear density field in Eulerian perturbation theory and for biased tracers. Since we assume that the only parameter in the theory is the variance of the density field, there are no large displacements in this ansatz. This can be achieved in practice by the appropriate choice of the power spectrum for which the velocity dispersion is small. 

With our assumptions the forward model can be inverted perturbatively on large scales, i.e.~
\be
\label{eq:backwards_model}
\delta(\k) = \sum_{n=1}^{+\infty} \int_{\p_1,\dots,\p_n} (2\pi)^3 \delta^{(3)}_{\rm D}(\k-\p_{1\cdots n} ) Y_{n}(\param; \p_1, \dots, \p_n) \deltaNL(\p_1) \cdots \deltaNL(\p_n) \equiv \sum_{n=1}^{+\infty} \Delta_g^{[n]} (\k) \,\,,
\ee
where the $Y_n$ kernels can be calculated in term of the original nonlinearities $X_n$. Up to cubic order, they are given by 
\begin{subequations}
\label{eq:formulas_for_inverse_kernels}
\begin{align}
Y_1(\param) & = X_1^{-1}(\param) \,\,, \label{eq:formulas_for_inverse_kernels-1} \\
Y_2(\param;\p_1,\p_2) & = {-X_1^{-3}(\param) X_2(\param;\p_1,\p_2)}\,\,, \label{eq:formulas_for_inverse_kernels-2} \\ 
Y_3(\param;\p_1,\p_2,\p_3) & = {\frac23 X_1^{-5}}(\param) \bigg[ X_2(\param;\p_1,\p_2+\p_3) X_2(\param;\p_2,\p_3) + X_2(\param;\p_2,\p_1+\p_3) X_2(\param;\p_1,\p_3) \nonumber \\
&\;\;\;\,\,\,\hphantom{{- X_1^{-1}} \bigg[ } + X_2(\param;\p_3,\p_1+\p_2) X_2(\param;\p_1,\p_2) - \frac 32 X_1(\param) X_3(\param; \p_1,\p_2,\p_3) \bigg] \,\,. \label{eq:formulas_for_inverse_kernels-3}
\end{align}
\end{subequations} 
Notice that in all the examples discussed in this paper we will consider only multiplicative parameters, for which the transfer functions, and hence the kernels $\smash{X_1}$ and $\smash{Y_1}$, are scale-independent. In more general cases, one has to keep track of the ratio of transfer functions as in \EQ{posterior_without_noise-6} when deriving the inverse kernels $\smash{Y_n}$. The inverse model defines simple operations on the galaxy density field (order by order in $\deltaNL$) which ensures the optimal combination of the data in order to recover the linear modes. At the few leading orders they are explicitly given by
\begin{subequations}
\begin{align}
\Delta_g^{[1]} (\k) & = Y_1(\param) \deltaNL(\k)  \,\,, \\
\Delta_g^{[2]} (\k) & = \int_{\p} Y_2(\param;\k-\p,\p) \deltaNL(\p)\deltaNL(\k-\p) \,\,,  \\ 
\Delta_g^{[3]} (\k) & = \int_{\p_1,\p_2} Y_3(\param;\k-\p_1-\p_2,\p_1,\p_2) \deltaNL(\p_1)\deltaNL(\p_2)\deltaNL(\k-\p_1-\p_2) \,\,,
\end{align}
\end{subequations} 
and they are simple convolutions of the data. Let us stress again that the whole dependence on parameters~$\param$ is in the kernels~$Y_n$. 

It is important to point out that in order to apply these formulas we will assume that the inverse model is valid up to the similar scale as the forward model. This is not a trivial assumption. To get some intuition about possible differences between these two scales, we can focus on a simple model of spherical collapse. In this case perturbative forward model can be written as\footnote{The usual perturbative expansion of the exact spherical collapse solution is in powers of the growth factor. We convert it here into the expansion in the linear density field evaluated at late times, such that $\delta_\ell= a \, \delta_{\rm initial}$. At the time of collapse, the critical linear overdensity is approximately $\delta_\ell(a_{\rm collapse})=\delta_{\rm cr}\approx1.68$.}
\be
\delta_{\rm sc} = \sum_{n=1}^\infty \nu_n \delta^n_{\ell} \,\,,
\ee
where $\nu_n$ are spherically averaged perturbation theory kernels and $\delta_{\ell}$ is the size of the spherical linear overdensity or underdensity in real space evolved to the present time. It is well-known that this series converges to the exact solution for $|\delta_\ell|<\delta_{\rm cr}\approx1.68$. While for $0<\delta_\ell<\delta_{\rm cr}$ the nonlinear overdensities $\delta_{\rm sc}$ can be arbitrarily large, for $\delta_\ell<0$ this series converges to the true answer only for voids with $\delta_{\rm sc}[-\delta_{\rm cr}]\gtrsim -0.7$. Let us now turn to the inverse perturbative model, which can be written as
\be
\delta_{\ell} = \sum_{n=1}^\infty c_n \delta_{\rm sc}^n \,\,,
\ee
where the coefficients $c_n$ can be derived from $\nu_n$. This series converges to the true linear solution for $|\delta_{\rm sc}|\lesssim 1$, which corresponds to any $\delta_\ell\lesssim 0.5$. Note that the inverse perturbative solution correctly predicts $\delta_\ell$ even for very empty voids, which the forward model cannot describe. We can see that the convergence properties for the forward and the inverse models in this simple example are rather different. 

However, in practice we are interested in the nonlinear model which has only a small number of terms in the perturbative expansion. Validity of such model is even more restricted compared to the full perturbative series, by the requirement that the variance of the density field is small and of order $\langle \delta_\ell^2 \rangle \lesssim \mathcal O(0.1)$. Using the spherical collapse as a toy example, we can explicitly check that with this requirement the forward and inverse models perform similarly. Up to cubic order we can write
\be
\delta_{\rm sc}^{[3]} = \delta_\ell+ \frac{17}{21} \delta_\ell^2 + \frac{341}{567} \delta_\ell^3 \,\,. 
\ee
Note that this is just the spherical average of our cubic forward model in~\EQ{perturbative_forward_model}, assuming the standard dark matter perturbation theory kernels instead of $X_n$. The inverse model is then given by
\be
\delta_\ell^{[3]} = \delta_{\rm sc} - \frac{17}{21} \delta_{\rm sc}^2 + \frac{2815}{3969} \delta_{\rm sc}^3 \,\,.
\ee
It can be verified explicitly that both forward and inverse model reproduce the correct answer to better than $5\%$, for $|\delta_\ell|\lesssim0.4$ and $|\delta_{\rm sc}|\lesssim0.4$ respectively. We use this as an indication that the inversion is valid to the similar scale as the forward model. The full check of this claim with the realistic 3D fields can be done only using numerical simulations. While the details can be different due to the mode coupling, we do not expect the conclusions to change dramatically, as long as the field-level analysis is performed on scales $k_{\rm max} \lesssim 0.1\; h/{\rm Mpc}$. 

With the forward and inverse models at hand, we can compute the prior and Jacobian contributions to the full posterior. In this paper we want to work at order equivalent to the one-loop calculations of the two-point function, and therefore we have to keep all nonlinearities up to cubic order. Note that this is equivalent to working with the one-loop power spectrum and the tree-level bispectrum in the conventional correlation functions approach. 

Let us start from the prior term $\chi^2_{\rm prior}$. Using the definitions from the previous section, it immediately follows that $\smash{\chi^2_{\rm prior}}$ can be written as the integral over various cross spectra of $\smash{\hat\Delta_g^{[n]}}$. Up to cubic order in the observed galaxy density field we get
\be
\label{eq:chi_squared_prior}
\chi^2_{\rm prior} = \int_{\k} \frac{\hat\Delta_g^{[1]}(\k) \hat\Delta_g^{[1]}(-\k) + 2\hat\Delta_g^{[2]}(\k) \hat\Delta_g^{[1]}(-\k) + \hat\Delta_g^{[2]}(\k) \hat\Delta_g^{[2]}(-\k) + 2\hat\Delta_g^{[3]}(\k) \hat\Delta_g^{[1]}(-\k) }{P(k)} \,\,.
\ee
As expected, this is a function of the data $\smash{\hat\delta_g}$ and parameters $\param$ through the kernels $Y_n$. In order to calculate the Jacobian we have to take the derivative of the inverse model with respect to $\deltaNL$ first and then set $\smash{\deltaNL=\hat\delta_g}$. In our setup this is simply given by
\be
\label{eq:jacobian_master_definition}
\begin{split}
J(\k,\k') &= {\underbrace{Y_1(\param)\,\dirac{\k-\k'}}_{\hp{J^{(0)}(\k,\k')\,}\equiv\,J^{(0)}(\k,\k')}} \\
&\;\;\;\; + \sum_{n=1}^{+\infty} (n+1) \int_{\p_1,\dots,\p_n} \dirac{\k-\k'-\p_1 - \cdots - \p_n}\,Y_{n+1}(\param; \p_1, \dots, \p_n,\k')\, \hat \delta_g(\p_1) \cdots \hat \delta_g (\p_n) \,\,.
\end{split}
\ee
Note that in the expression for $\chi^2_{\rm prior}$, cf.~\EQ{chi_squared_prior}, we have $P(k)$ in the denominator. This tells us that we have to keep only terms up to second order in $\smash{\hat \delta_g}$ in the Jacobian to have correct expressions at one-loop order. Ultimately, we are interested in ${\rm Tr} \ln J$ since this quantity appears in the log-posterior. Taking the logarithm and keeping all terms at second order, we find 
\be
\label{eq:jacobian_expression}
\begin{split}
[\ln\,& J(\k,\k') ]^{\text{$1$-loop}} = \ln J^{(0)}(\k,\k') + 2\int_{\p} \dirac{\k-\k'-\p}\,Y_1^{-1}(\param) Y_2(\param;\p,\k')\,\hat\delta_g (\p) \\ 
&\;\;\; + 3\int_{\p_1,\p_2} \dirac{\k-\k'-\p_1-\p_2}\, Y_1^{-1}(\param) Y_3(\param;\p_1,\p_2,\k') \, \hat\delta_g (\p_1) \hat\delta_g (\p_2) \\ 
&\;\;\; - 2 \int_{\p_1,\p_2,\k''}\dirac{\k-\k''-\p_1}\dirac{\k''-\k'-\p_2}\,Y_1^{-2}(\param) Y_2(\param;\p_1,\k'') Y_2(\param;\p_2,\k')\, \hat{\delta}_g(\p_1) \hat{\delta}_g(\p_2)\,\,. 
\end{split}
\ee 
We use the convention in which the ``$1$-loop'' label indicates that all terms up to one-loop order are taken into account. Finally, taking the trace leads to
\be
\label{eq:trace_of_jacobian}
\begin{split}
[{\rm Tr} \ln J[\hat\delta_g,\param]]^{\text{$1$-loop}} = N_{\rm pix}\ln Y_1(\param) & + 3\, Y_1^{-1}(\param) \int_{\k,\p} Y_3(\param;\p,-\p,\k) \hat\delta_g (\p) \hat\delta_g (-\p) \\
& - 2\, Y_1^{-2}(\param) \int_{\k,\p} Y_2(\param;\p,\k-\p) Y_2(\param;-\p,\k) \hat\delta_g (\p) \hat\delta_g (-\p) \,\,, 
\end{split}
\ee 
where we have defined the number of pixels\footnote{Notice that this is not exactly equal to $k^3_{\rm max}/k^3_{\rm min}$ as it would be in a box: we have $N_{\rm pix} = (4\pi/3)(k^3_{\rm max}/k^3_{\rm min})\approx 4k^3_{\rm max}/k^3_{\rm min}$.} as $\smash{N_{\rm pix}\equiv V\int_{\k}}$. 

In summary, we have derived the negative log-posterior for the perturbative forward model, in the limit of small noise and analytically marginalizing over the initial field $\delta$. This can be written as
\be
\label{eq:full_posterior}
-\log \mathcal P[\param|\hat \delta_g] = \frac 12 \chi^2_{\rm prior}[\hat\delta_g,\param] - {\rm Tr}\ln J[\hat\delta_g,\param] \,\, ,
\ee
where the prior and the Jacobian terms are given by by~\EQ{chi_squared_prior} and~\EQ{trace_of_jacobian} respectively. Our final result depends only on the observed galaxy density field $\hat\delta_g$, convolved with the kernels of the inverse model $Y_n$. In our convention, these kernels contain the {\em entire} dependence on cosmological and nuisance parameters $\param$, as well as the nonlinear dynamics. This posterior can be used for data analysis or to derive optimal estimators for a given cosmological parameter, and we will show several examples of this in the rest of the paper. Finally, one can also calculate the {\em averaged} log-posterior and the corresponding Fisher matrix, given some fiducial galaxy density field $\hat \delta_g$. We turn to that in the next section.

\subsection{Fisher matrix in the perturbative forward model}
\label{subsec:perturbative_forward_and_inverse-2}

\noindent In order to derive Fisher matrix for the perturbative forward modeling, we first have to compute the averaged negative log-posterior
\be
\label{eq:minus_averaged_log_posterior} 
\big\langle{-\ln \mathcal{P}[\param|\hat \delta_g]}\big\rangle = \frac 12 \big\langle \chi^2_{\rm prior}[\hat\delta_g,\param] \big\rangle - \big\langle {\rm Tr}\ln J[\hat\delta_g,\param] \big\rangle \,\,, 
\ee 
where the average is done over the fiducial initial conditions. We average the log-posterior assuming that the error on cosmological parameters does not vary a lot between realizations of the initial conditions. In the language of statistics, we are looking at the ``Cram{\'e}r-Rao'' bound. 
The average of $\chi^2_{\rm prior}$ can be written as
\be
\label{eq:avg_chi_squared_prior}
\langle \chi^2_{\rm prior} \rangle = V\int_{\k} \frac{Y_1^2(\param) \hat P_g(k) + P_{g,12}(\param;k) + P_{g,22}(\param;k) + P_{g,13}(\param;k) }{P(k)} \,\,,
\ee
where we have used $V = \dirac{\vec{0}}$, $\hat P_g(k)$ is the measured galaxy power spectrum and $P_{g,12}(\param;k)$, $P_{g,22}(\param;k)$ and $P_{g,13}(\param;k)$ are defined in terms of higher order correlation functions in the following way:
\begin{subequations} 
\label{eq:three_and_four_point_functions}
\begin{align}
P_{g,12}(\param;k) & \equiv 2Y_1(\param) \int_{\p} Y_2(\param;\p,\k-\p) \, \langle \hat\delta_g(\p) \hat\delta_g(\k') \hat\delta_g(\k-\p) \rangle' \,\,, \label{eq:chi_squared_prior_help} \\
P_{g,22}(\param;k) & \equiv \int_{\p_1,\p_2} Y_2(\param;\p_1,\k-\p_1) Y_2(\param;\p_2,-\k-\p_2) \, \langle \hat\delta_g(\p_1) \hat\delta_g(\k-\p_1) \hat\delta_g(\p_2) \hat\delta_g(\k'-\p_2) \rangle' \,\,, \label{eq:chi_squared_prior_help-bis} \\
P_{g,13}(\param;k) & \equiv 2 Y_1(\param) \int_{\p_1,\p_2} Y_3(\param;\p_1,\p_2,\k-\p_1-\p_2) \, \langle \hat\delta_g(\p_1) \hat\delta_g(\p_2) \hat\delta_g(\k-\p_1-\p_2) \hat\delta_g(\k') \rangle' \,\,. 
\end{align} 
\end{subequations} 
In these expressions, the prime on the $n$-point functions of the observed galaxy density field $\langle \hat\delta_g \cdots \hat\delta_g \rangle'$ indicates that the overall factor of $\smash{\dirac{\k+\k'}}$ should be removed from the final result. Note that these correlation functions include the disconnected pieces. In the Fisher matrix calculation we will assume that the observed galaxy $n$-point functions are evaluated at the fiducial initial conditions and the fiducial parameters $\hat\param$. Already from this part of the averaged log-posterior we can see that the Fisher matrix will depend on the higher order $n$-point functions, appropriately combined with the kernels of the inverse model. 

It is instructive to write down explicitly contributions to the prior at leading order in perturbation theory, which is sufficient for the $1$-loop forward model. Keeping the tree-level bispectrum and disconnected parts of the $4$-point function and using kernels for the forward model, we get
\begin{subequations} 
\begin{align}
\hat P_{g}^{\text{$1$-loop}}(k) & = X_1^2(\hat\param) P(k) + 2 \int_{\p} X_2^2(\hat\param;\p,\k-\p) P(p)P(|\k-\p|) + 6 X_1(\hat\param)P(k) \int_{\p} X_3(\hat\param;\p,-\p,\k) P(p) \,\,, \\
P_{g,12}^{\text{$1$-loop}}(\param;k) & = 4Y_1(\param) X_1^2(\hat\param) \int_{\p} Y_2(\param;\p,\k-\p) \, \left( 2X_2(\hat\param;\p,-\k) P(p) P(k) + X_2(\hat\param;\p,\k-\p)P(p)P(|\k-\p|) \right) \,\,, \\
P_{g,22}^{\text{$1$-loop}}(\param;k) & = 2 X_1^4(\hat\param) \int_{\p} Y_2^2(\param;\p,\k-\p) \, P(p)P(|\k-\p|) \,\,, \\
P_{g,13}^{\text{$1$-loop}}(\param;k) & = 6 Y_1(\param) X_1^4(\hat\param) P(k) \int_{\p} Y_3(\param;\p,-\p,\k) \, P(p) \,\,. 
\end{align} 
\end{subequations} 
Note that we do not include those contractions that lead to the results proportional to $\smash{\dirac{\k}}$, which contribute only to the unobservable zero mode. At leading order in perturbation theory the fiducial galaxy power spectrum was evaluated as~$\hat P_g(k)= X_1^2(\hat\param)P(k)$. We are going to use these equations to evaluate the Fisher matrix for the forward model for biased tracers in real space, up to $1$-loop order in perturbation theory. We next turn to evaluating $\langle \ln\det{J} \rangle = \langle {\rm Tr} \ln J\rangle $. Using results of the previous section, we get 
\be
\label{eq:avg_trace_of_jacobian}
\begin{split}
\langle {\rm Tr} \ln J\rangle^{\text{$1$-loop}} = N_{\rm pix}\ln Y_1(\param) & + 3V\, X_1^2(\hat\param) Y_1^{-1}(\param) \int_{\k,\p} Y_3(\param;\p,-\p,\k) P(p) \\
& - 2V\, X_1^2(\hat\param) Y_1^{-2}(\param) \int_{\k,\p} Y_2(\param;\p,\k-\p) Y_2(\param;-\p,\k) P(p) \,\,.
\end{split}
\ee  

This completes our derivation of the averaged log-posterior at one-loop and one can proceed by calculating the Fisher matrix. Before we do that, let us make two comments. First, note that a nontrivial check of the previous equations is unbiasedness. If the formulas are correct, then the derivative of the averaged log posterior at the fiducial values of parameters must be zero. Calculating this derivative explicitly and using~\EQ{formulas_for_inverse_kernels-1} to~\EQ{formulas_for_inverse_kernels-3}, we find
\be
\frac{\partial}{\partial \param} \langle {-\ln \mathcal{P} } \rangle^{\text{$1$-loop}} \bigg|_{\param=\hat{\param}} = \frac{2V}{Y_1^4(\hat \param)} \int_{\k,\p} \left( Y_2(\hat\param;\p,\k-\p) \frac{\partial}{\partial \param} Y_2(\param;-\p,\k) - Y_2(\hat\param;-\p,\k) \frac{\partial}{\partial \param} Y_2(\param;\p,\k-\p) \right) \bigg|_{\param=\hat{\param}} P(p) \,\,.
\ee
After a simple change of variables $\k-\p\to -\k$ in one of the terms under the integral, the right hand side vanishes. This implies that the estimate of the cosmological parameters at the field level is indeed unbiased. Note that in order be able to do the change of variables, it is crucial that both integrals run over all possible values of the momenta. On the other hand, we always have some maximum wavenumber up to which we can trust perturbation theory. In order to preserve the unbiasedness and implement this cutoff in practice, one can always apply the window function on the power spectrum, rather than changing the limits of integration. 

The second comment is about higher orders in perturbation theory. 
So far we have discussed only the leading nonlinearities, but it is important to stress how a clear loop expansion in $\smash{\langle{-\ln \mathcal{P}}\rangle}$ arises if one wants to go further. The negative log-posterior is expressed in terms of the fiducial initial conditions via a combination of \EQ{perturbative_forward_model} and \EQ{backwards_model}. Schematically, the solution of $\smash{\delta=\deltaNL^{-1}[\deltaNL,\param]}$ can be written in terms of the fiducial linear field $\smash{\hat 
\delta}$ as follows
\be
\begin{split}
\delta = \sum_{m=1}^{+\infty}Y_m(\param) \Bigg(\sum_{n=1}^{+\infty}{X}_n(\hat{\param})\hat{\delta}^n\Bigg)^m = \sum_{k=1}^{+\infty}Z_k(\param,\hat{\param})\hat{\delta}^k \,\,, 
\end{split}
\ee 
where, by construction, the new kernels $Z_k(\param,\hat{\param})$ satisfy the following properties
\begin{equation}
Z_1(\hat\param,\hat{\param}) = 1 \,\,, \qquad {\rm and} \qquad Z_{k>1}(\hat\param,\hat{\param}) = 0 \,\,. 
\end{equation}
In other words, evaluated at the fiducial values of parameters, the linear field in the inverse model must be equal to the fiducial linear field. One can explicitly check that this is indeed the case for the inverse kernels derived above (at $1$-loop order). Using \EQ{jacobian_master_definition}, a similar expression in terms of $\smash{\hat\delta}$ can be found for the Jacobian as well. Therefore, the negative log-posterior is naturally organized as a perturbative series in the fiducial linear density field $\smash{\hat\delta}$. It follows, that the expectation value $\smash{\langle{-\ln \mathcal{P}}\rangle}$ has a clear loop expansion, the same one as in the standard perturbation theory. To evaluate the averaged log-posterior at the given order in the loop expansion, only a finite number of terms in the forward and inverse model have to be kept. Note, however, that due to the linear power spectrum in the denominator in the expression for $\smash{\langle \chi^2_{\rm prior} \rangle}$ and the functional derivative in the Jacobian, the leading order term in $\smash{\langle{-\ln \mathcal{P}}\rangle}$ starts at zeroth order in $\smash{P(k)}$. This is expected, since in the linear theory, as we will see shortly, the posterior depends only on the number of pixels. 

With all these results at hand, it is straightforward to calculate the Fisher matrix $F$, given some fiducial galaxy density field $\smash{\hat \delta_g}$. In practice, we evaluate
\be
\label{eq:field_Fisher_1loop}
F^{\text{$1$-loop}}_{\alpha\beta} =
\frac{\partial^2}{\partial \theta_\alpha \partial \theta_\beta} \left( \frac 12 \big\langle \chi^2_{\rm prior}[\hat\delta_g,\param] \big\rangle^{\text{$1$-loop}} - \big\langle {\rm Tr}\ln J[\hat\delta_g,\param] \big\rangle^{\text{$1$-loop}} \right) \Big |_{\param = \hat \param} \,\,,
\ee
where the first term is given by~\EQ{avg_chi_squared_prior} and the second term by~\EQ{avg_trace_of_jacobian}, both evaluated up to one-loop order, and the expectation value refers to averaging over fiducial initial conditions as discussed under \EQ{minus_averaged_log_posterior}. It is important to point out that this formula makes the connection between perturbative forward modeling and the standard analyses manifest. As long as the variance of the density field is the only relevant parameter, the two approaches are equivalent, order by order in perturbation theory. Before considering more interesting situations in which new, potentially large parameters play an important role, we show how the filed-level Fisher matrix works in practice in several examples of interest.

\section{Applications of the field-level posterior and Fisher matrix }
\label{sec:b1_marginalization}

\noindent In this section we apply~\EQ{full_posterior} and~\EQ{field_Fisher_1loop} and show explicitly in some simple setups that the field-level analysis leads to the same errors on cosmological parameters as the standard analysis based on the $n$-point correlation functions, as long as the only relevant parameter in the theory is the nonlinear scale $k_{\rm NL}$ and the only expansion parameter is the variance of the density field.

\subsection{Simple examples of the field-level Fisher matrix: linear and nonlinear dark matter} 
\label{subsec:biased_tracers_real_space}

\noindent Let us begin with two very simple examples: linear theory and nonlinear dark matter field. In order to keep expressions as clear as possible, we will focus on the amplitude of the linear density field~$A$ as the only cosmological parameter of interest.

Let us first consider the simplest possible scenario, in which evolution of density fluctuations is linear. In this example the kernels of the forward model are trivial
\be
\label{eq:linear_theory-1}
\text{$X_1 = A$ \quad and \quad $X_{n>1} = 0\,\,,$} 
\ee
which implies 
\be
\label{eq:linear_theory-2}
\text{$Y_1 = 1/A$ \quad and \quad $Y_{n>1} = 0\,\,.$} 
\ee
This simplifies the form of the posterior significantly and we can write (note that we keep using $\deltaNL$ for the nonlinear field even though we are not considering galaxies here) 
\be
-\ln \mathcal P[A|\hat \delta_g] = \frac1{2A^2} \int_{\k} \frac{\hat\delta_g(\k) \hat\delta_g(-\k) }{P(k)} + N_{\rm pix} \ln A \,\,.
\ee
Finding the maximum of the posterior, we get the optimal estimator for $A$ 
\be
{\cal E}
= \frac{1}{N_{\rm pix}} \int_{\k} \frac{\hat\delta_g(\k) \hat\delta_g(-\k) }{P(k)} \,\,.
\ee
Not surprisingly, the optimal estimator for the amplitude of fluctuations in linear theory coincides with the estimator of the power spectrum. 
The averaged negative log-posterior is given by 
\be
\langle {-\ln \mathcal{P}} \rangle = \frac{N_{\rm pix}}{2A^2} + N_{\rm pix} \ln A \,\,,
\ee
where we have used $\hat P_g(k) = \hat A^2 P(k)$ and the fiducial value of the amplitude is set to one, $\hat A=1$. In the absence of the nonlinear evolution, $\langle {-\ln \mathcal{P}} \rangle$  depends only on the number of pixels $N_{\rm pix}$. Using this equation one can explicitly show that the estimate of the amplitude is unbiased and that the error is given by the well-known formula for the linear theory\footnote{The reader is perhaps more familiar with the formula for the error of the amplitude of the power spectrum $A_{\rm s}$. Changing variables from $A$ to $A_{\rm s} \equiv A^2$ we find
\be
\sigma_{A_{\rm s}}^2 = \frac{2}{N_{\rm pix}} 
\ee 
as expected.}
\be
\frac{1}{\sigma_A^2} = \frac{\partial^2 \langle {-\ln \mathcal{P}} \rangle}{\partial A^2}\bigg|_{A=\hat{A}} = 2 N_{\rm pix} \,\,.
\ee 

A slightly more nontrivial example is the nonlinear evolution, where the amplitude of the linear field $A$ is still the only unknown parameter. It is easy to see that in this case the kernels are given by
\be
X_n(A) = \frac{A^n}{\hat A^n} X_n(\hat A) \qquad {\rm and} \qquad Y_n(A) = \frac{\hat A}{A} Y_n(\hat A) \,\,.
\ee
Note that all inverse kernels scale as $1/A$. This has two important consequences. One is that $\chi^2_{\rm prior}$ in~\EQ{chi_squared_prior} scales exactly as $1/A^2$. The other is that only the first term in~\EQ{trace_of_jacobian} depends on $A$. The full posterior can be then written as (setting $\hat A=1$)
\be
\label{eq:chi_squared_linear_theory_help}
{-\ln \mathcal P[A|\hat \delta_g]} = \frac 1{2A^2} \int_{\k} \frac{|\delta[\hat\delta_g,\hat A=1](\k)|^2 }{P(k)} + N_{\rm pix} \ln A + (\text{$A$-independent terms}) \,\,.
\ee
The optimal estimator for the amplitude of density fluctuations is 
\be
{\cal E}
= \frac{1}{N_{\rm pix}} \int_{\k} \frac{|\delta[\hat\delta_g,\hat A=1](\k)|^2 }{P(k)} \,\,.
\ee
We can see that in this case the estimator for the amplitude does not depend only on the nonlinear power spectrum, but also various other combinations of data which enter the numerator and which are explicitly written in~\EQ{chi_squared_prior}. However, by definition of the inverse model, when evaluated at the fiducial values of parameters, it gives the initial Fourier modes of~$\hat\delta$ given the observed nonlinear density field. This means that the amplitude of the matter fluctuations can be still optimally measured from the power spectrum only, but of the initial field reconstructed from~$\smash{\hat{\delta}_g}$. Given this, we expect the error on~$A$ to be the same as in the linear theory. Indeed, the averaged log-posterior is given by
\be
\langle {-\ln \mathcal{P}} \rangle = \frac{N_{\rm pix}}{2A^2} + N_{\rm pix} \ln A + (\text{$A$-independent terms}) \,\,,
\ee
which implies the same error as before
\be
\frac{1}{\sigma_A^2} = \frac{\partial^2 \langle {-\ln \mathcal{P}} \rangle}{\partial A^2}\bigg|_{A=\hat{A}} = 2 N_{\rm pix} \,\,.
\ee 
Such result makes sense since for a fixed nonlinear model and in the absence of noise the field-level posterior should contain all available information on cosmological parameters from the data optimally combined. In such ideal setup, this amount of information saturates the bound given by the linear theory. While it is difficult to prove this statement for generic cosmological parameters following the approach of this section, this can be done using the non-perturbative formulation of the posterior at the field level. We defer this general analysis to Appendix~\ref{app:from_likelihood_to_posterior-2}.

Let us make a comment about the derivation above. In the case of imperfect inverse model, we can imagine that the $Y_n$ kernels are different from their expression of \eqsI{formulas_for_inverse_kernels}, but with the same overall scaling with $A$ intact. The argument leading to \EQ{chi_squared_linear_theory_help} behaving as $1/A^2$ still holds, with the crucial difference that now all the nonlinear terms in the numerator under the integral will \emph{not} combine to the reconstruction of the linear field. As a consequence, the momentum integral in \EQ{avg_chi_squared_prior} multiplied by the volume will not lead to $N_{\rm pix}$, and the error bar on $A$ would be different than in the linear theory, as expected if the inversion is wrong.\footnote{Interestingly, things are different for the Jacobian part of $\smash{\langle{-\ln \mathcal{P}}\rangle}$. Even if the $Y_n$ kernels are wrong but their overall scaling with $A$ is correct, $\smash{{- {\rm Tr}\ln J
} = N_{\rm pix}\ln A + (\text{$A$-independent terms})}$ will continue to hold. This is a coincidence for the amplitude of the density fluctuations, and it is not true in general for other cosmological parameters.}

\subsection{Including the linear bias}
\label{subsec:including_the_linear_bias}

\noindent Going beyond these idealized examples, things become more complicated. In the real universe, our ability to measure cosmological parameters depends on their degeneracies in the linear power spectrum, peculiarities of the nonlinear evolution and complexity of galaxy formation which is on large scales encoded in a number of nuisance parameters one must marginalize over. It is interesting to show how the field-level inference boils down to the conventional analysis with $n$-point functions even in this case. 

In the simplest and observationally most relevant setup in which one can still gain some intuition in analytically tractable way, we allow two parameters: the amplitude $A$ of the linear density field and the linear bias $b_1$. We keep the product $Ab_1$ fixed (for simplicity, we assume that $\smash{\hat{A}}$ and $\smash{\hat{b}_1}$ are both equal to~$1$). This is inspired by the fact that the overall amplitude of the power spectrum of biased tracers is usually very well measured on large scales and such assumption does not affect our conclusions. As it is well known, since this particular combination multiplies the linear power spectrum, all information on the amplitude $A$ in this example must come from the nonlinearities. Here we compute the posterior and the Fisher matrix for $A$ in forward modeling. For $Ab_1=1$, perturbation theory kernels scale as
\be
\text{$X_1(A) = 1$ \quad and \quad $X_n(A; \p_1, \ldots, \p_n) \equiv A^{n-1} \; X_n (\p_1, \ldots, \p_n) \,\,,$} 
\ee
and
\be
\text{$Y_1(A) = 1$ \quad and \quad $Y_n (A; \p_1, \ldots, \p_n) \equiv A^{n-1} \; Y_n (\p_1, \ldots, \p_n) \,\,.$} 
\ee
Note that here (and the rest of the paper) the kernels without the explicit dependence on $A$ are evaluated at the fiducial value $\hat A =1$. The explicit form of the posterior for $A$ is a bit more complicated in this example. The prior part is given by
\begin{equation}
\label{eq:prior_part_b1A_fixed}
\begin{split}
\frac 12 \chi^2_{\rm prior} &= \frac 12 \int_{\k} \frac{\hat\delta_g(\k)\hat\delta_g(-\k)}{P(k)} + A  \int_{\k,\p} Y_2(\p,\k-\p) \frac{\hat\delta_g(\p) \hat\delta_g(\k-\p) \hat\delta_g(-\k)}{P(k)} \\
&\;\;\;\;+ \frac 12 A^2 \int_{\k,\p_1,\p_2} Y_2(\k-\p_1,\p_1) Y_2(-\k-\p_2,\p_2) \frac{\hat\delta_g(\p_1)\hat\delta_g(\k-\p_1) \hat\delta_g(\p_2) \hat\delta_g(-\k-\p_2) }{P(k)} \\
&\;\;\;\;+ A^2 \int_{\k,\p_1,\p_2} Y_3(\k-\p_1-\p_2,\p_1,\p_2) \frac{\hat\delta_g(-\k) \hat\delta_g(\p_1) \hat\delta_g(\p_2) \hat\delta_g(\k-\p_1-\p_2) }{P(k)} \,\,.
\end{split}
\end{equation}
The Jacobian term is simpler and it has the following form 
\be
{\rm Tr}\ln J = 3A^2 \int_{\k,\p} Y_3(\p,-\p,\k) \hat\delta_g(\p) \hat\delta_g(-\p) -2A^2 \int_{\k,\p} Y_2(\p,\k-\p) Y_2(-\p,\k) \hat\delta_g(\p) \hat\delta_g(-\p) \,\,.
\ee
Note that the leading terms that correspond to the linear theory do not depend on $A$, in agreement with the expectation that all information on the amplitude of $\delta$ comes from the nonlinearities. Given the simple dependence of the log-posterior on $A$, it is possible to explicitly write down the optimal estimator
\be
\label{eq:full_optimal_estimator}
\mathcal{E}
= {-\frac{1}{
{2}\,{\rm quadr}}} \int_{\k,\p} Y_2(\p,\k-\p) \frac{\hat\delta_g(\p) \hat\delta_g(\k-\p) \hat\delta_g(-\k)}{P(k)} \,\,.
\ee
The numerator is the only term in the log-posterior linear in $A$, while in the denominator we collect all other quadratic and quartic combinations of data $\hat\delta_g$ which are all proportional to $A^2$. More precisely, we write 
\begin{equation}
    \label{eq:precision}
    ({-\ln{\cal P}}) = ({-\ln{\cal P}})_{(0)}+A\,({-\ln{\cal P}})_{(1)} + A^2{\underbrace{({-\ln{\cal P}})_{(2)}}_{\hp{\text{quadr}\,}\equiv\,\text{quadr}}}+\cdots
\end{equation}
Given the realization of data both the numerator and the denominator can be computed easily since they are just numbers. This means that in practice one can use the exact posterior to find the constraints on cosmological parameters. 

However, in order to get a better understanding of the optimal estimator and simplify equations, it is convenient to replace the denominator by its average, assuming that it does not vary significantly between the different realizations of data. The modified estimator is given by
\be
\label{eq:simple_est_A}
\tilde{\cal E}
= -\frac{1}{
{2}\,\langle{\rm quadr} \rangle} \int_{\k,\p} Y_2(\p,\k-\p) \frac{\hat\delta_g(\p) \hat\delta_g(\k-\p) \hat\delta_g(-\k)}{P(k)} \,\,.
\ee
From now on, we will use the tilde to denote such ``simplified'' estimators. In our example, the explicit form of the denominator evaluated at leading order in perturbation theory is 
\be
\label{eq:master_formula_for_marginalized_bias-quadr_definition}
\langle{\rm quadr} \rangle = 
V\int_{\k,\p} \left[ Y_2^2 (\p,\k-\p)  \frac{P(p)P(|\k-\p|)}{P(k)} + 2 Y_2(\p,\k-\p) Y_2(-\p,\k) P(p) \right] \,\, .
\ee
Note that the contributions to prior and Jacobian with cubic kernels $Y_3$ exactly cancel when taking the average. 
It is then easy to explicitly check that the estimator is unbiased, 
$\smash{\langle\tilde{\cal E}\rangle = 1}$, calculating the tree-level galaxy bispectrum and remembering that $X_2 = -Y_2$ in this example. Finally, computing the variance of $\smash{\tilde{\cal E}
}$ or using the one-loop expression for the Fisher matrix from the previous section, we find that the error on $A$ is given by
\be
\label{eq:master_formula_for_marginalized_bias}
\frac{1}{\sigma_A^2} = 2V\int_{\k,\p} \bigg[ X_2^2(\p,\k-\p) \frac{P(p) P(|\k-\p|)}{P(k)} + 2 X_2(\p,\k-\p) X_2(-\p,\k) P(p) \bigg]\,\,. 
\ee 
Note that $1/\sigma_A^2=
{2}\langle{\rm quadr} \rangle$. This is expected since $\langle{\rm quadr} \rangle$ is the expectation value of all the terms in the negative log-posterior which are proportional to $A^2$ and therefore equal to the Fisher matrix for the amplitude $A$.

Three comments are in order. First, it is clear from the expression for the error that the degeneracy between $b_1$ and $A$ is broken only by nonlinearities. The right hand side of~\EQ{master_formula_for_marginalized_bias} has the typical size of $V\int_{\k} P^{\text{$1$-loop}}(k)/P(k)$. This is an explicit example in which we can see that the loop counting in forward modeling works the same way as in the conventional analyses, as discussed above. Since we are always working at the one-loop order, we expect $\sigma_A^2$ to be exactly the same as the error in the standard joint one-loop power spectrum and tree-level bispectrum analysis, as we will demonstrate soon. 

Second, the $X_2$ kernel in this setup is equal to the $F_2$ kernel of standard perturbation theory. It follows that the first of the two terms in \EQ{master_formula_for_marginalized_bias} is nothing but the usual $P_{22}(k)$ diagram (divided by $P(k)$ and integrated in $\dif^3k/(2\pi)^3$). It is well-known that $P_{22}(k)$ has very large contributions from soft loops at high $k$. This is due to the shifts in $F_2$, whose variance is large in a $\Lambda$CDM-like cosmology. However, the same shifts would be present even in a more generic example, where other cosmological or biased parameters are considered, and the conclusion below would not change in such more general setup. Na{\"i}vely, large contributions to $P_{22}$ from soft loops would lead to a very small error bar on $A$. In the standard calculation of the one-loop power spectrum these large contributions are usually cancelled by the $P_{13}$ diagram. For the field-level posterior we expect similar cancellation to happen. In our case, this is ensured by the second term in~\EQ{master_formula_for_marginalized_bias}.
To see this explicitly, let us remember that the two infrared contributions ($p\to 0$ and $\abs{\k-\p}\to 0$) from $P_{22}(k)$ give 
\be
\label{eq:unsafe_P22} 
F_2^2(\p,\k-\p) \frac{P(p) P(|\k-\p|)}{P(k)} \;\; \to \;\; 2 \frac{(\p\cdot\k)^2}{4p^4} P(p) , \;\; p\ll k \,\,. 
\ee
The only infrared contribution from the Jacobian part in \EQ{master_formula_for_marginalized_bias}, instead, is 
\be
2 F_2(\p,\k-\p) X_2(-\p,\k) P(p) \;\; \to \;\; 2 \frac{(\p\cdot\k)}{2p^2}\frac{(-\p\cdot\k)}{2p^2} P(p) = - 2 \frac{(\p\cdot\k)^2}{4p^4} P(p) , \;\; p\ll k \,\,,
\ee 
which exactly cancels~\EQ{unsafe_P22}. One important lesson that we learn from this result is that one cannot use large displacements to break degeneracies between cosmological and nuisance parameters, e.g.~linear bias and $A$, even though at the map level large displacements produce a large effect and by the equivalence principle they are proportional only to the amplitude of the fluctuations for any type of tracer. The reason is that the posterior is expressed as a combination of data which on average are equal to the summary statistics of the observed nonlinear field. Therefore, the effects off the displacements cancel in the final result for the error on cosmological parameters in the same way they do in equal-time correlators~\cite{Peloso:2013zw,Kehagias:2013yd,Creminelli:2013mca,Creminelli:2013poa,Creminelli:2013nua,Mirbabayi:2014gda}. On the other hand, if the initial conditions are fixed and known, as is the case in simulations, the amplitude of the linear density field can be indeed measured from the displacements~\cite{Elsner:2019rql}. Note that this is not only the consequence of the cosmic variance cancellation for the known initial conditions. Even if the biases are unknown, the nontrivial dynamics involving large shifts is what allows to measure $A$.

\begin{figure}[h!]
\centering
\includegraphics[width=0.75\textwidth]{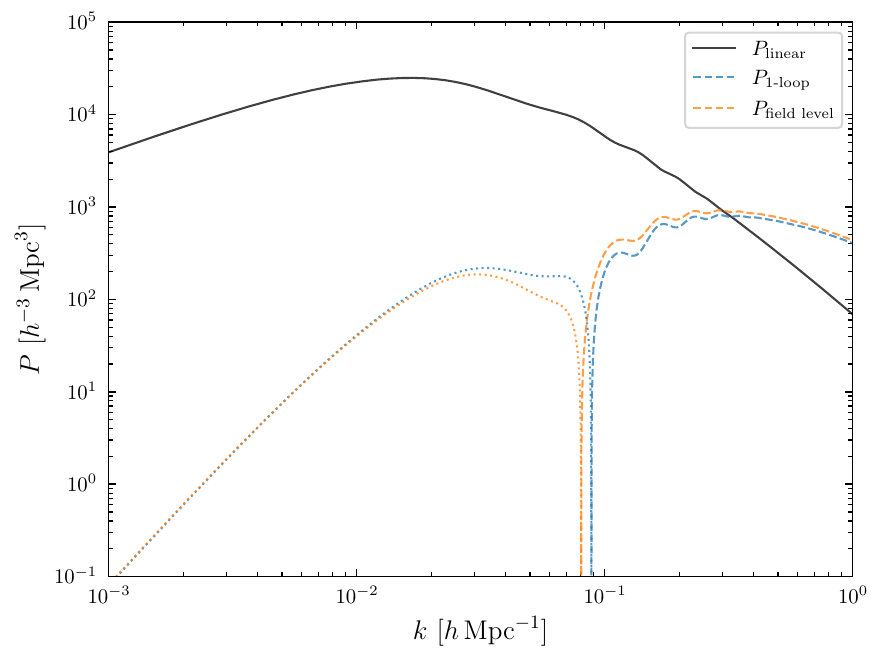}
\caption{Integrand in the square brackets of \EQ{master_formula_for_marginalized_bias} (``$P_{\text{field level}}$'') compared with the one-loop SPT power spectrum for a standard $\Lambda$CDM cosmology.} 
\label{fig:error_field_level} 
\end{figure} 

The third comment regards the positivity of ${1}/{\sigma_A^2}$. In Fig.~\ref{fig:error_field_level} we show that the square bracket in \EQ{master_formula_for_marginalized_bias} is very similar to the one-loop matter power spectrum, $P_{\text{$1$-loop}}(k)$. This quantity is negative at low $k$, so one might worry that this affects the sign of ${1}/{\sigma_A^2}$. However, we must recall that all our derivations in the previous section assume integration over all the modes. Therefore, we only need to show that the large-$k$ limit of the square bracket in \EQ{master_formula_for_marginalized_bias} is positive, whatever the form of the power spectrum. If this is true, then ${1}/{\sigma_A^2}$ is positive-definite. The discussion about infrared safety helps us to confirm this. Let us expand the square bracket at next-to-leading order in $1/k$, since the leading order vanishes. More precisely, let us consider only the (angle-averaged) expansion of the kernels. We have 
\begin{subequations}
\label{eq:positivity}
\begin{align}
4\int\dif\mu\,F^2_2(\p,\k-\p) &= \frac{2}{3}\frac{k^2}{q^2} + \frac{1138}{735} + {\cal O}\bigg(\frac{q^2}{k^2}\bigg)\,\,, \label{eq:positivity-1} \\
4\int\dif\mu\,F_2(\p,\k-\p)F_2({-\p},\k) &= {-\frac{2}{3}\frac{k^2}{q^2}} + \frac{152}{147} + {\cal O}\bigg(\frac{q^2}{k^2}\bigg)\,\,, \label{eq:positivity-2}
\end{align}
\end{subequations}
where $\mu = \vers{k}\cdot\vers{p}$. Hence we see that the contribution to the square bracket at large $k$ takes the form 
\be
\frac{1898}{735}P(k)\int\frac{\dif p\,p^2}{(2\pi)^2} P(p)\,\,, 
\ee 
which is manifestly positive for any cosmology. 

In practice, one always has to use some finite cutoff~$k_{\rm max}$. However, in that case the expression in square bracket in \EQ{master_formula_for_marginalized_bias} is also modified, since the same cutoff has to be applied in each integration over momenta. One can show that the error remains positive for any choice of~$k_{\rm max}$ as long as the cutoff is implemented consistently.

\subsection{Comparison to the joint power spectrum and bispectrum analysis}

\noindent It is instructive to compere results of the previous section with the conventional joint power spectrum and bispectrum analysis and show explicitly that the likelihood and the Fisher matrix for cosmological parameters are the same. To that end, we can compute the posterior and the Fisher matrix for the standard analysis and compare it with the field-level results in~\EQ{simple_est_A} and~\EQ{master_formula_for_marginalized_bias}. Let us begin with the power spectrum. In the setup where $Ab_1=1$, the galaxy power spectrum at leading order in perturbation theory can be schematically written as 
\be
P_g(k) = P(k) + A^2 P_{\text{$1$-loop}}(k) \,\,.
\ee
As expected, the only information on the amplitude of the density fluctuations comes from the nonlinearities. How well can we measure $A$ this way? We can estimate this using the Fisher matrix. Using the Gaussian covariance, it is given by the well-known expression
\be
\left( \frac{1}{\sigma_A^2} \right)_P = \frac V2 \int_{\k} \left( \frac{\partial P_g(k)}{\partial A} \frac{1}{P^2(k)} \frac{\partial P_g(k)}{\partial A} \right)\Bigg|_{A=1}  = 2V \int_{\k} \left( \frac{P_{\text{$1$-loop}}(k)}{P(k)} \right)^2 \,\,.
\ee
This result is rather different from the error in~\EQ{master_formula_for_marginalized_bias}, which schematically looks like
\be
\frac{1}{\sigma_A^2} \approx 2V \int_{\k} \frac{P_{\text{$1$-loop}}(k)}{P(k)} \,\,.
\ee
Clearly, the signal in the power spectrum analysis is suppressed by the variance of the density field compared to the field-level result. In the perturbative setup this means that the one-loop power spectrum does not carry significant information about the amplitude of the density fluctuations once we marginalize over $b_1$. This can be checked numerically choosing $k_{\rm max}$ to be in the mildly nonlinear regime and computing the Fisher matrix. As a consequence, the leading information in this setup is expected to be only in the bispectrum.

In order to confirm this expectation, we can explicitly calculate the bispectrum likelihood. Using the Gaussian covariance we can write\footnote{Note that introducing the finite size bins of width $\Delta k$ and replacing the integrals with the sum over triangles, we can rewrite the likelihood in a more familiar form
\be
\chi^2_B = \sum_{T} \frac{(\hat B_g(T)- B_g(T))^2 } {{\cal C}_{B}(T) }\,\,,
\ee
where the sum runs over all triangles $T = \{k_1, k_2, k_3\}$ such that $k_1\geq k_2 \geq k_3$. The Gaussian covariance for the bispectrum is 
\be 
\label{eq:bispectrum_covariance}
{\cal C}_B(k_1,k_2,k_3) = \frac{(2\pi)^6S_{\rm shape}}{VV_{123}}P(k_1)P(k_2)P(k_3)\,\,, 
\ee 
where $V_{123} = 8\pi^2k_1k_2k_3\Delta k^3$ and $S_{\rm shape}=6,2,1$ for equilateral, isosceles and scalene triangles, respectively.} 
\be
\chi^2_B = \frac{V}{6} \int_{\q_1,\q_2} \frac{[\hat B_g (\q_1,\q_2,-\q_1-\q_2)-B_g(\q_1,\q_2,-\q_1-\q_2)]^2}{P(q_1)P(q_2)P(|\q_1+\q_2|)} \,\,,
\ee
where the bispectrum estimator is
\be
\hat B_g (\q_1,\q_2,-\q_1-\q_2) = \frac 1V \; \hat\delta_g(\q_1) \hat\delta_g(\q_2) \hat\delta_g(-\q_1-\q_2) \,\,,
\ee
and the theoretical model (remembering that $Ab_1=1$) is
\be
B_g(\q_1,\q_2,\q_3) = 2A \; X_2(\q_1,\q_2) P(q_1) P(q_2) + \text{$2$ perms.} 
\ee
Note that the bispectrum likelihood is Gaussian in $A$, with the variance
\be
\left( \frac{1}{\sigma_A^2} \right)_B = \frac{V}{6} \int_{\q_1,\q_2} \frac{B_g^2(\q_1,\q_2,-\q_1-\q_2)}{P(q_1)P(q_2)P(|\q_1+\q_2|)} \,\, .
\ee
Using the explicit form of the theoretical model and keeping all the permutations, it is easy to show that
\be
\left( \frac{1}{\sigma_A^2} \right)_B = 2V\int_{\k,\p} \bigg[ X_2^2(\p,\k-\p) \frac{P(p) P(|\k-\p|)}{P(k)} + 2 X_2(\p,\k-\p) X_2(-\p,\k) P(p) \bigg]\,\,.
\ee
This precisely agrees with~\EQ{master_formula_for_marginalized_bias}, confirming the expectation that the error bar in the field-analysis is the same as in the (power spectrum and) bispectrum analysis at leading order in perturbation theory. Furthermore, this equivalence can be checked for the value of the best fit parameter as well. Using the bispectrum likelihood we get
\begin{align}
\mathcal{E}_B
&= \sigma_A^2 \cdot \frac{V}{6} \int_{\q_1,\q_2} \frac{\hat B_g(\q_1,\q_2,-\q_1-\q_2)\cdot B_g(\q_1,\q_2,-\q_1-\q_2)}{P(q_1)P(q_2)P(|\q_1+\q_2|)} \nonumber \\
& = \sigma_A^2 \cdot \int_{\k,\p} X_2(\p,\k-\p) \frac{\hat\delta_g(\p) \hat\delta_g(\k-\p) \hat\delta_g(-\k)}{P(k)}\,\, ,
\end{align}
which is again identical to the field level result for the modified estimator 
$\smash{\tilde{\mathcal{E}}
}$ in~\EQ{simple_est_A}. In conclusion, we have shown that the perturbative forward modeling recovers the same information about the amplitude of density fluctuations as the few leading correlation functions, computed at the same order in perturbation theory. 

Even though this result was derived in a simple and analytically tractable case of $Ab_1=1$, our conclusions hold more generally. If the variance of the density field is the only expansion parameter in the theory, one can show, order-by-order in perturbation theory, that the field-level and the $n$-point function based inferences of cosmological parameters are equivalent for any cosmological parameter, although the explicit demonstration is more involved. If other, potentially large parameters are present in the theory, this simple picture can change. We turn to these more interesting situations next.

\section{Beyond the simple perturbative model}
\label{sec:beyond_simple_PT}

\noindent So far we have focused on a universe in which the only relevant scale for the nonlinear evolution is the nonlinear scale $k_{\rm NL}$ and the only small expansion parameter is the variance of the density field. However, the real universe can be more complicated, and depending on the shape of the linear power spectrum other scales can play an important role in the nonlinear dynamics. The most well-known example is the parameter related to the velocity dispersion which is responsible for the broadening of the BAO peak. This parameter is given by~\cite{Baldauf:2015xfa,Blas:2016sfa}
\be
\Sigma^2 = \frac{1}{6\pi^2}\int_0^{k_{\rm NL}} {\rm d}q\, P(q)\, [1-j_0(q\ell_{\rm BAO}) +2j_2(q\ell_{\rm BAO})] \,\,,
\ee
where $\ell_{\rm BAO}$ is the BAO scale. Note that the combination of spherical Bessel functions in the square brackets is such that it scales as $q^2$ in the limit $q\ll \ell_{\rm BAO}^{-1}$. Therefore, neglecting the contribution given by the variance of the density field on the BAO scale, we can approximate the previous expression as
\be
\Sigma^2 \approx \frac{1}{6\pi^2}\int_{\ell_{\rm BAO}^{-1}}^{k_{\rm NL}} {\rm d}q\, P(q) \;.
\ee 
The integral is dominated by the peak of the power spectrum, which in a $\Lambda$CDM-like cosmology is at the equality scale $k_{\rm eq}$. However, given that for our universe $\ell_{\rm BAO}^{-1} \approx k_{\rm eq}$, we will keep the BAO scale as the lower boundary of the integral, having in mind a more general power spectrum for which the maximum in principle can be at much smaller $k$. Approximating the linear power spectrum as a power law such that $P(k) \approx P(k_{\rm NL}) (k/k_{\rm NL})^{-n}$, we can estimate $\Sigma^2$ as 
\be
\Sigma^2 \approx \frac{1}{k_{\rm NL}^2} \frac13 \left( k_{\rm NL} \ell_{\rm BAO} \right)^{n-1} \;. 
\ee
Depending on the slope of the power spectrum and position of the BAO peak, the enhancement $(k_{\rm NL} \ell_{\rm BAO})^{n-1}$ can be large. In $\Lambda$CDM this is not dramatic, since we have $n\approx 1.8$ and therefore $\Sigma \approx 2 k^{-1}_{\rm NL}$ at redshift zero. However, note that 
\be
\frac13 \left( k_{\rm NL} \ell_{\rm BAO} \right)^{n-1} \approx 4 \left( \frac{D^2(z)}{D^2(0)} \right)^{\tfrac{1-n}{3-n}} \;,
\ee
such that $\Sigma$ compared to $k_{\rm NL}^{-1}$ grows at larger redshifts for $1<n<3$. 

The existence of a large parameter controlled by some infrared scale implies a possible breakdown of the simple Eulerian-like forward model that we discussed in the previous section. Indeed, a simple one-loop calculation is known to poorly describe the shape of the BAO peak in the nonlinear two-point correlation function. It is well-understood that in order to circumvent this issue one has to employ either Lagrangian perturbation theory~\cite{Carlson:2012bu,Porto:2013qua,Vlah:2015sea,Baldauf:2015tla,Vlah:2016bcl} or modify predictions of the Eulerian perturbation theory through the so-called infrared resummation~\cite{Senatore:2014via,Baldauf:2015xfa,Vlah:2015zda,Blas:2016sfa,Senatore:2017pbn,Ivanov:2018gjr}. Therefore, the measurement of the BAO scale is a well-known counterexample to our statement in the previous section. The field-level inference of $\ell_{\rm BAO}$ is indeed more optimal compared to the measurement from the two-point correlation function~\cite{Babic:2022dws}. 

However, we will not further discuss this example here for two reasons. First, it is rather special, since large displacements affect only features in the two-point correlation function and the only parameter that is impacted is $\ell_{\rm BAO}$. All other cosmological parameters are unaffected and our general conclusions still apply. More explicitly, the average log-posterior is expressed in terms of the $n$-point functions of the data, whose smooth part is not impacted by the large displacements~\cite{Peloso:2013zw,Kehagias:2013yd,Creminelli:2013mca,Creminelli:2013poa,Creminelli:2013nua,Mirbabayi:2014gda}. The second reason is that the simple BAO reconstruction schemes~\cite{Eisenstein:2006nk,Padmanabhan:2008dd,Schmittfull:2017uhh} recover almost optimal information on~$\ell_{\rm BAO}$, making the full forward modelling unnecessary. One may still do the forward modelling of the reconstructed field, but since in this case the significant fraction of large displacements is cancelled, this is much closer to the regime that we discussed in the previous section and our conclusions remain valid. It would be interesting to check this explicitly in numerical simulations and we leave it for future work. 

For the rest of this section we will focus instead on different type of situations in which forward modeling can be more optimal. Unlike the BAO where the long-short interactions dilute the signal, in these examples these interactions increase the error. More precisely, they lead to large covariance matrices for the $n$-point functions, making the standard analyses suboptimal. Importantly, this affects {\em all} cosmological parameters. We will show how these situations arise and what are the new large parameters associated to them.

\subsection{Large covariance matrix from long-wavelength fluctuations}
\label{sec:large_covariance_IR}

\noindent In order to see how the standard analysis can be suboptimal, we can already use the simplest example of nonlinear dark matter field. We have shown in the previous section that in this case the optimal estimator for $A$ is 
\be
\label{full_estimator}
\mathcal{E}
= \frac{1}{N_{\rm pix}} \int_{\k} \frac{|\delta[\hat\delta_g,\hat A=1](\k)|^2 }{P(k)} \,\,,
\ee
where the numerator has various combinations of the data~$\hat\delta_g$ dictated by the inverse model. In order to highlight new relevant parameters, in this section we will assume that the nonlinearities controlled by the variance of the density field are very small. In this limit the inverse model is well approximated by the linear term $\delta=Y_1\delta_g$ and the higher order loop contributions are expected to be small. The approximate estimator valid in such regime is given by 
\be
\label{eq:simple_estimator_nonlinear_DM}
\tilde{\cal E}
= \frac{1}{N_{\rm pix}} \int_{\k} \frac{|\hat\delta_g(\k)|^2 }{P(k)} \,\,,
\ee
assuming $Y_1=1$. Such result is not surprising. This is the estimator of the nonlinear power spectrum, which in the limit of small nonlinearities gives the correct estimate of~$A$. 

While such simplified estimator (which we will interchangeably call ``na{\"i}ve'' and ``simplified'' in the following) may lead to correct amplitude of the linear power spectrum, its variance may be large. We can easily compute that 
\be
\label{eq:total_variance_As}
{\rm var}(\tilde{\cal E}) 
= \frac{2V}{N^2_{\rm pix}} \int_{\k} \frac{P^2_g(k) }{P^2(k)} + \frac{V}{N^2_{\rm pix}} \int_{\k,\k'} \frac{T_g(\k,-\k,\k',-\k') }{P(k)P(k')} \,\,,
\ee
where $T_g$ is connected $4$-point function of the data {(in the following, we will drop the hat on correlation functions of the data for simplicity of notation)}. Going beyond the leading result given by the linear theory, a simple estimate of the one-loop power spectrum in the first term and tree-level trispectrum in the second term both lead to
\be
{\rm var}(\tilde{\cal E}) = \frac{2}{N_{\rm pix}} \left[ 1 + {\mathcal O}\left(\Delta^2(k_{\rm max})\right) \right] \,\,,
\ee
where~$\Delta^2(k_{\rm max})$ is the variance of the density field at~$k_{\rm max}$ used in the analysis. As one may expect, corrections to the linear theory error bars are small at leading order in perturbation theory. However, going to the one-loop trispectrum in $\smash{{\rm var}(\tilde{\cal E})}$ something unexpected happens. Due to the particular momentum configuration there are one-loop contributions that schematically look like 
\be
\text{$T_g(\k,-\k,\k',-\k') \supset P(k)P(k') \int_{\q}P^2(q)$ 
\quad in the limit $q\ll k,k'$\,\,.}
\ee
This is a well-known result that can be explicitly derived in perturbation theory~\cite{Rimes:2005xs,Hamilton:2005dx,Sefusatti:2006pa,Mohammed:2016sre,Barreira:2017kxd} and holds even in the nonlinear regime if one uses nonlinear responses to compute the covariance matrix~\cite{Barreira:2017sqa,Barreira:2017kxd}. Importantly, such contributions to the covariance matrix are {\em not} controlled by the variance of the density field as one may naively expect. Instead, they are proportional to the new parameter---the variance of~$\delta^2$. In general, we can define the following dimensionless quantity
\be
\sigma^2_{n,-} \equiv \left(\frac{k_{\rm NL}^3}{2\pi^2} \right)^{n-1} \int_{\q<k_{\rm NL}} P^n(q) \,\,,
\ee
such that the ratio of the one-loop and the tree-level trispectrum contributions to the covariance is given by
\be
\frac{T_g^{\text{$1$-loop}}}{T_g^{\rm tree}} \approx \sigma_{2,-}^2 \,\,,
\ee
for~$k\sim k_{\rm NL}$.
The minus sign in our notation for~$\sigma^2_{n,-}$ indicates that momenta which~$P^n(k)$ is integrated over are smaller than some scale, which we chose to be~$k_{\rm NL}$. 

The existence of a new parameter that controls the loop expansion is indeed surprising, but it is the consequence of specific momentum configuration of the trispectrum that contributes to the covariance matrix. If all momenta in an $n$-point function are different, these parameters never appear. Importantly,~$\sigma_{2,-}^2$ can be very large, even when the variance of the density field is small. To see this explicitly, let us consider a power-law universe with the IR cutoff $k_*$, which can be given by the size of the survey or can mimic the equality scale in a $\Lambda$CDM-like universe. In this simplified setup, the power spectrum is given by
\begin{equation}
\text{$P(k) = \frac{2\pi^2(3-n)}{k_{\rm NL}^3} \left( \frac{k}{k_{\rm NL}} \right)^{-n} \theta(k-k_*) \,\,,$ \quad with \quad ${3/2}<n<{3} \,\,.$} 
\end{equation} 
Note that the slope of the power spectrum in the~$\Lambda$CDM cosmology at the nonlinear scale is approximately~$n\approx 2$, which is in the range we consider. 
The variance of the density field is given by
\be
\Delta^2(k) = \frac{2\pi^2(3-n)}{k_{\rm NL}^3} \frac 1{2\pi^2} \int_{k_{\rm min}}^{k}\dif q\,q^2\,\left( \frac{q}{k_{\rm NL}} \right)^{-n} \approx \left( \frac{k}{k_{\rm NL}} \right)^{3-n} \,\,, 
\ee 
where for $3/2<n<3$ we have neglected the lower bound of the integral. As usual, the power spectrum is normalized such that $\Delta^2(k_{\rm NL})=1$. Also, the variance is smaller than 1 on perturbative scales and it is growing with~$k$ for our choice of~$n$. We can now explicitly compute~$\sigma_{2,-}^2$ and find
\be
\sigma_{2,-}^2 = \frac{(3-n)^2}{2n-3} \left( \frac{k_*}{k_{\rm NL}} \right)^{3-2n} \,\,.
\ee
Note that for our choice~$3/2<n<3$ the integral in~$\smash{\sigma_{2,-}^2}$ is dominated in the infrared and in this estimate we neglected the upper bound. For~$n\approx 2$ we have~$\smash{\sigma_{2,-}^2 \approx k_{\rm NL}/k_*}$, which can be much larger than $1$. This large parameter can significantly modify the variance of the estimator for~$A$. Following~\EQ{total_variance_As} we have
\be
\label{eq:var_epsilon_s2}
{\rm var}(\tilde{\cal E}) = \frac{2}{N_{\rm pix}} \left( 1 +\frac{N_{\rm pix}}{2V}\int_{\q}P^2(q) \right) = \frac{2}{N_{\rm pix}} \left( 1 +\pi^2\frac{N_{\rm pix}}{Vk_{\rm NL}^3}\, \sigma_{2,-}^2 \right) \,\,.
\ee
Using~$N_{\rm pix} = Vk_{\rm max}^3/6\pi^2$, we can estimate the correction to the linear theory error as
\be
\frac{N_{\rm pix}}{2V}\int_{\q}P^2(q) = \pi^2\frac{N_{\rm pix}}{Vk_{\rm NL}^3}\, \sigma_{2,-}^2 = \frac16 \left( \frac{k_{\rm max}}{k_{\rm NL}} \right)^3 \sigma_{2,-}^2 \,\,.
\ee  
For a $\Lambda$CDM-like cosmology where $k_*\approx k_{\rm eq} \approx 0.02 \; h/{\rm Mpc}$ and $k_{\rm NL}\approx0.3 \; h/{\rm Mpc}$, this number at redshift zero is
\be
\frac{N_{\rm pix}}{2V}\int_{\q}P^2(q) = \mathcal O(1) \times \left( \frac{k_{\rm max}}{0.3\;h/{\rm Mpc}} \right)^3 \,\,,
\ee
assuming~$n\approx2$. We can see that for a~$\Lambda$CDM-like universe the correction is not large on perturbative scales and in a realistic data analysis would be subdominant to other sources of error, such as marginalization over nuisance parameters~\cite{Wadekar:2020hax}. However, it is easy to imagine a universe where the situation is very different. For instance, setting~$k_*$ to be the fundamental mode of the survey, we can write the correction to the linear theory errors as 
\be
\label{scaling_variance_to_recall}
\frac{N_{\rm pix}}{2V}\int_{\q}P^2(q) = \frac{(3-n)^2}{6(2n-3)} \left( \frac{k_{\rm max}}{k_{\rm NL}} \right)^{6-2n} \left( \frac{3N_{\rm pix}}{4\pi} \right)^{\tfrac{2n}{3}-1} \,\,.
\ee
If the power-law power spectrum extends to arbitrarily large scales, for any~$k_{\rm max}$ there is a volume big enough when the number of Fourier modes makes this number bigger than one. Setting $n\approx2$ and $k_{\rm max}\approx0.1 \; h/{\rm Mpc}$ and $k_{\rm NL}\approx0.3 \; h/{\rm Mpc}$ we find 
\be
{\rm var}(\tilde{\cal E}) = \frac{2}{N_{\rm pix}} \left( 1 + \mathcal O(0.01)\times N_{\rm pix} ^{1/3} \right) \,\,.
\ee
For a hypothetical survey with more than one million pixels, the second term would dominate the covariance. The errors would then scale only as~$\smash{N_{\rm pix}^{2/3}}$. In the extreme case of the nearly scale-invariant power spectrum with~$n\approx 3$, the error bars would improve only logarithmically as in the case of local non-Gaussianities~\cite{Creminelli:2006gc}. 
This clearly indicates that the na{\"i}ve estimator is suboptimal. 
Let us point out that correcting the na{\"i}ve estimator is not necessary in order to get a detection of $A$: for $n\approx2$, the variance of $\smash{\tilde{\cal E}}$ is much less than one for the typical number of pixels in modern galaxy surveys. However, we see that by using $\smash{\tilde{\cal E}}$ instead of $\smash{\cal E}$ we could be paying a very high price on the error bars. In contrast, as we have shown in the previous section, the full field level analysis leads to the expected optimal result.

The appearance of large parameters~$\sigma^2_{n,-}$ in the variance of ``na{\"i}ve'' estimators, where one takes a simplified version of the optimal estimators, is quite generic. For example, let us consider a case of a linearly-biased tracer with unknown linear bias~$b_1$ discussed in Section~\ref{subsec:including_the_linear_bias}. The simplified estimator is given by Eq.~\eqref{eq:simple_est_A}: 
\be
\label{eq:reminder_simple_est_A}
\tilde{\cal E} = {-\sigma^2_A}\int_{\k,\p} Y_2(\p,{-\k}-\p) \frac{\hat\delta_g(\p) \hat\delta_g({-\k}-\p) \hat\delta_g(\k)}{P(k)}\,\,, 
\ee
where we have used Eqs.~\eqref{eq:master_formula_for_marginalized_bias-quadr_definition}, \eqref{eq:master_formula_for_marginalized_bias} and the simple relation between $Y_2$ and $X_2$. 
Computing $\smash{\langle\tilde{\cal E}^2-1\rangle}$, we see that the variance of $\smash{\tilde{\cal E}}$ contains a piece 
\be
{\rm var}(\tilde{\cal E})
\supset(\sigma^2_A)^2V\int_{\k,\k',\p}\frac{P(p)}{P(k)P(k')}
T_g(\k,\k',{-\k}-\p,{-\k'}+\p){X}_2({-\k}-\p,\p){X}_2({-\k'}+\p,{-\p})\,\,. 
\ee
The two relevant contributions in the trispectrum are
\be
\label{eq:contribution_we_focus_on_for_simplicity}
T_g(\k,\k',{-\k}-\p,{-\k'}+\p)\supset P(|\k+\p|)P(|\k'-\p|)P(p)X_2(\k+\p,\p)X_2(\k'-\p,\p)
\ee
and 
\be
\label{eq:other_contribution}
T_g(\k,\k',{-\k}-\p,{-\k'}+\p)\supset P(k)P(k')P(p)X_2({-\k},{-\p})X_2({-\k'},\p)\,\,. 
\ee
We see that both, in the limit $p\ll k, k'$, give rise to the large parameter $\sigma^2_{2,-}$, which can make the variance of the na{\"i}ve estimator much larger than expected. 

It is instructive to discuss another possible contribution to the variance of the na{\"i}ve estimator, i.e.~the one coming from the fully connected six-point function 
\be
{\rm var}(\tilde{\cal E})\supset(\sigma^2_A)^2\int_{\vec{k},\vec{k}',\vec{p},\vec{p}'}\frac{V\braket{\hat\delta_g(\vec{p}) \hat\delta_g(\vec{p}') \hat\delta_g({-\vec{k}}-\vec{p})\hat\delta_g({-\vec{k}'}-\vec{p}') \hat\delta_g(\vec{k}) \hat\delta_g({\vec{k}'})}'}{P(k)P(k)'} \,\,.
\ee
For a particular momentum configuration where the pairs of momenta are opposite ($\p'=-\p$ and $\k'=-\k$), we expect that the one-loop contribution to the six-point function is controlled by $\sigma^2_{3,-}$. In the universe where large parameters~$\smash{\sigma^2_{n,-}}$ are infrared-dominated, we expect that~$\smash{\sigma^2_{3,-}\gg \sigma^2_{2,-}}$ and this would be the leading contribution to the variance. This expectation is, however, wrong since~$\smash{\sigma^2_{3,-}}$ appears only for special arrangements of momenta, unlike in the case of the estimator in~\EQ{simple_estimator_nonlinear_DM} where the large parameter~$\sigma^2_{2,-}$ exists for {\em all} momentum configurations of the four-point function. We can check this by explicitly evaluating the six-point function. Taking $X_2=1$ for simplicity (this does not change our conclusions), we get  
\be
\begin{split}
{\rm var}(\tilde{\cal E}) &
\supset \frac{(\sigma^2_A)^2}{V}\int_{\vec{k},\vec{k}',\vec{p},\vec{p}'}\frac{(2\pi)^3\delta^{(3)}_{\rm D}(\vec{k}+\vec{k}')(2\pi)^3\delta^{(3)}_{\rm D}(\vec{p}+\vec{p}')}{P(k)P(k')}P(p)P(|\vec{k}+\vec{p}|)P(k)\int_{\vec{q}}P^3(q) \\ 
&\approx \frac{(\sigma^2_A)^2}{V}{\underbrace{\int_{\vec{k},\vec{p}} \frac{P(p)P(|\vec{k}+\vec{p}|)}{P(k)}}_{\hphantom{\frac{1}{V\sigma^2_A}}\approx\,\frac{1}{V\sigma^2_A}}}\int_{\vec{q}}P^3(q)\approx\frac{\sigma^2_A}{V^2}\int_{\vec{q}}P^3(q)\,\,, 
\end{split}
\ee
where we have isolated the part of the six-point function that can give rise to $\sigma^2_{3,-}$ with two Dirac delta functions~$\smash{\dirac{\k+\k'}/V}$ and~$\smash{\dirac{\p+\p'}/V}$. Crucially, the presence of the factor $\smash{1/V^2}$ from these two Dirac deltas guarantees that even in the worst-case scenario of a scale-invariant power spectrum with $n=-3$, this contribution to the variance is never large: indeed we would have $\smash{\sigma^2_{3,-}\sim V^2}$ so that the variance would scale as $\smash{1/N_{\rm pix}}$, i.e.~much faster than the $\smash{\sigma^2_{2,-}}$ contribution discussed above, see e.g.~Eq.~\eqref{scaling_variance_to_recall}.

While this argument shows that $\smash{\sigma^2_{3,-}}$ does not produce the leading contribution to the variance of $\smash{\tilde{\cal E}}$, it also suggests us that it will enter its skewness, where the connected six-point function automatically appears in the configuration where all momenta are ``pinched'' two-by-two. We can see this already for the simple estimator of Eq.~\eqref{eq:simple_estimator_nonlinear_DM} for the nonlinear dark matter, i.e.~
\be
\tilde{\cal E} = \frac{1}{N_{\rm pix}} \int_{\vec{k}} \frac{|\hat\delta_g(\vec{k})|^2 }{P(k)} \,\,.
\ee
The skewness of the estimator is given by 
\be
\text{skewness}(\tilde{\cal E})=\frac{\braket{( \tilde {\cal E}-1)^3}}{\sigma^3_A}\,\,,
\ee 
where
\be
\begin{split}
\braket{( \tilde {\cal E}-1)^3} &= \frac{8}{N_{\rm pix}^2} + \frac{1}{N_{\rm pix}^3} \int_{\vec{k},\vec{k}',\vec{k}''} \frac{\braket{\hat{\delta}_g(\vec{k})\hat{\delta}_g({-\vec{k}})\hat{\delta}_g(\vec{k}')\hat{\delta}_g({-\vec{k}'})\hat{\delta}_g(\vec{k}'')\hat{\delta}_g({-\vec{k}''})}_{\rm c}}{P(k)P(k')P(k'')} \,\, .
\end{split}
\ee
Note that in the second term we have to keep only the connected six-point function. Naively, this contribution is suppressed by~$\Delta^4(k_{\rm max})$, but similarly to the case of the variance, there is one-loop contribution to the connected six-point function which is much bigger and can be estimated as 
\be
\begin{split}
\braket{( \tilde {\cal E}-1)^3} &= \frac{8}{N_{\rm pix}^2} \left( 1 + \frac{1}{8} \int_{\vec{k},\vec{k}'} \int_{\vec{q}}P^3(q) \right ) =  \frac{8}{N_{\rm pix}^2} \left( 1 + \frac{1}{72} \left( \frac{k_{\rm max}}{k_{\rm NL} }\right)^6 \sigma^2_{3,-}  \right )  \,\, .
\end{split}
\ee
This is the analogue of~\EQ{var_epsilon_s2} for the variance of~$\tilde {\cal E}$. We can see that the new large parameter
\be
\sigma_{3,-}^2 = \frac{(3-n)^3}{3n-3} \left( \frac{k_*}{k_{\rm NL}} \right)^{3-3n} \,\,
\ee
appears, making the second contribution potentially much bigger than the first one. Assuming the perfect power-law power spectrum and setting~$k_*$ to be the fundamental mode of the survey, we can estimate the skewness as  
\be
\text{skewness}(\tilde{\cal E}) =  \frac{2\sqrt{2}}{\sqrt{N_{\rm pix}}} \left ( 1 + \mathcal{O}(10^{-5}) \times N_{\rm pix} \right ) \,\, ,
\ee
for typical numbers we used before: $n\approx2$ and $k_{\rm max}\approx0.1 \; h/{\rm Mpc}$ and $k_{\rm NL}\approx0.3 \; h/{\rm Mpc}$. In this hypothetical universe, even for $\mathcal{O}(10^5)$ pixels, corrections to the na{\"i}ve result become significant. For steeper power spectra, the corrections are even more pronounced. However, as before, in $\Lambda$CDM-like cosmology, these effects are never large on perturbative scales.  

It is worth emphasising that so far we have considered the case where~$k_{\rm max}<k_{\rm NL}$, such that for $\Lambda$CDM-like cosmologies the impact of~$\sigma^2_{2,-}$ or~$\sigma^2_{3,-}$ is not dramatic. However, if $k_{\rm max}>k_{\rm NL}$, even in $\Lambda$CDM there can be large contributions to the variance of the power spectrum estimator. Weak lensing is the natural place where we would expect this effect to show up.\footnote{Notice that here we are disregarding other effects that could prove to dominate the error budget in realistic scenarios, e.g.~the super-sample covariance.} In that case it is worth exploring alternative, more optimal estimators. We will come back to this in Section~\ref{subsec:optimal_estimators}.

\subsection{Large noise from mildly nonlinear scales}
\label{sec:disconnected_diagrams}

\noindent In previous section we have shown how long-wavelength fluctuations can lead to possibly large variance of na{\"i}ve estimators on small scales. It is interesting to ask whether the opposite can happen. For dark matter the answer is no, since due to momentum conservation the impact of small-scale fluctuations on large scales is strongly suppressed. However, for biased tracers the situation is different. Let us consider a galaxy density field where nonlinearities induced by biasing and the shot noise~$P_\epsilon$ are very small. In this limit the variance of the na{\"i}ve estimator of Eq.~\eqref{eq:total_variance_As} for the amplitude of the linear power spectrum can be written as
\be
{\rm var}(\tilde{\cal E}) = \frac{2V}{N^2_{\rm pix}} \int_{\k} \frac{P^2_g(k) }{P^2(k)}\,\,, 
\ee
where we are neglecting the trispectrum contribution assuming that the power spectrum is such that~$\sigma^2_{2,-}$ is not a large number. The galaxy power spectrum is given by (assuming fiducial values of the amplitude and~$b_1$ to be 1)
\be
P_g(k) = P(k)+P_\epsilon+\frac{b^2_2}{2}\int_{\q}P(q)P(|\vec{k}-\vec{q}|) + \cdots \,\,,
\ee 
where the ellipses denote the other one-loop terms that are all suppressed by~$\smash{\Delta^2(k_{\rm max})}$. Even in the limit~$P_\epsilon \to 0$ that we consider in this paper, there is an effective noise at large scales given by the third term in previous equation. Defining 
\be
{\cal I}_{\delta^2\delta^2}(k) \equiv 2\int_{\q}P(q)P(|\vec{k}-\vec{q}|) \,\, ,
\ee
we can see that the large-scale limit of the effective noise is given by the variance of~$\delta^2$. It is important to stress that this term can dominate the total noise even in a vanilla $\Lambda$CDM-like cosmology. Some well-known examples are small-mass dark matter halos~\cite{Modi:2016dah,Schmittfull:2018yuk} or neutral hydrogen~\cite{Modi:2019hnu,Obuljen:2022cjo,Villaescusa-Navarro:2018vsg}. In these examples the amplitude of the noise on large scales can be even ten times larger than the na{\"i}ve Poisson expectation and using the field level methods one can show that this noise comes exactly from the quadratic bias nonlinearities we discuss here~\cite{Schmittfull:2018yuk,Modi:2019hnu,Obuljen:2022cjo}. In a conventional analysis such large noise contributes to the covariance matrix and it can lead to larger errors. On the other hand, the field-level analysis remains optimal.

While it is possible to have a tracer such that~${\cal I}_{\delta^2\delta^2}(k)\gg P_\epsilon$ on all scales of interest, this does not automatically implies a dramatic difference in the errors between conventional analyses and forward modeling. The reason is that the errors can be still dominated by the cosmic variance. To see this explicitly, we can write
\be
{\cal I}_{\delta^2\delta^2}(k) = 2P(k)\int_{\q<k} P(q) + 2\int_{\q>k} P^2(q) = 2P(k)\int_{\q<k} P(q) + \frac{\sigma^2_{2,+}(k)}{k_{\rm NL}^3}\,\,,
\ee
where we have defined 
\be
\sigma^2_{2,+}(k) \equiv 2 k_{\rm NL}^3\int_{\q>k} P^2(q) \,\, .
\ee
This new parameter is similar to~$\sigma^2_{2,-}$ but with some very important differences. For instance, it is important to note that this parameter depends on scale. While the integral is still infrared dominated, the range of integration has a natural IR cutoff at the scale of interest~$k$. This is a consequence of the fact that for each~$k$ we are considering the effects of shorter modes that combine to produce the effective noise on larger scales. The~$+$ sign in the definition indicates that one has to calculate contribution to the variance of~$\delta^2$ coming only from modes larger than~$k$. In a simple power-law universe, we can estimate 
\be
\sigma^2_{2,+}(k) = 2 k_{\rm NL}^3 P(k) \frac{3-n}{2n-3} \left( \frac{k}{k_{\rm NL}} \right)^{3-n} \approx 2 k_{\rm NL}^3 P(k) \Delta^2(k) \,\, .
\ee
Note that this implies that 
\be
{\cal I}_{\delta^2\delta^2}(k) = 2P(k)\int_{\q<k} P(q) + \frac{\sigma^2_{2,+}(k)}{k_{\rm NL}^3} \approx 4 P(k) \Delta^2(k) \,\,,
\ee
While~$\sigma^2_{2,+}(k)$ can be much bigger than one, its contribution to the variance of the estimator is controlled by the ratio of~${\cal I}_{\delta^2\delta^2}(k)$ and~$P(k)$, which is small. More precisely
\be
{\rm var}(\tilde{\cal E}) \approx \frac{2V}{N^2_{\rm pix}} \int_{\k} \left( 1+ b_2^2 \Delta^2(k) \right)^2 \,\,.
\ee
Depending on the value of~$b_2$, the linear theory variance can be increased by~$\mathcal{O}(1)$ for~$k_{\rm max}\approx k_{\rm NL}$. While these effects are not parametrically large, it would be worth exploring the possibility to reduce the effective noise using the field level analysis in the future. 

One might wonder whether higher-order terms in the bias expansion can contribute with new large parameters to the effective noise. For example, including $(b_3/6)\times\delta^3$ in the forward model leads to the following two-loop contribution to the shot noise:
\be
\frac{b_3^2}{6}\int_{\p,\q}P(|\k-\p-\q|)P(p)P(q)\,\,. 
\ee
However, we see that in the low-$k$ limit this contribution is suppressed with respect to $\smash{{\cal I}_{\delta^2\delta^2}(k)}$ by the variance of the density field on large scales, which is a small parameter. This is generically true for higher-order contributions. 

Let us finish this section by pointing out one interesting result that follows from this discussion. Given that at the field level we can predict the realization of the long-wavelength fluctuations that lead to the effective noise, we can also use these fluctuations in the data to infer cosmological and nuisance parameters. In particular, we can measure~$b_2$ from the amplitude of the effective noise. We show this explicitly in Appendix~\ref{app:shot_noise_renormalization}. Note that this is very different from the conventional power spectrum analysis. There, following the standard prescription for renormalized bias,~$b_2$ can be measured only from a shape of loop corrections which is different from the flat power spectrum of the noise. This suggest that there is a different, potentially more optimal renormalization scheme, in which information on bias parameters can be obtained from the amplitude of the effective noise even in the power spectrum analysis. We leave exploration of this interesting possibility for future work.

\subsection{Simple nearly optimal estimators}
\label{subsec:optimal_estimators}

\noindent We have seen in previous sections how in different situations standard power spectrum and bispectrum estimators can have large variance due to various large parameters which emerge from averaging over interactions of long-wavelength and short-wavelength modes. On the other hand, in forward modelling, all such interactions are explicitly taken into account and the analysis is always optimal. We have shown that for the simple perturbative model and in the limit of small noise, the posterior can be calculated analytically and for any cosmological parameter of interest one can find the optimal estimator in terms of simple operations on the data~$\hat\delta_g$. However, our equations do not apply in all regimes of interest. One example where the variance of na{\"i}ve estimators can be large due to~$\sigma^2_{2,-}$ and where one would benefit from doing the full field-level analysis is measurement of~$A_{\rm s}$ from small scales beyond~$k_{\rm NL}$. Since our perturbative equations do not apply there, one would have to do the full forward modeling which is technically challenging. In this section we would like to show that it is possible to find a middle ground and keep the simplicity of the conventional analyses by using modified, forward-model-inspired estimators that are nearly optimal. 

In order to see how to construct these nearly optimal estimators, we will first take a closer look at how the large parameters cancel in the field-level analysis. We will focus on $\sigma^2_{2,-}$ since it can have the largest impact in practice. First, in the case of the nonlinear dark matter, the variance of the optimal estimator of Eq.~\eqref{full_estimator} scales as 
\be
{\rm var}({\cal E}) = \frac{2}{N_{\rm pix}} \left[ 1 + \mathcal O\left( \Delta^4(k_{\rm max}) \right) \right]\,\,, 
\ee
i.e.~without~$\smash{\sigma^2_{2,-}}$ appearing. This is manifest by the virtue of the inverse model which is correct at one-loop order. We emphasize that including $Y_3$ in the inverse model is crucial:~$\smash{\sigma^2_{2,-}}$ can appear in the one-loop trispectrum also from diagrams involving the~$X_3$ kernel of the forward model. 

A similar cancellation of~$\sigma^2_{2,-}$ must happen in the case of a linearly-biased tracer discussed in Section~\ref{subsec:including_the_linear_bias}, even though it is a bit more difficult to show it explicitly. For simplicity, we focus on the contribution of Eq.~\eqref{eq:contribution_we_focus_on_for_simplicity} to the variance of the simplified estimator {$\smash{{\rm var}(\tilde{\cal E})}$ for the amplitude of the power spectrum}. {Let us recall that this contribution makes the variance of the estimator behave as (for a power-law universe with the spectral index~$n\approx2$)} 
\be
{\rm var}(\tilde{\cal E}) = \frac{2}{N_{\rm pix}} \left( 1 + \mathcal O(0.01)\times N_{\rm pix} ^{1/3} \right) \,\,.
\ee
{That is, while it is always a perturbative correction ($\smash{{\rm var}(\tilde{\cal E})}$ remains $\smash{\ll 1}$ for sufficiently large number of pixels), this contribution can make the variance suboptimal and scale as $\smash{1/N_{\rm pix}^{2/3}}$ if $\smash{N_{\rm pix}\gtrsim 10^6}$.}
To understand better this contribution {and how it is cancelled when using the optimal estimator,} it is useful to expand the simplified estimator in terms of the realization of the initial conditions $\smash{\hat{\delta}}$ from which the observed tracer field $\smash{\hat{\delta}_g}$ is generated \cite{Creminelli:2006gc}: 
\be
\begin{split}
    \tilde{\cal E} &\supset \tilde{\cal E}_{(1)} + \tilde{\cal E}_{(2)}
    = 
    {\sigma^2_{A}}\int_{\k}\frac{1}{P(k)}\hat{\delta}(\k)\hat{\delta}^{(2)}_g({-\k}) 
    + \sigma^2_A\int_{\k}\frac{|\hat{\delta}^{(2)}_g(\k)|^2}{P(k)}\,\,, 
\end{split}
\ee
where we have used the relation $Y_2={-X_2}$ and by the $\supset$ symbol we mean that we work at leading and next-to-leading order in $\smash{\hat{\delta}}$, but consider only terms that come from expanding $\smash{\hat\delta_g(\k)}$ in Eq.~\eqref{eq:reminder_simple_est_A} at second order. The term that gives rise to~$\sigma^2_{2,-}$ in the variance of the estimator is~$\smash{\langle\tilde{\cal E}_{(2)}^2\rangle}$. More precisely, we have 
\be
\label{eq:large_variance}
\begin{split}
    {\rm var}\bigg(\int_{\k}\frac{|\hat{\delta}^{(2)}_g(\k)|^2}{P(k)}\bigg)_{\sigma^2_{2,-}} &= 48 \int_{\k,\k',\p}\frac{P^2(p)P(|\k+\p|)P({\k'}-\p)}{P(k)P(k')} \\
    &\;\;\;\;\hphantom{\int_{\k,\k',\p}}\times X_2(\k+\p,\p)X_2(\k'-\p,\p)X_2({-\k}-\p,\p)X_2({-\k'}+\p,{-\p}) \\ &\sim N^2_{\rm pix} N_{\rm pix}^{-\frac{2}{3}(3+n)}\,\,, 
    \end{split}
\ee
where by the subscript we indicate that we are focusing only on the contribution that gives rise to $\sigma^2_{2,-}$. In the last line we have used the assumption of an exact power-law universe without the IR cutoff~$k_*$ to estimate the scaling of the variance with the number of pixels. 
On the other hand, the full field-level estimator of Eq.~\eqref{eq:full_optimal_estimator} is given by 
\be
\label{eq:reminder_full_optimal_estimator}
{\cal E} = \frac{\tilde{\cal E}}{1+\delta{\cal E}}\,\,,
\ee
where we have defined 
\be
\delta{\cal E}\equiv\frac{{\rm quadr} - \langle{\rm quadr}\rangle}{\langle{\rm quadr}\rangle}
\ee
and used Eqs.~\eqref{eq:master_formula_for_marginalized_bias-quadr_definition}, \eqref{eq:master_formula_for_marginalized_bias}. It is now straightforward to see that in the variance of the full estimator $\cal E$ the contribution from $\smash{{\rm var}(\tilde{\cal E}_{(2)})}$ is cancelled, i.e.~there is no appearance of $\smash{\sigma^2_{2,-}}$. Indeed, $\smash{\delta{\cal E}}$ has zero mean so its typical value is given by its variance. The contribution to its variance that contains the parameter $\smash{\sigma^2_{2,-}}$ comes from the second line of Eq.~\eqref{eq:prior_part_b1A_fixed}, and is precisely given by 
\be
\label{eq:help_deltaE_variance}
{\rm var}(\delta{\cal E})_{\sigma^2_{2,-}} = (\sigma^2_A)^2\,{\rm var}\bigg(\int_{\k}\frac{|\hat{\delta}^{(2)}_g(\k)|^2}{P(k)}\bigg)_{\sigma^2_{2,-}}\sim N_{\rm pix}^{-\frac{2}{3}(3+n)}\,\,. 
\ee
Hence we see that for large $\smash{N_{\rm pix}}$ the typical value of $\smash{\delta{\cal E}}$ is small, and we can approximate 
\be
\label{eq:true_estimator_approximation}
{\cal E} \approx \frac{\tilde{\cal E}}{1+\sqrt{{\rm var}(\delta{\cal E})}} \approx \tilde{\cal E} - \tilde{\cal E}\sqrt{{\rm var}(\delta{\cal E})}\,\,. 
\ee 
Let us stress again that the smallness of $\smash{\delta{\cal E}}$ does not imply that the estimators~$\tilde{\cal E}$ and~$\cal E$ are the same. As we saw above, the naive estimator can have the suboptimal variance, even though all contributions decay with the number of pixels~$N_{\rm pix}$. Computing the variance of~$\smash{\cal E}$ in this approximation is easy. First, it is straightforward to see that at leading order
\be
{\rm var}(\tilde{\cal E}_{(1)}) = \sigma^2_A\,\,. 
\ee
Then, using this leading order result~$\smash{{\rm var}(\tilde{\cal E}) = {\rm var}(\tilde{\cal E}_{(1)})}$, we see that the part of the variance of $\smash{\tilde{\cal E}_{(2)}}$ that contains the large parameter $\smash{\sigma^2_{2,-}}$ cancels with the contribution coming from the variance of $\smash{\delta{\cal E}}$. 
We can treat the contribution from Eq.~\eqref{eq:other_contribution} in a similar way. This time it cancels with the part of ``$\rm quadr$'' that contains the cubic interactions $Y_3$. More precisely, the cancellation comes from the terms~$\smash{Y_3\sim {X_2X_2}}$. 
In conclusion, approximation to the true estimator given by~\EQ{true_estimator_approximation} is good enough to ensure that the estimator is nearly optimal. This is the analogue of the estimator for local primordial non-Gaussianities derived in~\cite{Creminelli:2006gc}. 

We can also construct a simple estimator for the case where~$A_{\rm s}$ is measured from nonlinear modes where our formulas do not apply. This is particularly relevant for weak lensing surveys where most of the signal comes from~$k>k_{\rm NL}$. In order to do so, we can start from observation that the large covariance matrix for the naive estimator in~\EQ{simple_estimator_nonlinear_DM} was coming from two long modes modulating the short scale power spectrum. Indeed, the leading contribution to the covariance matrix even in the nonlinear regime can be expressed through the so-called response of~$P_g(k)$ to the two long modes~\cite{Barreira:2017sqa}.\footnote{More generally, responses can be defined and measured for the nonlinear field rather than the power spectrum~\cite{Taruya:2021jhg}.} This response is defined as 
\be
\mathcal{R}_2(\k,\p_1,\p_2) = \frac{1}{2} \frac{1}{P_g(k)} \frac{\partial^2P_g(k)}{\partial \delta(\p_1) \partial\delta(\p_2) } \Big|_{\delta(\p_i)=0} \,\, ,
\ee
where~$\p_i\ll k$, and it can be in principle measured in simulations~\cite{Barreira:2017sqa}. What we want to achieve is to ``remove'' this modulation of the power spectrum by the long modes. We can define the following modified estimator for the power spectrum
\be
\hat P_g^{\rm new}(k) \equiv \hat P_g(k) - Y_1^2 \int_{\p\ll k} \mathcal{R}_2(\k,\p,-\p) \hat\delta_g(\p) \hat\delta_g(-\p) \hat P_g(k) \,\, ,
\ee
where we have used~$\smash{\delta(\p)=Y_1\delta_g(\p)}$ on very large scales. It is easy to show that large covariance of the standard power spectrum estimator~$\smash{\hat P_g(k)}$ is exactly canceled by the second term. For this reason we expect this new estimator to be nearly optimal. Some indirect evidence for this can be found for instance in Ref.~\cite{Sefusatti:2006pa} where it is shown that constraints on cosmological parameters for a joint power spectrum and bispectrum analysis improve once the cross-covariance between the two is taken into account. The reason is that the bispectrum partially takes into account the effects of the long modes on the small-scale power spectrum that we discussed here. To include the whole information at leading order one would also have to include the trispectrum. It is also important to note that~$\hat P_g^{\rm new}(k)$ is not unbiased. However, the bias is proportional to the variance of the field on large scales, and therefore it can be easily computed either in perturbation theory or simulations. In its essence, the proposed new estimator is a version of reconstruction where the scatter in~$\hat P_g(k)$ induced by two long modes is reduced by using the knowledge of realization of large-scale galaxy density field. This is in spirit very similar to the standard BAO reconstruction, but with the aim of undoing real gravitational nonlinearities rather than displacements. It would be interesting to explore this strategy in more details and test it on simulations or real data. We leave this interesting investigation for future work. 

Finally, let us briefly comment the case of large noise from mildly nonlinear scales that we discussed in the previous section. There too we can define 
\be
\hat P_g^{\rm new}(k) \equiv \hat P_g(k) - \frac{b_2Y_1^2}{2} \int_{k<\q< k_{\rm NL}} \hat\delta_g(\p) \hat\delta_g(\k-\p) \,\, ,
\ee
and check again that the second term cancels the large contribution to the effective noise in~$P_g(k)$. Even though we have argued that the noise contribution coming from quadratic galaxy does not lead to parametrically large error bars, it would be interesting to explore this in more details and check if even a moderate improvement of cosmological constraints is possible for very dense tracers such as neutral hydrogen. We leave this for future work. 

In summary, we have shown in this section a few examples of new simple estimators for the power spectrum and bispectrum. Their form is inspired by computing the full perturbative posterior and identifying relevant long-short interactions that can make the standard analyses suboptimal. The variance of new estimators does not suffer from large parameters such as~$\sigma^2_{2,-}$ or~$\sigma^2_{2,+}(k)$. Therefore, these estimators are nearly optimal, but still much simpler to implement in practice than the full forward modeling. This is particularly true in regimes where the simple perturbative model does not hold, such as the nonlinear regime.

\section{Conclusions}
\label{sec:conclusions}

\noindent In this work we have studied how well perturbative forward modeling can constrain cosmological parameters compared to conventional analyses based on $n$-point functions. We have focused on the case where cosmic variance dominates the error budget. In this limit it is easy to derive the field-level posterior for cosmological parameters. We have shown that perturbative forward modeling is equivalent to conventional analyses: once the field is predicted at a given order in perturbation theory, an analysis with all the correlation functions one can correctly predict at that order will achieve the same errors on cosmological parameters. 

As all theorems, our result relies on several assumptions, the main one being that the only relevant parameter for the nonlinear evolution is the nonlinear scale. While this is true in most situations of interest, even in~$\Lambda$CDM-like cosmologies we know that there are other relevant parameters that change this simplified picture. In the conventional analyses these parameters can lead ether to depletion of the signal or increase of errors, making them suboptimal. We discussed some examples, such as broadening of the BAO peak or large contributions to the covariance matrix due to the long modes. On the other hand, in all such cases the field-level analysis remains optimal. Given these counterexamples to our general claim, one may argue that the field-level inference is the only way to harvest all cosmological information. However, we have argued that in all relevant cases one can do a simple reconstruction and use simple modified estimators that are nearly optimal. 

Do the results of this work mean that perturbative forward modeling should not be pursued further? We think it is still an interesting direction to follow. However, we argue that one should carry out forward modeling fully perturbatively, i.e.~the marginalization over the initial conditions should also be carried out in perturbation theory. It is only at this level that: 1) the non-Gaussianity of the likelihood and the modulation of the noise by matter fluctuations can be consistently included; 2) the correct comparison to analyses based on correlation functions can be carried out; 3) one can have a systematic understanding of how to implement the theoretical error~\cite{Baldauf:2016sjb,Chudaykin:2020hbf} at the field level. In such setup the usefulness of perturbative forward modeling over standard analyses would not come from obtaining better constraints on cosmology, but from having an alternative way to include information from higher-order correlation functions. 
Importantly, we believe that if this direction is pursued further and perturbative forward modeling is to be successfully applied to data, a full understanding of renormalization at the field level and how it compares with renormalization of correlation functions must be achieved. While here we have not discussed this topic in detail, being it far from the scope of this work, an interesting application of the formalism we used in this work is presented in Appendix~\ref{app:shot_noise_renormalization}, where we extend our formulas beyond the zero-noise, cosmic-variance-limited case discussed in the main text and apply them to show that the ``effective noise'' discussed in Section~\ref{sec:disconnected_diagrams} carries information on the amplitude of the quadratic bias $b_2$ of the tracer under consideration, even if the amplitude of the shot noise $P_\epsilon$ is marginalized over. This goes against the usual lore, where it is assumed that the low-$k$ limit of $\smash{{\cal I}_{\delta^2\delta^2}(k)}$ is fully degenerate with $P_\epsilon$, and hence it is reabsorbed by it after renormalization. We leave a more detailed investigation to future work. 

There are many other aspects of our work that require further investigation. For example, a key ingredient in building the posterior was the inverse model. It would be very important to test the inverse model in simulations and check its range of validity. Related to this, it would be also interesting to evaluate our perturbative posterior given some data~$\hat\delta_g$ and compare it to the full forward modeling. As we have explained, constraints on the BAO scale are the only ones we expect to differ significantly. Therefore, it would be also interesting to apply our methods to the reconstructed galaxy field, where the large displacements are largely removed and the agreement with the optimal analysis is expected to be much better. Along the same lines, one could try to implement our new nearly optimal estimators in practice and check if they lead to tightening of the error bars, particularly in the case of weak lensing. All our results can be straightforwardly generalized to redshift space. Finally, it would be interesting to do it explicitly, test the inverse model in redshift space and develop a pipeline for a realistic spectroscopy survey.

We conclude with some words on primordial non-Gaussianity. Ref.~\cite{Baumann:2021ykm} discussed constraints on primordial non-Gaussianity of the equilateral type at the field level. The results of this work can be straightforwardly applied to the case of non-Gaussian initial conditions:\footnote{For recent constraints on primordial non-Gaussianity from BOSS data see Refs.~\cite{Cabass:2022wjy,DAmico:2022gki,Cabass:2022ymb}.} the only difference in the formulas of Section~\ref{sec:from_likelihood_to_posterior-1} will be that the prior is not anymore a Gaussian but depends on the primordial bispectrum. The posterior at the field level will still be a combination of suitably-weighted correlation functions of the data: hence we expect that, for non-Gaussianity of the equilateral and orthogonal type, working at a given loop order at the field level will give the same constraints on $\smash{f_{\rm NL}}$ as the correlation functions that we correctly capture at that order. It is interesting that local-type primordial non-Gaussianities are enhanced in the infrared: however, unlike the case of CMB anisotropies discussed in Ref.~\cite{Creminelli:2006gc}, the presence of the transfer function never allows to achieve scale invariance, and consequently the effects discussed in Section~\ref{sec:large_covariance_IR} will never be dramatic.

\section*{Acknowledgements}

\noindent It is a pleasure to thank Kazuyuki Akitsu, Alex Barreira, Stephen Chen, Misha Ivanov, Andrija Kosti{\'c}, Matt Lewandowski, Chirag Modi, Minh Nguyen, Andrej Obuljen, Oliver Philcox, Fabian Schmidt, Blake Sherwin and Zvonimir Vlah for useful discussions. G.C. acknowledges support from the Institute for Advanced Study. Part of the work of G.C. was carried out at the Aspen Center for Physics, which is supported by National Science Foundation grant PHY-1607611.  MZ is supported by NSF 2209991 and NSF-BSF 2207583.

\appendix

\section{From likelihood to posterior -- non-perturbative inversion} 
\label{app:from_likelihood_to_posterior-2}

\noindent In this Appendix we show how we can extend the results of Section~\ref{sec:from_likelihood_to_posterior-1} to all orders in perturbation theory. The difficulty of the calculation lies in how to treat the Dirac delta functional to which the likelihood reduces to in the limit of small noise. For this purpose we introduce the following compact notation. 
\begin{itemize}[leftmargin=*]
\item We use Greek indices to denote functionals of momentum $\vec{k}$. The initial conditions $\delta(\vec{k})$ become the ``coordinates'' $\delta^\mu$. Similarly, the fiducial initial conditions $\smash{\hat{\delta}(\vec{k})}$ become $\smash{\hat{\delta}^\mu}$. We will need to take functional derivatives with respect to $\smash{\hat{\delta}(\vec{k})}$, and {not} with respect to $\delta(\vec{k})$: since there is no possibility of confusion we denote these simply by $\partial_\mu$, without any superscript. 
\item Latin indices are used for the remaining parameters, i.e.~$\vec{\theta}$. We use $\partial_i$ to denote derivatives with respect to these parameters. These derivatives are {always} evaluated at the fiducial values $\vers{\theta}$ of these parameters: hence, we do not need to use any further symbol. 
\item We define the following quantities (we use the letter ``$F$'' as in ``forward model'') 
\begin{subequations}
\label{eq:from_likelihood_to_posterior-1}
\begin{align}
F^\mu &= \deltaNL[\delta,\vec{\theta}](\vec{k})\,\,, \label{eq:from_likelihood_to_posterior-1-1} \\ 
\tilde{F}^\mu &= F[\hat{\delta},\vec{\theta}](\vec{k})\,\,, \label{eq:from_likelihood_to_posterior-1-2} \\ 
\hat{F}^\mu &= F[\hat{\delta},\vers{\theta}](\vec{k})\,\,, \label{eq:from_likelihood_to_posterior-1-3} \\
X^\mu &= \hat{F}^\mu - \tilde{F}^\mu\,\,, \label{eq:from_likelihood_to_posterior-1-4} \\ 
\tilde{{\cal D}}^\mu_{\hp{\mu}\nu} &= \partial_\nu\tilde{F}^\mu\,\,, \label{eq:from_likelihood_to_posterior-1-5} \\ 
\tilde{{\cal D}}^\mu_{\hp{\mu}\nu_1\nu_2\cdots\nu_n} &= \partial_{\nu_1}\partial_{\nu_2}\cdots\partial_{\nu_n}\tilde{F}^\mu\,\,. 
\label{eq:from_likelihood_to_posterior-1-6} 
\end{align}
\end{subequations} 
Then, we define $\tilde{\cal I}^\mu_{\hp{\mu}\nu}$ as the matrix inverse of \EQ{from_likelihood_to_posterior-1-5}. That is, 
\begin{equation}
\label{eq:from_likelihood_to_posterior-2}
\tilde{\cal I}^\mu_{\hp{\mu}\rho}\tilde{{\cal D}}^\rho_{\hp{\rho}\nu} = \delta^\mu_{\hp{\mu}\nu} = 
\delta_\nu^{\hp{\nu}\mu} = \tilde{{\cal D}}^\rho_{\hp{\rho}\nu}\tilde{\cal I}^\mu_{\hp{\mu}\rho}\,\,. 
\end{equation} 
\item Similar definitions hold for derivatives of $\hat{F}^\mu$. That is, we have 
\begin{equation}
\label{eq:from_likelihood_to_posterior-3}
\text{$\hat{{\cal D}}^\mu_{\hp{\mu}\nu} = \partial_\nu\hat{F}^\mu\,\,,$ etc.} 
\end{equation} 
\item The remaining definitions we need are those for the prior. We have 
\begin{equation}
\label{eq:from_likelihood_to_posterior-4}
\hat{\cal P} = {\cal P}[\hat{\delta}]\,\,. 
\end{equation} 
Moreover, we also define the symmetric matrices 
\begin{subequations}
\label{eq:from_likelihood_to_posterior-5}
\begin{align}
P(k)\, \dirac{\vec{k}+\vec{k}'} &= P^{\mu\nu}\,\,, \label{eq:from_likelihood_to_posterior-5-1} \\ 
P^{-1}(k)\, \dirac{\vec{k}+\vec{k}'} &= P_{\mu\nu}\,\,, \label{eq:from_likelihood_to_posterior-5-2} 
\end{align}
\end{subequations}
so that 
\begin{equation}
\label{eq:from_likelihood_to_posterior-6}
\text{$\partial_\mu\ln\hat{\cal P} = {-P_{\mu\nu}\hat{\delta}^\nu}$ \quad and \quad 
$\partial_\mu\partial_\nu\ln\hat{\cal P} = {-P_{\mu\nu}}\,\,.$} 
\end{equation}
\end{itemize}

After cleaning up we will arrive at expressions involving only $\hat{\cal P}$ together with its derivatives $\partial_\mu\ln\hat{\cal P}$ and $\partial_\mu\partial_\nu\ln\hat{\cal P}$, the derivatives of $\smash{\hat{F}^\mu}$ with respect to the parameters $\vec{\theta}$ at their fiducial values $\smash{\vers{\theta}}$, the matrix $\smash{\hat{\cal I}_{\hp{\mu}\nu}^\mu}$, and the matrices $\smash{\hat{{\cal D}}^\mu_{\hp{\mu}\nu_1\nu_2\cdots\nu_n}}$ with their derivatives $\partial_i\hat{{\cal D}}^\mu_{\hp{\mu}\nu_1\nu_2\cdots\nu_n}$. At the end of Section~\ref{subsec:second_derivatives_log_posterior} we will drop out all ``hat'' superscripts to make the notation as compact as possible. We will also need to take averages of various expressions over $\hat{\delta}$: we denote these simply by $\braket{\cdots}$.

\subsection{Marginalization over initial conditions}
\label{subsec:marginalization_over_initial_conditions}

\noindent As discussed in the Section~\ref{subsec:posterior_without_noise}, in the cosmic-variance-limited case the conditional likelihood becomes a Dirac delta functional. We recall \EQ{posterior_without_noise-7}, i.e.~ 
\begin{equation}
\label{eq:marginalization_over_initial_conditions-1}
{\cal P}[\hat{\delta}_g|\vec{\theta}] = \int{\cal D}\delta\,\delta^{(\infty)}_{\rm D}(\hat{F}- F)\,{\cal P}[\delta]\,\,. 
\end{equation} 
In Section~\ref{sec:from_likelihood_to_posterior-1} we carried out this integral by solving perturbatively the delta functional for $\delta$ in terms of $\delta_g$. Now we will see how to carry out this integral at all orders in $\delta$. Before proceeding notice that we have only considered a dependence of the forward model, and not of the prior, on $\smash{\vec{\theta}}$. As discussed in Section~\ref{sec:from_likelihood_to_posterior-1} there is no loss of generality in doing this. 

Back to \EQ{marginalization_over_initial_conditions-1}. We do the change of variables 
\begin{equation}
\label{eq:marginalization_over_initial_conditions-2}
\delta^\mu = \hat{\delta}^\mu + \Delta^\mu\,\,. 
\end{equation} 
The advantage is that we expect that the integrand peaks at $\Delta^\mu=0$, so it pays to write 
\begin{equation}
\label{eq:marginalization_over_initial_conditions-2-bis}
F^\mu - \hat{F}^\mu = {\underbrace{\tilde{{\cal D}}^\mu_{\hp{\mu}\nu}\Delta^\nu + \frac{1}{2!}\tilde{{\cal D}}^\mu_{\hp{\mu}\nu\rho}\Delta^\nu\Delta^\rho + \cdots}_{
\hp{G[\Delta]^\mu\,}\equiv\,G[\Delta]^\mu}} - X^\mu\,\,,
\end{equation} 
and after changing variables from $\Delta^\mu$ to $Y^\mu = G[\Delta]^\mu$ we get 
\begin{equation}
\label{eq:marginalization_over_initial_conditions-3}
{-\ln{\cal P}[\hat{\delta}_g|\vec{\theta}]} = {-{\ln\cal P}}\big[\hat{\delta}+G^{-1}[X]\big] - \ln\abs[\bigg]{\frac{\partial G^{-1}[X]}{\partial X}}\,\,. 
\end{equation} 
It is important to emphasize that here we are only consider only the solution of $Y^\mu = G[\Delta]^\mu$ connected to linear theory, as we have done throughout the rest of the paper. If the forward model is built from a filtered field, and is itself cut at a finite momentum (or if equivalently we are working with a coarse enough lattice in real space) this is a good assumption. 

What we need to do now is to compute derivatives $\smash{{\partial_{i_1}}\cdots\partial_{i_n}}$: the key point that comes to our help is that $\smash{X^\mu}$ vanishes if $\smash{\vec{\theta}=\vers{\theta}}$. Let us first check that the average of the first derivative vanishes once we average over the fiducial initial conditions (Section~\ref{subsec:unbiasedness}), and then obtain expressions for the second derivatives (Section~\ref{subsec:second_derivatives_log_posterior}).

\subsection{Unbiasedness}
\label{subsec:unbiasedness}

\noindent Showing unbiasedness is now straightforward irrespectively of what kind of parameter we are looking at. First, we need an expression for $G^{-1}[X]^\mu$. Luckily we only need this as a power series in $X^\mu$. Given the definition of \EQ{marginalization_over_initial_conditions-2-bis}, it is easy to see that 
\begin{equation}
\label{eq:unbiasedness-1-A} 
\begin{split} 
G^{-1}[X]^\mu &= \tilde{\cal I}^\mu_{\hp{\mu}\nu}X^\nu 
- \frac{1}{2}\tilde{\cal I}_{\hp{\nu}\beta}^\nu\tilde{\cal I}_{\hp{\mu}\rho}^\mu 
\tilde{\cal I}_{\hp{\sigma}\alpha}^\sigma\tilde{\cal D}^\rho_{\hp{\rho}\sigma\nu} X^\alpha X^\beta + \cdots\,\,. 
\end{split} 
\end{equation} 
To obtain this equation we have used the relation 
\begin{equation}
\label{eq:unbiasedness-1-B}
\partial_\nu\tilde{\cal I}_{\hp{\mu}\alpha}^\mu = 
{-\tilde{\cal I}_{\hp{\mu}\rho}^\mu\tilde{\cal D}^\rho_{\hp{\rho}\sigma\nu}\tilde{\cal I}_{\hp{\sigma}\alpha}^\sigma} 
\end{equation} 
and the tensor multiplying $X^\alpha X^\beta$ is symmetric in $\alpha\leftrightarrow\beta$ because 
$\tilde{\cal D}^\rho_{\hp{\rho}\sigma\nu} = \tilde{\cal D}^\rho_{\hp{\rho}\nu\sigma}$. 
Notice that we need the expansion up to second order because of the Jacobian in \EQ{marginalization_over_initial_conditions-3}. Being careful about the sign in the definition of $X^\mu$ in \EQ{from_likelihood_to_posterior-1-4}, we obtain 
\begin{equation}
\label{eq:unbiasedness-2}
{-\partial_i\ln{\cal P}[\hat{\delta}_g|\vers{\theta}]} = (\partial_\nu\ln\hat{\cal P})\,\hat{\cal I}_{\hp{\nu}\mu}^\nu\partial_i\hat{F}^\mu 
+ \partial_\nu\big\{\hat{\cal I}_{\hp{\nu}\mu}^\nu\partial_i\hat{F}^\mu\big\}\,\,. 
\end{equation} 

It is easy to see that the average of \EQ{unbiasedness-2} over the fiducial 
initial conditions vanishes. We can -- somewhat suggestively -- rewrite it as 
\begin{equation}
\label{eq:unbiasedness-3}
{-\partial_i\ln{\cal P}[\hat{\delta}_g|\vers{\theta}]} = {\underbrace{(\partial_\nu\ln\hat{\cal P})}_{
\hp{\Gamma^\rho_{\rho\nu}\,}=\,\Gamma^\rho_{\rho\nu}}}\,\hat{\cal I}_{\hp{\nu}\mu}^\nu\partial_i\hat{F}^\mu 
+ \partial_\nu\big\{{\underbrace{\hat{\cal I}_{\hp{\nu}\mu}^\nu\partial_i\hat{F}^\mu}_{
\hp{V^\nu_i\,}=\,V^\nu_i}}\big\} = \nabla_\mu V^\mu_i\,\,, 
\end{equation} 
where averaging over $\hat{\delta}$ is equivalent to integrating over $\hat{\delta}$ 
with a measure given by a diagonal metric with determinant equal to $\hat{{\cal P}}$. Using Stokes' 
theorem, and the fact that the measure vanishes exponentially fast on the ``boundary'', we get 
\begin{equation}
\label{eq:unbiasedness-4}
\big\langle{-\partial_i\ln{\cal P}[\hat{\delta}_g|\vers{\theta}]}\big\rangle = 0\,\,. 
\end{equation}

\subsection{Second derivatives of log-posterior}
\label{subsec:second_derivatives_log_posterior}

\noindent When looking at second derivatives the algebra becomes more complicated. It is helpful to define the matrix $Z^\mu_{\hp{\mu}\nu}$ as 
\begin{equation}
\label{eq:second_derivatives_log_posterior-1}
Z^\mu_{\hp{\mu}\nu} = \frac{\partial G^{-1}[X]^\mu}{\partial X^\nu}\,\,, 
\end{equation}
with $(Z^{-1})_{\hp{\mu}\nu}^\mu$ as its matrix inverse. Then, we have that ${-\partial_i\partial_j\ln{\cal P}[\delta_g|\vers{\theta}]}$ 
contains four terms: two come from the prior and two come from the Jacobian. More precisely, we have 
\begin{equation}
\label{eq:second_derivatives_log_posterior-2}
\begin{split}
{-\partial_i\partial_j\ln{\cal P}[\hat{\delta}_g|\vers{\theta}]} &= 
{-(\partial_\mu\partial_\nu\ln\hat{\cal P})\,\partial_iG^{-1}[X]^\mu\,\partial_jG^{-1}[X]^\nu}
- (\partial_\mu\ln\hat{\cal P})\,\partial_i\partial_jG^{-1}[X]^\mu \\
&\;\;\;\; + {\underbrace{(Z^{-1})^\nu_{\hp{\nu}\rho}(Z^{-1})^\sigma_{\hp{\sigma}\mu}\partial_iZ^\mu_{\hp{\mu}\nu}\,\partial_jZ^\rho_{\hp{\rho}\sigma} 
 - (Z^{-1})^\nu_{\hp{\nu}\mu}\partial_i\partial_jZ^\mu_{\hp{\mu}\nu}}_{
\hp{J^{11}_{ij}+J^{02}_{ij}\,}\equiv\,J^{11}_{ij}+J^{02}_{ij}}}\,\,, 
\end{split}
\end{equation} 
where we have used the relation 
\begin{equation}
\label{eq:second_derivatives_log_posterior-3}
\partial_i(Z^{-1})_{\hp{\mu}\alpha}^\mu = 
{-(Z^{-1})_{\hp{\mu}\rho}^\mu (\partial_i Z^\rho_{\hp{\rho}\sigma})(Z^{-1})_{\hp{\sigma}\alpha}^\sigma}\,\,. 
\end{equation}

In order to compute the last term in \EQ{second_derivatives_log_posterior-2} 
we need to extend \EQ{unbiasedness-1-A} to third order. We get 
\begin{equation}
\label{eq:second_derivatives_log_posterior-4}
\begin{split} 
G^{-1}[X]^\mu &= \tilde{\cal I}^\mu_{\hp{\mu}\nu}X^\nu 
- \frac{1}{2}\tilde{\cal I}_{\hp{\nu}\beta}^\nu\tilde{\cal I}_{\hp{\mu}\rho}^\mu 
\tilde{\cal I}_{\hp{\sigma}\alpha}^\sigma\tilde{\cal D}^\rho_{\hp{\rho}\sigma\nu} X^\alpha X^\beta 
+ \tilde{\cal I}_{\hp{\mu}\nu}^\mu M^\nu_{\hp{\nu}\alpha\beta\gamma}X^\alpha X^\beta X^\gamma + \cdots\,\,, 
\end{split} 
\end{equation} 
where 
\begin{equation}
\label{eq:second_derivatives_log_posterior-5} 
\begin{split} 
M^\nu_{\hp{\nu}\alpha\beta\gamma} &= {-\frac{1}{6}}\tilde{\cal D}^\nu_{\hp{\nu}\rho\sigma\lambda} 
\tilde{\cal I}^\rho_{\hp{\rho}\alpha}\tilde{\cal I}^\sigma_{\hp{\sigma}\beta}\tilde{\cal I}^\lambda_{\hp{\lambda}\gamma} 
+ \frac{1}{2}\tilde{\cal D}^\nu_{\hp{\nu}\rho\sigma}\tilde{\cal I}^\rho_{\hp{\rho}\alpha} 
\tilde{\cal I}^\sigma_{\hp{\sigma}\lambda}\tilde{\cal D}^\lambda_{\hp{\lambda}\kappa\eta} 
\tilde{\cal I}^\kappa_{\hp{\kappa}\beta}\tilde{\cal I}^\eta_{\hp{\eta}\gamma}\,\,. 
\end{split} 
\end{equation} 
We then have all the ingredients to compute all four terms of \EQ{second_derivatives_log_posterior-2}. 
From now on we drop all ``hat'' superscripts: all quantities are intended as evaluated at $\smash{\vers{\theta}}$. 
\begin{itemize}[leftmargin=*] 
\item The first prior term, involving the first derivative squared of the forward model, is given by 
\begin{equation}
\label{eq:second_derivatives_log_posterior-6}
\partial_iG^{-1}[X]^\mu\,\partial_jG^{-1}[X]^\nu = {\cal I}^\mu_{\hp{\mu}\rho}{\cal I}^\nu_{\hp{\nu}\sigma}\partial_iF^\rho\partial_jF^\sigma\,\,. 
\end{equation}
\item The second prior term, which instead involves the second derivative of the forward model, is given by 
\begin{equation}
\label{eq:second_derivatives_log_posterior-7}
\begin{split}
\partial_i\partial_jG^{-1}[X]^\mu &= {-{\cal I}^\mu_{\hp{\mu}\nu}}\partial_i\partial_jF^\nu 
+ 2\,{\cal I}^\mu_{\hp{\mu}\alpha}{\cal I}^\beta_{\hp{\beta}\nu}\partial_{(i}F^\nu\partial_{j)}{\cal D}^\alpha_{\hp{\alpha}\beta} 
- {\cal I}^\nu_{\hp{\nu}\beta}{\cal I}^\mu_{\hp{\mu}\rho}{\cal I}^\sigma_{\hp{\sigma}\alpha}{\cal D}^\rho_{\hp{\rho}\sigma\nu} 
\partial_{(i}F^\alpha\partial_{j)}F^\beta\,\,. 
\end{split} 
\end{equation} 
\item The first of the two Jacobian terms is 
\begin{equation}
\label{eq:second_derivatives_log_posterior-8}
\begin{split}
J^{11}_{ij} &= 
{\cal I}^\nu_{\hp{\nu}\rho}{\cal I}^\sigma_{\hp{\sigma}\mu}\partial_i{\cal D}^\rho_{\hp{\rho}\sigma}\,\partial_j{\cal D}^\mu_{\hp{\mu}\nu} 
- 2\,{\cal I}^\mu_{\hp{\mu}\nu}{\cal D}^\rho_{\hp{\rho}\sigma\mu}{\cal I}^\sigma_{\hp{\sigma}\lambda}{\cal I}^\gamma_{\hp{\gamma}\rho} 
\partial_{(i}F^\nu\partial_{j)}{\cal D}^\lambda_{\hp{\lambda}\gamma} \\ 
&\;\;\;\; + {\cal I}^\gamma_{\hp{\gamma}\lambda}{\cal D}^\sigma_{\hp{\sigma}\rho\gamma}{\cal D}^\mu_{\hp{\mu}\nu\alpha} 
{\cal I}^\rho_{\hp{\rho}\mu}{\cal I}^\alpha_{\hp{\alpha}\beta}{\cal I}^\nu_{\hp{\nu}\sigma}\partial_iF^\lambda\partial_jF^\beta\,\,. 
\end{split}
\end{equation}
\item Unsurprisingly, ${-(Z^{-1})^\nu_{\hp{\nu}\mu}\partial_i\partial_jZ^\mu_{\hp{\mu}\nu}}$ is by far the most complicated term. We have 
\begin{equation}
\label{eq:second_derivatives_log_posterior-9}
\begin{split}
J^{02}_{ij} &= {\cal I}^\beta_{\hp{\beta}\alpha}\partial_i\partial_j{\cal D}^\alpha_{\hp{\alpha}\beta} 
- 2\,\iD{\sigma}{\alpha}\iD{\beta}{\rho}\del_i\D{\rho}{\sigma}\del_j\D{\alpha}{\beta} 
+ 4\,\iD{\rho}{\sigma}\iD{\gamma}{\alpha}\D{\alpha}{\beta\rho}\iD{\beta}{\lambda}\del_{(i}F^\sigma\del_{j)}\D{\lambda}{\gamma} \\ 
&\;\;\;\; + \iD{\rho}{\lambda}\iD{\gamma}{\sigma}\D{\alpha}{\beta\rho}\iD{\beta}{\alpha}\del_{(i}F^\sigma\del_{j)}\D{\lambda}{\gamma} 
+ \iD{\rho}{\alpha}\D{\alpha}{\beta\rho}\iD{\beta}{\lambda}\iD{\gamma}{\sigma}\del_{(i}F^\sigma\del_{j)}\D{\lambda}{\gamma} \\ 
&\;\;\;\; - 2\,\iD{\rho}{\alpha}\iD{\beta}{\rho}\del_{(i}F^\sigma\del_{j)}\D{\alpha}{\beta\rho} 
- \iD{\rho}{\sigma}\iD{\beta}{\alpha}\D{\alpha}{\beta\rho}\del_i\del_jF^\sigma 
+ \D{\mu}{\nu\rho\sigma}\iD{\nu}{\mu}\iD{\rho}{\alpha}\iD{\sigma}{\beta}\del_iF^\beta\del_jF^\alpha \\ 
&\;\;\;\; - \D{\rho}{\gamma\lambda}\D{\mu}{\eta\sigma}\iD{\lambda}{\mu}\iD{\gamma}{\rho}\iD{\eta}{\alpha}\iD{\sigma}{\beta}\del_iF^\beta\del_jF^\alpha 
- 2\,\D{\rho}{\gamma\lambda}\D{\mu}{\eta\sigma}\iD{\lambda}{\mu}\iD{\gamma}{\alpha}\iD{\eta}{\beta}\iD{\sigma}{\rho}\del_iF^\beta\del_jF^\alpha\,\,. 
\end{split}
\end{equation}
\end{itemize} 
In the next sections we use these expressions to show that, if one knows the full matrix $\smash{\I{\mu}{\nu}}$, 
the field level reproduces the linear-theory errors for parameters that appear only in the linear power spectrum. 

Before proceeding, we emphasize that there is actually a great simplification in \EQ{second_derivatives_log_posterior-2}. 
By direct calculation one can show that $\smash{{-(\partial_\mu\ln\hat{\cal P})}\,\partial_i\partial_jG^{-1}[X]^\mu}$ and 
$\smash{{-(Z^{-1})^\nu_{\hp{\nu}\mu}\partial_i\partial_jZ^\mu_{\hp{\mu}\nu}}}$ combine to give 
\begin{equation}
\label{eq:second_derivatives_log_posterior-10}
{-(\partial_\mu\ln\hat{\cal P})}\,\partial_i\partial_jG^{-1}[X]^\mu - (Z^{-1})^\nu_{\hp{\nu}\mu}\partial_i\partial_jZ^\mu_{\hp{\mu}\nu} = 
\nabla_\mu\big\{\partial_i\partial_jG^{-1}[X]^\mu\big\}\,\,, 
\end{equation}
where the ``covariant derivative'' $\nabla_\mu$ is defined in the same way as in \EQ{unbiasedness-3}. 

Also, notice that a relation of the sort of \EQ{second_derivatives_log_posterior-10} 
should have been expected. It essentially shows that, on average, the error is controlled 
by ``first derivatives squared'' of the forward model, and not second derivatives. This is 
in line with the ``standard'' Fisher matrix expressions for marginalization over parameters, 
see e.g.~Eq.~(62) of Ref.~\cite{Heavens:2009nx} for a review.

\subsection{Reproducing linear theory}
\label{subsec:reproducing_linear_theory}

\noindent Let us now show that, for parameters appearing in the linear matter power spectrum, 
the forward model reproduces linear theory if the inverse matrix $\smash{\I{\mu}{\nu}}$ is known exactly. Recall the discussion in 
Section~\ref{subsec:posterior_without_noise}: whatever the parameter we are looking at, we can make the change of variables 
\begin{equation}
\label{eq:reproducing_linear_theory-1}
\text{$\delta(\k)\to\tau(k,\vec{\theta})\,\delta(\k)\,\,,$ \quad with \quad $\tau(k,\vec{\theta}) = \frac{{\cal M}(k,\vec{\theta})}{{\cal M}(k,\vers{\theta})}\,\,.$} 
\end{equation} 
The prior for the new $\delta$ is a Gaussian with power spectrum equal to the fiducial linear power spectrum at that redshift, and 
all the dependence on the parameters $\vec{\theta}$ is now in the forward model. We can then use the results of 
Section~\ref{subsec:second_derivatives_log_posterior}. 

For simplicity of notation we denote derivatives with respect to the single cosmological parameter we are focusing on 
(evaluated at their fiducial) via $'$ and $\partial_{i_0}$. It then proves useful to define the field $\gamma(\k)$ 
and the matrix $\Gamma(\k,\k')$ as 
\begin{equation}
\label{eq:reproducing_linear_theory-2}
\gamma(\k) = \frac{\tau'(k,\vers{\theta})}{\tau(k,\vers{\theta})}\,\delta(\k)\,\,,\quad\Gamma(\k,\k') 
= \frac{\partial\gamma(\k)}{\partial\delta(\k')} = \frac{\tau'(k,\vers{\theta})}{\tau(k,\vers{\theta})}\,\dirac{\k-\k'}
\end{equation} 
Then, the key result is that derivatives with respect to any parameter can be rewritten 
in terms of derivatives with respect to the initial conditions. For example, it is easy to see that 
\begin{equation}
\label{eq:reproducing_linear_theory-3}
\text{$\partial_{i_0} F^\mu = \gamma^\nu\D{\mu}{\nu}\,\,,$ \quad i.e.~\quad 
$F'(\k,\vers{\theta}) = \int_{\k'}\frac{\tau'(k',\vers{\theta})}{\tau(k',\vers{\theta})}\,\delta(\k')\, 
\frac{\partial F(\k,\vers{\theta})}{\partial\delta(\k')}\,\,.$} 
\end{equation} 
One can see why this formula works by expanding the forward model in a power series in $\tau(k,\vec{\theta})\,\delta(\k)$: 
\EQ{reproducing_linear_theory-3} is a consequence of the fact that $\tau(k,\vec{\theta})$ and $\delta(\k)$ enter always 
in this particular combination in this power series. 

With the chain rule, and the fact that $\partial_{i_0}\partial_\mu = \partial_\mu\partial_{i_0}$ 
we can obtain similar formulas for other derivatives appearing in \EQ{second_derivatives_log_posterior-8}. More precisely, we have 
\begin{equation}
\label{eq:reproducing_linear_theory-4}
\partial_{i_0}\D{\mu}{\nu} = \partial_\nu\partial_{i_0}F^\mu = 
\partial_\nu(\gamma^\rho\D{\mu}{\rho}) = \M{\Gamma}{\rho}{\nu}\D{\mu}{\rho} + \gamma^\rho\D{\mu}{\rho\nu}\,\,. 
\end{equation} 
We can then use \eqsII{reproducing_linear_theory-3}{reproducing_linear_theory-4} in \EQ{second_derivatives_log_posterior-2}. 
Recalling \EQ{from_likelihood_to_posterior-6}, all terms combine to give 
\begin{equation}
\label{eq:reproducing_linear_theory-5}
{-\partial^2_{i_0}\ln{\cal P}[\hat{\delta}_g|\vers{\theta}]} = P_{\mu\nu}\gamma^\mu\gamma^\nu + \M{\Gamma}{\alpha}{\beta}\M{\Gamma}{\beta}{\alpha} 
+ \text{total divergence}\,\,. 
\end{equation} 
Using the relations 
\begin{equation}
\label{eq:reproducing_linear_theory-6}
\text{$\gamma^\mu = \M{\Gamma}{\mu}{\nu}\delta^\nu$ \quad and \quad $\braket{\delta^\mu\delta^\nu} = P^{\mu\nu}\,\,,$} 
\end{equation}
we see that the average error per $\dif^3 k/(2\pi)^3$ and per unit volume is 
\begin{equation}
\label{eq:reproducing_linear_theory-7}
2\bigg(\frac{\tau'(k,\vers{\theta})}{\tau(k,\vers{\theta})}\bigg)^2\,\,, 
\end{equation}
i.e.~the same as in linear theory. 

We also emphasize that, while to arrive at \EQ{reproducing_linear_theory-5} we have dropped 
the total divergence, we have explicitly checked that if we consider also all the terms coming from 
\eqsII{second_derivatives_log_posterior-7}{second_derivatives_log_posterior-9} 
we obtain\footnote{Notice that, in order to derive this result, one needs also quantities like $\smash{\partial^2_{i_0}F^\mu}$ and 
$\smash{\partial_{i_0}\D{\mu}{\nu\rho}}$. These are straightforwardly obtained from \eqsII{reproducing_linear_theory-3}{reproducing_linear_theory-4}.} 
\begin{equation}
\label{eq:reproducing_linear_theory-8}
{-\partial^2_{i_0}\ln{\cal P}[\hat{\delta}_g|\vers{\theta}]} = 3P_{\mu\nu}\gamma^\mu\gamma^\nu - \M{\Gamma}{\alpha}{\beta}\M{\Gamma}{\beta}{\alpha}\,\,,
\end{equation}
whose average is clearly the same as \EQ{reproducing_linear_theory-5}.

\section{Including finite shot noise (and shot noise renormalization)}
\label{app:shot_noise_renormalization}

\noindent In this appendix we want to address the following question. In Section~\ref{sec:disconnected_diagrams} we have seen how it is the ``effective noise'' 
\be
P_\epsilon + \frac{b^2_2}{4}\lim_{k\to 0}{\cal I}_{\delta^2\delta^2}(k)
\ee
and not only the shot noise $P_\epsilon$ that controls the variance of the ``na{\"i}ve'' estimator for the amplitude of the linear power spectrum. There we were focusing on the scenario where $P_\epsilon\ll P(k)$, the quadratic bias $b_2$ is fixed, and we want to measure $A_{\rm s}$. But what if we instead want to measure $b_2$? And what if we do not know what $P_\epsilon$ is, and want to marginalize over it? Is there some information on $b_2$ contained in the effective noise? 

To answer this question we need to include a finite $P_\epsilon$ in our calculation of the posterior. It is sufficient, however, to assume: 1) the limit $P_\epsilon\ll P(k)$; 2) that the fiducial noise is zero. In this case, we can derive a perturbative expansion of the posterior via\footnote{There are many ways to derive this expansion, e.g.~via the Fourier transform. For one-dimensional integrals in ${\rm d}x$ one can write 
\be
\begin{split}
\frac{1}{\sqrt{2\pi\sigma^2}}{\rm e}^{-\frac{[d-f(x)]^2}{2\sigma^2}} &= \int_{-\infty}^{+\infty}{\rm d}J\,{\rm e}^{-\frac{\sigma^2J^2}{2}}{\rm e}^{{\rm i}J[d-f(x)]} = \int_{-\infty}^{+\infty}{\rm d}J\,\Bigg(\sum_{n=0}^{+\infty}\frac{({-1})^n\sigma^{2n}J^{2n}}{2^nn!}\Bigg){\rm e}^{{\rm i}J[d-f(x)]} \\
&=\int_{-\infty}^{+\infty}{\rm d}J\,\Bigg(\sum_{n=0}^{+\infty}\frac{\sigma^{2n}}{2^nn!}\Bigg)\frac{\partial^{2n}}{\partial d^{2n}}{\rm e}^{{\rm i}J[d-f(x)]} = \sum_{n=0}^{+\infty}\frac{\sigma^{2n}}{2^nn!}\frac{\partial^{2n}}{\partial d^{2n}}\delta^{(1)}_{\rm D}\big(d-f(x)\big)\,\,, 
\end{split}
\ee 
where the Dirac delta is normalized such that $\int_{-\infty}^{+\infty}{\rm d}x\,\delta^{(1)}_{\rm D}(x)=1$.} 
\be
\label{eq:dirac_delta_expansion}
{\cal L}[\hat{\delta}_g|\delta,\vec{\theta}] = \bigg(1 + \frac{P_\epsilon}{2}\int_{\k} \frac{\partial^2}{\partial\hat{\delta}_g(\k)\partial\hat{\delta}_g({-\k})} + \frac{P_\epsilon^2}{8}\int_{\k,\p}\frac{\partial^4}{\partial\hat{\delta}_g(\k)\partial\hat{\delta}_g({-\k})\partial\hat{\delta}_g(\p)\partial\hat{\delta}_g({-\p})} + \cdots\bigg)\delta^{(\infty)}_{\rm D}\Big(\hat{\delta}_g - \deltaNL[\delta,\param]\Big)\,\,. 
\ee

At this point it is straightforward to compute the posterior at each order in $P_\epsilon/P(k)$ by bring the series of derivatives out of the functional integral of \EQ{posterior_without_noise-5}. We obtain 
\be
\label{eq:useful_for_appendix}
\begin{split}
{-\ln \mathcal{P}[\param|\hat \delta_g]} &= {\underbrace{\frac 12 \chi^2_{\rm prior}[\hat\delta_g,\param] - {\rm Tr}\ln J[\hat\delta_g,\param]}_{\hphantom{{{-\ln \mathcal{P}[\param|\hat \delta_g]_0}}}\equiv\,{{-\ln \mathcal{P}[\param|\hat \delta_g]_0}}}} \\
&\;\;\;\; + \frac{P_\epsilon}{2}\int_{\k}\bigg( \frac{\partial^2\langle{-\ln \mathcal{P}[\param|\hat \delta_g]}\rangle_0}{\partial\hat{\delta}_g(\k)\partial\hat{\delta}_g({-\k})} + \frac{\partial\langle{-\ln \mathcal{P}[\param|\hat \delta_g]}\rangle_0}{\partial\hat{\delta}_g(\k)}\frac{\partial\langle{-\ln \mathcal{P}[\param|\hat \delta_g]}\rangle_0}{\partial\hat{\delta}_g({-\k})}\bigg) \\ 
&\;\;\;\; + \cdots\,\,, 
\end{split}
\ee
where the higher orders in $P_\epsilon$ are straightforwardly obtained from \EQ{dirac_delta_expansion} using 
\be
\frac{\partial\mathcal{P}[\param|\hat \delta_g]_0}{\partial\delta_g(\k)} = {-\mathcal{P}[\param|\hat \delta_g]_0}\,\frac{\partial({{-\ln \mathcal{P}[\param|\hat \delta_g]_0}})}{\partial\delta_g(\k)}\,\,. 
\ee

After averaging over the fiducial initial conditions, we see that the leading order in this expansion is still given by the formulas of Section~\ref{sec:from_likelihood_to_posterior-1}. More precisely, we find 
\be
\big\langle{{-\ln \mathcal{P}[\param|\hat \delta_g]_0}}\big\rangle = \frac{\delta b^2_2}{4}V\int_{\k,\p}\frac{P(p)P(|\k-\p|)}{P(k)}\,\,,
\ee
which is a one-loop term. Notice that here we have already expanded up to second order in $\delta\param=\delta b_2$. The first order in $\delta b_2$ vanishes, consistently with unbiasedness. For the same reason, also the linear order in $P_\epsilon$ vanishes (one can use the formalism of Appendix~\ref{app:from_likelihood_to_posterior-2} to provide a non-perturbative proof of this). One can then derive, in a straightforward albeit tedious way, the orders $\delta b_2\times P_\epsilon$ and $P_\epsilon^2$, both at tree level and at one-loop order. 
Let us however discuss the expression for the error on $b_2$ after marginalization over $P_\epsilon$. By inverting the matrix of second derivatives of minus the log posterior with respect to $b_2$ and $P_\epsilon$, we find 
\be
\frac{1}{\sigma^2_{b_2}} = \frac{\partial^2\langle{{-\ln \mathcal{P}[\param|\hat \delta_g]}}\rangle}{\partial b^2_2} - \bigg(\frac{\partial^2\langle{{-\ln \mathcal{P}[\param|\hat \delta_g]}}\rangle}{\partial b_2\partial P_\epsilon}\bigg)^2\bigg({\frac{\partial^2\langle{{-\ln \mathcal{P}[\param|\hat \delta_g]}}\rangle}{\partial P_\epsilon^2}}\bigg)^{-1}\,\,. 
\ee
This error can be expanded in loops, i.e.~in powers of $\smash{\Delta^2(k)}$. The leading-order (LO) contribution will be a one-loop term, the next-to-leading-order contribution (NLO) a two-loop term, and so on. From this expression we can now see that the explicit form of minus the log posterior at orders $\delta b_2\times P_\epsilon$ and $\smash{P^2_\epsilon}$ is not important. What is crucial is that at tree level only the order $\smash{P^2_\epsilon}$ is not vanishing, being equal to 
\begin{align}
\text{${\frac{\partial^2\langle{{-\ln \mathcal{P}[\param|\hat \delta_g]}}\rangle}{\partial P_\epsilon^2}} = \frac{V}{2}\int_{\k}\frac{1}{P^2(k)}$ at tree level\,\,.}
\end{align}
This leads to 
\be
{-\bigg(\frac{\partial^2\langle{{-\ln \mathcal{P}[\param|\hat \delta_g]}}\rangle}{\partial b_2\partial P_\epsilon}\bigg)^2\bigg({\frac{\partial^2\langle{{-\ln \mathcal{P}[\param|\hat \delta_g]}}\rangle}{\partial P_\epsilon^2}}\bigg)^{-1}} = \text{NLO}\,\,,
\ee
and consequently 
\be
\frac{1}{\sigma^2_{b_2}}\bigg|_{\rm LO} = \frac{\partial^2\langle{{-\ln \mathcal{P}[\param|\hat \delta_g]_0}}\rangle}{\partial b^2_2} = \frac{V}{2}\int_{\k,\p}\frac{P(p)P(|\k-\p|)}{P(k)}\,\,.
\ee
This tells us that at the field level not only we can still constrain $b_2$ after marginalizing over $P_\epsilon$ and that there is information on $b_2$ in the amplitude of the effective noise, but also that the error on $b_2$ becomes smaller the larger the low-$k$ limit of ${\cal I}_{\delta^2\delta^2}(k)$ is. This suggests that one should not let $P_\epsilon$ absorb after renormalization the full $\smash{\lim_{k\to 0}{\cal I}_{\delta^2\delta^2}(k)}$, but only the part of the loop integral in $\lim_{k\to 0}{\cal I}_{\delta^2\delta^2}(k)$ from $\sim k_{\rm NL}$ to very $\rm UV$ modes. This would be similar to what happens with the speed of sound and the low-$k$ limit of $P_{13}$ for the nonlinear dark matter. A more detailed investigation of this is left to future work.



\bibliographystyle{utphys}
\bibliography{refs}

\providecommand{\href}[2]{#2}\begingroup\raggedright\begin{thebibliography}{100}

\bibitem{Scoccimarro:2000sn}
R.~Scoccimarro, ``{The bispectrum: from theory to observations},''
  \href{http://dx.doi.org/10.1086/317248}{{\em Astrophys. J.} {\bfseries 544}
  (2000) 597}, \href{http://arxiv.org/abs/astro-ph/0004086}{{\ttfamily
  arXiv:astro-ph/0004086}}.

\bibitem{Sefusatti:2006pa}
E.~Sefusatti, M.~Crocce, S.~Pueblas, and R.~Scoccimarro, ``{Cosmology and the
  Bispectrum},'' \href{http://dx.doi.org/10.1103/PhysRevD.74.023522}{{\em Phys.
  Rev. D} {\bfseries 74} (2006) 023522},
  \href{http://arxiv.org/abs/astro-ph/0604505}{{\ttfamily
  arXiv:astro-ph/0604505}}.

\bibitem{Baldauf:2014qfa}
T.~Baldauf, L.~Mercolli, M.~Mirbabayi, and E.~Pajer, ``{The Bispectrum in the
  Effective Field Theory of Large Scale Structure},''
  \href{http://dx.doi.org/10.1088/1475-7516/2015/05/007}{{\em JCAP} {\bfseries
  05} (2015) 007}, \href{http://arxiv.org/abs/1406.4135}{{\ttfamily
  arXiv:1406.4135 [astro-ph.CO]}}.

\bibitem{Angulo:2014tfa}
R.~E. Angulo, S.~Foreman, M.~Schmittfull, and L.~Senatore, ``{The One-Loop
  Matter Bispectrum in the Effective Field Theory of Large Scale Structures},''
  \href{http://dx.doi.org/10.1088/1475-7516/2015/10/039}{{\em JCAP} {\bfseries
  10} (2015) 039}, \href{http://arxiv.org/abs/1406.4143}{{\ttfamily
  arXiv:1406.4143 [astro-ph.CO]}}.

\bibitem{Gil-Marin:2014sta}
H.~Gil-Mar\'\i{}n, J.~Nore\~na, L.~Verde, W.~J. Percival, C.~Wagner, M.~Manera,
  and D.~P. Schneider, ``{The power spectrum and bispectrum of SDSS DR11 BOSS
  galaxies \textendash{} I. Bias and gravity},''
  \href{http://dx.doi.org/10.1093/mnras/stv961}{{\em Mon. Not. Roy. Astron.
  Soc.} {\bfseries 451} no.~1, (2015) 539--580},
  \href{http://arxiv.org/abs/1407.5668}{{\ttfamily arXiv:1407.5668
  [astro-ph.CO]}}.

\bibitem{Slepian:2015hca}
Z.~Slepian {\em et~al.}, ``{The large-scale three-point correlation function of
  the SDSS BOSS DR12 CMASS galaxies},''
  \href{http://dx.doi.org/10.1093/mnras/stw3234}{{\em Mon. Not. Roy. Astron.
  Soc.} {\bfseries 468} no.~1, (2017) 1070--1083},
  \href{http://arxiv.org/abs/1512.02231}{{\ttfamily arXiv:1512.02231
  [astro-ph.CO]}}.

\bibitem{Gil-Marin:2016wya}
H.~Gil-Mar\'\i{}n, W.~J. Percival, L.~Verde, J.~R. Brownstein, C.-H. Chuang,
  F.-S. Kitaura, S.~A. Rodr\'\i{}guez-Torres, and M.~D. Olmstead, ``{The
  clustering of galaxies in the SDSS-III Baryon Oscillation Spectroscopic
  Survey: RSD measurement from the power spectrum and bispectrum of the DR12
  BOSS galaxies},'' \href{http://dx.doi.org/10.1093/mnras/stw2679}{{\em Mon.
  Not. Roy. Astron. Soc.} {\bfseries 465} no.~2, (2017) 1757--1788},
  \href{http://arxiv.org/abs/1606.00439}{{\ttfamily arXiv:1606.00439
  [astro-ph.CO]}}.

\bibitem{Slepian:2016kfz}
Z.~Slepian {\em et~al.}, ``{Detection of baryon acoustic oscillation features
  in the large-scale three-point correlation function of SDSS BOSS DR12 CMASS
  galaxies},'' \href{http://dx.doi.org/10.1093/mnras/stx488}{{\em Mon. Not.
  Roy. Astron. Soc.} {\bfseries 469} no.~2, (2017) 1738--1751},
  \href{http://arxiv.org/abs/1607.06097}{{\ttfamily arXiv:1607.06097
  [astro-ph.CO]}}.

\bibitem{Pearson:2017wtw}
D.~W. Pearson and L.~Samushia, ``{A Detection of the Baryon Acoustic
  Oscillation features in the SDSS BOSS DR12 Galaxy Bispectrum},''
  \href{http://dx.doi.org/10.1093/mnras/sty1266}{{\em Mon. Not. Roy. Astron.
  Soc.} {\bfseries 478} no.~4, (2018) 4500--4512},
  \href{http://arxiv.org/abs/1712.04970}{{\ttfamily arXiv:1712.04970
  [astro-ph.CO]}}.

\bibitem{Eggemeier:2018qae}
A.~Eggemeier, R.~Scoccimarro, and R.~E. Smith, ``{Bias Loop Corrections to the
  Galaxy Bispectrum},''
  \href{http://dx.doi.org/10.1103/PhysRevD.99.123514}{{\em Phys. Rev. D}
  {\bfseries 99} no.~12, (2019) 123514},
  \href{http://arxiv.org/abs/1812.03208}{{\ttfamily arXiv:1812.03208
  [astro-ph.CO]}}.

\bibitem{Oddo:2019run}
A.~Oddo, E.~Sefusatti, C.~Porciani, P.~Monaco, and A.~G. S\'anchez, ``{Toward a
  robust inference method for the galaxy bispectrum: likelihood function and
  model selection},''
  \href{http://dx.doi.org/10.1088/1475-7516/2020/03/056}{{\em JCAP} {\bfseries
  03} (2020) 056}, \href{http://arxiv.org/abs/1908.01774}{{\ttfamily
  arXiv:1908.01774 [astro-ph.CO]}}.

\bibitem{MoradinezhadDizgah:2020whw}
A.~Moradinezhad~Dizgah, M.~Biagetti, E.~Sefusatti, V.~Desjacques, and
  J.~Nore\~na, ``{Primordial Non-Gaussianity from Biased Tracers: Likelihood
  Analysis of Real-Space Power Spectrum and Bispectrum},''
  \href{http://dx.doi.org/10.1088/1475-7516/2021/05/015}{{\em JCAP} {\bfseries
  05} (2021) 015}, \href{http://arxiv.org/abs/2010.14523}{{\ttfamily
  arXiv:2010.14523 [astro-ph.CO]}}.

\bibitem{Eggemeier:2021cam}
A.~Eggemeier, R.~Scoccimarro, R.~E. Smith, M.~Crocce, A.~Pezzotta, and A.~G.
  S\'anchez, ``{Testing one-loop galaxy bias: Joint analysis of power spectrum
  and bispectrum},'' \href{http://dx.doi.org/10.1103/PhysRevD.103.123550}{{\em
  Phys. Rev. D} {\bfseries 103} no.~12, (2021) 123550},
  \href{http://arxiv.org/abs/2102.06902}{{\ttfamily arXiv:2102.06902
  [astro-ph.CO]}}.

\bibitem{Alkhanishvili:2021pvy}
D.~Alkhanishvili, C.~Porciani, E.~Sefusatti, M.~Biagetti, A.~Lazanu, A.~Oddo,
  and V.~Yankelevich, ``{The reach of next-to-leading-order perturbation theory
  for the matter bispectrum},''
  \href{http://dx.doi.org/10.1093/mnras/stac567}{{\em Mon. Not. Roy. Astron.
  Soc.} {\bfseries 512} no.~4, (2022) 4961--4981},
  \href{http://arxiv.org/abs/2107.08054}{{\ttfamily arXiv:2107.08054
  [astro-ph.CO]}}.

\bibitem{Oddo:2021iwq}
A.~Oddo, F.~Rizzo, E.~Sefusatti, C.~Porciani, and P.~Monaco, ``{Cosmological
  parameters from the likelihood analysis of the galaxy power spectrum and
  bispectrum in real space},''
  \href{http://dx.doi.org/10.1088/1475-7516/2021/11/038}{{\em JCAP} {\bfseries
  11} (2021) 038}, \href{http://arxiv.org/abs/2108.03204}{{\ttfamily
  arXiv:2108.03204 [astro-ph.CO]}}.

\bibitem{Baldauf:2021zlt}
T.~Baldauf, M.~Garny, P.~Taule, and T.~Steele, ``{Two-loop bispectrum of
  large-scale structure},''
  \href{http://dx.doi.org/10.1103/PhysRevD.104.123551}{{\em Phys. Rev. D}
  {\bfseries 104} no.~12, (2021) 123551},
  \href{http://arxiv.org/abs/2110.13930}{{\ttfamily arXiv:2110.13930
  [astro-ph.CO]}}.

\bibitem{Ivanov:2021kcd}
M.~M. Ivanov, O.~H.~E. Philcox, T.~Nishimichi, M.~Simonovi\'c, M.~Takada, and
  M.~Zaldarriaga, ``{Precision analysis of the redshift-space galaxy
  bispectrum},'' \href{http://dx.doi.org/10.1103/PhysRevD.105.063512}{{\em
  Phys. Rev. D} {\bfseries 105} no.~6, (2022) 063512},
  \href{http://arxiv.org/abs/2110.10161}{{\ttfamily arXiv:2110.10161
  [astro-ph.CO]}}.

\bibitem{Philcox:2021kcw}
O.~H.~E. Philcox and M.~M. Ivanov, ``{BOSS DR12 full-shape cosmology:
  \ensuremath{\Lambda}CDM constraints from the large-scale galaxy power
  spectrum and bispectrum monopole},''
  \href{http://dx.doi.org/10.1103/PhysRevD.105.043517}{{\em Phys. Rev. D}
  {\bfseries 105} no.~4, (2022) 043517},
  \href{http://arxiv.org/abs/2112.04515}{{\ttfamily arXiv:2112.04515
  [astro-ph.CO]}}.

\bibitem{Philcox:2020zyp}
O.~H.~E. Philcox, M.~M. Ivanov, M.~Zaldarriaga, M.~Simonovic, and
  M.~Schmittfull, ``{Fewer Mocks and Less Noise: Reducing the Dimensionality of
  Cosmological Observables with Subspace Projections},''
  \href{http://dx.doi.org/10.1103/PhysRevD.103.043508}{{\em Phys. Rev. D}
  {\bfseries 103} no.~4, (2021) 043508},
  \href{http://arxiv.org/abs/2009.03311}{{\ttfamily arXiv:2009.03311
  [astro-ph.CO]}}.

\bibitem{Philcox:2020vbm}
O.~H.~E. Philcox, ``{Cosmology without window functions: Quadratic estimators
  for the galaxy power spectrum},''
  \href{http://dx.doi.org/10.1103/PhysRevD.103.103504}{{\em Phys. Rev. D}
  {\bfseries 103} no.~10, (2021) 103504},
  \href{http://arxiv.org/abs/2012.09389}{{\ttfamily arXiv:2012.09389
  [astro-ph.CO]}}.

\bibitem{Philcox:2021ukg}
O.~H.~E. Philcox, ``{Cosmology without window functions. II. Cubic estimators
  for the galaxy bispectrum},''
  \href{http://dx.doi.org/10.1103/PhysRevD.104.123529}{{\em Phys. Rev. D}
  {\bfseries 104} no.~12, (2021) 123529},
  \href{http://arxiv.org/abs/2107.06287}{{\ttfamily arXiv:2107.06287
  [astro-ph.CO]}}.

\bibitem{Pardede:2022udo}
K.~Pardede, F.~Rizzo, M.~Biagetti, E.~Castorina, E.~Sefusatti, and P.~Monaco,
  ``{Bispectrum-window convolution via Hankel transform},''
  \href{http://dx.doi.org/10.1088/1475-7516/2022/10/066}{{\em JCAP} {\bfseries
  10} (2022) 066}, \href{http://arxiv.org/abs/2203.04174}{{\ttfamily
  arXiv:2203.04174 [astro-ph.CO]}}.

\bibitem{Rizzo:2022lmh}
F.~Rizzo, C.~Moretti, K.~Pardede, A.~Eggemeier, A.~Oddo, E.~Sefusatti,
  C.~Porciani, and P.~Monaco, ``{The halo bispectrum multipoles in redshift
  space},'' \href{http://dx.doi.org/10.1088/1475-7516/2023/01/031}{{\em JCAP}
  {\bfseries 01} (2023) 031}, \href{http://arxiv.org/abs/2204.13628}{{\ttfamily
  arXiv:2204.13628 [astro-ph.CO]}}.

\bibitem{Ivanov:2023qzb}
M.~M. Ivanov, O.~H.~E. Philcox, G.~Cabass, T.~Nishimichi, M.~Simonovi\'c, and
  M.~Zaldarriaga, ``{Cosmology with the galaxy bispectrum multipoles: Optimal
  estimation and application to BOSS data},''
  \href{http://dx.doi.org/10.1103/PhysRevD.107.083515}{{\em Phys. Rev. D}
  {\bfseries 107} no.~8, (2023) 083515},
  \href{http://arxiv.org/abs/2302.04414}{{\ttfamily arXiv:2302.04414
  [astro-ph.CO]}}.

\bibitem{DAmico:2022ukl}
G.~D'Amico, Y.~Donath, M.~Lewandowski, L.~Senatore, and P.~Zhang, ``{The
  one-loop bispectrum of galaxies in redshift space from the Effective Field
  Theory of Large-Scale Structure},''
  \href{http://arxiv.org/abs/2211.17130}{{\ttfamily arXiv:2211.17130
  [astro-ph.CO]}}.

\bibitem{DAmico:2022osl}
G.~D'Amico, Y.~Donath, M.~Lewandowski, L.~Senatore, and P.~Zhang, ``{The BOSS
  bispectrum analysis at one loop from the Effective Field Theory of
  Large-Scale Structure},'' \href{http://arxiv.org/abs/2206.08327}{{\ttfamily
  arXiv:2206.08327 [astro-ph.CO]}}.

\bibitem{Pardede:2023ddq}
K.~Pardede, E.~Di~Dio, and E.~Castorina, ``{Wide-angle effects in the galaxy
  bispectrum},'' \href{http://dx.doi.org/10.1088/1475-7516/2023/09/030}{{\em
  JCAP} {\bfseries 09} (2023) 030},
  \href{http://arxiv.org/abs/2302.12789}{{\ttfamily arXiv:2302.12789
  [astro-ph.CO]}}.

\bibitem{Kitaura:2007pe}
F.~S. Kitaura and T.~A. Ensslin, ``{Bayesian reconstruction of the cosmological
  large-scale structure: methodology, inverse algorithms and numerical
  optimization},''
  \href{http://dx.doi.org/10.1111/j.1365-2966.2008.13341.x}{{\em Mon. Not. Roy.
  Astron. Soc.} {\bfseries 389} (2008) 497},
  \href{http://arxiv.org/abs/0705.0429}{{\ttfamily arXiv:0705.0429
  [astro-ph]}}.

\bibitem{Jasche:2012kq}
J.~Jasche and B.~D. Wandelt, ``{Bayesian physical reconstruction of initial
  conditions from large scale structure surveys},''
  \href{http://dx.doi.org/10.1093/mnras/stt449}{{\em Mon. Not. Roy. Astron.
  Soc.} {\bfseries 432} (2013) 894},
  \href{http://arxiv.org/abs/1203.3639}{{\ttfamily arXiv:1203.3639
  [astro-ph.CO]}}.

\bibitem{Wang:2013ep}
H.~Wang, H.~J. Mo, X.~Yang, and F.~C. v.~d. Bosch, ``{Reconstructing the
  Initial Density Field of the Local Universe: Methods and Tests with Mock
  Catalogs},'' \href{http://dx.doi.org/10.1088/0004-637X/772/1/63}{{\em
  Astrophys. J.} {\bfseries 772} (2013) 63},
  \href{http://arxiv.org/abs/1301.1348}{{\ttfamily arXiv:1301.1348
  [astro-ph.CO]}}.

\bibitem{Ata:2014ssa}
M.~Ata, F.-S. Kitaura, and V.~M\"uller, ``{Bayesian inference of cosmic density
  fields from non-linear, scale-dependent, and stochastic biased tracers},''
  \href{http://dx.doi.org/10.1093/mnras/stu2347}{{\em Mon. Not. Roy. Astron.
  Soc.} {\bfseries 446} no.~4, (2015) 4250--4259},
  \href{http://arxiv.org/abs/1408.2566}{{\ttfamily arXiv:1408.2566
  [astro-ph.CO]}}.

\bibitem{Seljak:2017rmr}
U.~Seljak, G.~Aslanyan, Y.~Feng, and C.~Modi, ``{Towards optimal extraction of
  cosmological information from nonlinear data},''
  \href{http://dx.doi.org/10.1088/1475-7516/2017/12/009}{{\em JCAP} {\bfseries
  1712} (2017) 009},
\href{http://arxiv.org/abs/1706.06645}{{\ttfamily arXiv:1706.06645
  [astro-ph.CO]}}.

\bibitem{Modi:2018cfi}
C.~Modi, Y.~Feng, and U.~Seljak, ``{Cosmological Reconstruction From Galaxy
  Light: Neural Network Based Light-Matter Connection},''
  \href{http://dx.doi.org/10.1088/1475-7516/2018/10/028}{{\em JCAP} {\bfseries
  10} (2018) 028}, \href{http://arxiv.org/abs/1805.02247}{{\ttfamily
  arXiv:1805.02247 [astro-ph.CO]}}.

\bibitem{Alsing:2018eau}
J.~Alsing, B.~Wandelt, and S.~Feeney, ``{Massive optimal data compression and
  density estimation for scalable, likelihood-free inference in cosmology},''
  \href{http://dx.doi.org/10.1093/mnras/sty819}{{\em Mon. Not. Roy. Astron.
  Soc.} {\bfseries 477} no.~3, (2018) 2874--2885},
  \href{http://arxiv.org/abs/1801.01497}{{\ttfamily arXiv:1801.01497
  [astro-ph.CO]}}.

\bibitem{Alsing:2019xrx}
J.~Alsing, T.~Charnock, S.~Feeney, and B.~Wandelt, ``{Fast likelihood-free
  cosmology with neural density estimators and active learning},''
  \href{http://dx.doi.org/10.1093/mnras/stz1960}{{\em Mon. Not. Roy. Astron.
  Soc.} {\bfseries 488} no.~3, (2019) 4440--4458},
  \href{http://arxiv.org/abs/1903.00007}{{\ttfamily arXiv:1903.00007
  [astro-ph.CO]}}.

\bibitem{Jeffrey:2020xve}
N.~Jeffrey, J.~Alsing, and F.~Lanusse, ``{Likelihood-free inference with neural
  compression of DES SV weak lensing map statistics},''
  \href{http://dx.doi.org/10.1093/mnras/staa3594}{{\em Mon. Not. Roy. Astron.
  Soc.} {\bfseries 501} no.~1, (2021) 954--969},
  \href{http://arxiv.org/abs/2009.08459}{{\ttfamily arXiv:2009.08459
  [astro-ph.CO]}}.

\bibitem{Jasche:2009hz}
J.~Jasche and F.~S. Kitaura, ``{Fast Hamiltonian sampling for large scale
  structure inference},''
  \href{http://dx.doi.org/10.1111/j.1365-2966.2010.16897.x}{{\em Mon. Not. Roy.
  Astron. Soc.} {\bfseries 407} (2010) 29},
  \href{http://arxiv.org/abs/0911.2496}{{\ttfamily arXiv:0911.2496
  [astro-ph.CO]}}.

\bibitem{Jasche:2018oym}
J.~Jasche and G.~Lavaux, ``{Physical Bayesian modelling of the non-linear
  matter distribution: new insights into the Nearby Universe},''
  \href{http://dx.doi.org/10.1051/0004-6361/201833710}{{\em Astron. Astrophys.}
  {\bfseries 625} (2019) A64},
  \href{http://arxiv.org/abs/1806.11117}{{\ttfamily arXiv:1806.11117
  [astro-ph.CO]}}.

\bibitem{Dai:2022dso}
B.~Dai and U.~Seljak, ``{Translation and rotation equivariant normalizing flow
  (TRENF) for optimal cosmological analysis},''
  \href{http://dx.doi.org/10.1093/mnras/stac2010}{{\em Mon. Not. Roy. Astron.
  Soc.} {\bfseries 516} no.~2, (2022) 2363--2373},
  \href{http://arxiv.org/abs/2202.05282}{{\ttfamily arXiv:2202.05282
  [astro-ph.CO]}}.

\bibitem{Robnik:2022bzs}
J.~Robnik, G.~B. De~Luca, E.~Silverstein, and U.~Seljak, ``{Microcanonical
  Hamiltonian Monte Carlo},'' \href{http://arxiv.org/abs/2212.08549}{{\ttfamily
  arXiv:2212.08549 [stat.CO]}}.

\bibitem{Modi:2022pzm}
C.~Modi, Y.~Li, and D.~Blei, ``{Reconstructing the universe with variational
  self-boosted sampling},''
  \href{http://dx.doi.org/10.1088/1475-7516/2023/03/059}{{\em JCAP} {\bfseries
  03} (2023) 059}, \href{http://arxiv.org/abs/2206.15433}{{\ttfamily
  arXiv:2206.15433 [astro-ph.IM]}}.

\bibitem{McDonald:2009dh}
P.~McDonald and A.~Roy, ``{Clustering of dark matter tracers: generalizing bias
  for the coming era of precision LSS},''
  \href{http://dx.doi.org/10.1088/1475-7516/2009/08/020}{{\em JCAP} {\bfseries
  0908} (2009) 020},
\href{http://arxiv.org/abs/0902.0991}{{\ttfamily arXiv:0902.0991
  [astro-ph.CO]}}.

\bibitem{Baumann:2010tm}
D.~Baumann, A.~Nicolis, L.~Senatore, and M.~Zaldarriaga, ``{Cosmological
  Non-Linearities as an Effective Fluid},''
  \href{http://dx.doi.org/10.1088/1475-7516/2012/07/051}{{\em JCAP} {\bfseries
  1207} (2012) 051},
\href{http://arxiv.org/abs/1004.2488}{{\ttfamily arXiv:1004.2488
  [astro-ph.CO]}}.

\bibitem{Carrasco:2012cv}
J.~J.~M. Carrasco, M.~P. Hertzberg, and L.~Senatore, ``{The Effective Field
  Theory of Cosmological Large Scale Structures},''
  \href{http://dx.doi.org/10.1007/JHEP09(2012)082}{{\em JHEP} {\bfseries 09}
  (2012) 082},
\href{http://arxiv.org/abs/1206.2926}{{\ttfamily arXiv:1206.2926
  [astro-ph.CO]}}.

\bibitem{Porto:2013qua}
R.~A. Porto, L.~Senatore, and M.~Zaldarriaga, ``{The Lagrangian-space Effective
  Field Theory of Large Scale Structures},''
  \href{http://dx.doi.org/10.1088/1475-7516/2014/05/022}{{\em JCAP} {\bfseries
  05} (2014) 022}, \href{http://arxiv.org/abs/1311.2168}{{\ttfamily
  arXiv:1311.2168 [astro-ph.CO]}}.

\bibitem{Senatore:2014vja}
L.~Senatore and M.~Zaldarriaga, ``{Redshift Space Distortions in the Effective
  Field Theory of Large Scale Structures},''
  \href{http://arxiv.org/abs/1409.1225}{{\ttfamily arXiv:1409.1225
  [astro-ph.CO]}}.

\bibitem{2014JCAP...08..056A}
V.~{Assassi}, D.~{Baumann}, D.~{Green}, and M.~{Zaldarriaga}, ``{Renormalized
  halo bias},'' \href{http://dx.doi.org/10.1088/1475-7516/2014/08/056}{{\em
  \jcap} {\bfseries 8} (Aug., 2014) 056},
  \href{http://arxiv.org/abs/1402.5916}{{\ttfamily 1402.5916}}.
  \url{http://adsabs.harvard.edu/abs/2014JCAP...08..056A}.

\bibitem{Senatore:2014eva}
L.~Senatore, ``{Bias in the Effective Field Theory of Large Scale
  Structures},'' \href{http://dx.doi.org/10.1088/1475-7516/2015/11/007}{{\em
  JCAP} {\bfseries 1511} no.~11, (2015) 007},
\href{http://arxiv.org/abs/1406.7843}{{\ttfamily arXiv:1406.7843
  [astro-ph.CO]}}.

\bibitem{Lewandowski:2014rca}
M.~Lewandowski, A.~Perko, and L.~Senatore, ``{Analytic Prediction of Baryonic
  Effects from the EFT of Large Scale Structures},''
  \href{http://dx.doi.org/10.1088/1475-7516/2015/05/019}{{\em JCAP} {\bfseries
  05} (2015) 019}, \href{http://arxiv.org/abs/1412.5049}{{\ttfamily
  arXiv:1412.5049 [astro-ph.CO]}}.

\bibitem{Mirbabayi:2014zca}
M.~Mirbabayi, F.~Schmidt, and M.~Zaldarriaga, ``{Biased Tracers and Time
  Evolution},'' \href{http://dx.doi.org/10.1088/1475-7516/2015/07/030}{{\em
  JCAP} {\bfseries 1507} no.~07, (2015) 030},
\href{http://arxiv.org/abs/1412.5169}{{\ttfamily arXiv:1412.5169
  [astro-ph.CO]}}.

\bibitem{Angulo:2015eqa}
R.~Angulo, M.~Fasiello, L.~Senatore, and Z.~Vlah, ``{On the Statistics of
  Biased Tracers in the Effective Field Theory of Large Scale Structures},''
  \href{http://dx.doi.org/10.1088/1475-7516/2015/9/029}{{\em JCAP} {\bfseries
  09} (2015) 029}, \href{http://arxiv.org/abs/1503.08826}{{\ttfamily
  arXiv:1503.08826 [astro-ph.CO]}}.

\bibitem{Lewandowski:2015ziq}
M.~Lewandowski, L.~Senatore, F.~Prada, C.~Zhao, and C.-H. Chuang, ``{EFT of
  large scale structures in redshift space},''
  \href{http://dx.doi.org/10.1103/PhysRevD.97.063526}{{\em Phys. Rev. D}
  {\bfseries 97} no.~6, (2018) 063526},
  \href{http://arxiv.org/abs/1512.06831}{{\ttfamily arXiv:1512.06831
  [astro-ph.CO]}}.

\bibitem{Abolhasani:2015mra}
A.~A. Abolhasani, M.~Mirbabayi, and E.~Pajer, ``{Systematic Renormalization of
  the Effective Theory of Large Scale Structure},''
  \href{http://dx.doi.org/10.1088/1475-7516/2016/05/063}{{\em JCAP} {\bfseries
  1605} no.~05, (2016) 063},
\href{http://arxiv.org/abs/1509.07886}{{\ttfamily arXiv:1509.07886 [hep-th]}}.

\bibitem{Vlah:2015sea}
Z.~Vlah, M.~White, and A.~Aviles, ``{A Lagrangian effective field theory},''
  \href{http://dx.doi.org/10.1088/1475-7516/2015/09/014}{{\em JCAP} {\bfseries
  09} (2015) 014}, \href{http://arxiv.org/abs/1506.05264}{{\ttfamily
  arXiv:1506.05264 [astro-ph.CO]}}.

\bibitem{Blas:2015qsi}
D.~Blas, M.~Garny, M.~M. Ivanov, and S.~Sibiryakov, ``{Time-Sliced Perturbation
  Theory for Large Scale Structure I: General Formalism},''
  \href{http://dx.doi.org/10.1088/1475-7516/2016/07/052}{{\em JCAP} {\bfseries
  1607} no.~07, (2016) 052},
\href{http://arxiv.org/abs/1512.05807}{{\ttfamily arXiv:1512.05807
  [astro-ph.CO]}}.

\bibitem{Vlah:2016bcl}
Z.~Vlah, E.~Castorina, and M.~White, ``{The Gaussian streaming model and
  convolution Lagrangian effective field theory},''
  \href{http://dx.doi.org/10.1088/1475-7516/2016/12/007}{{\em JCAP} {\bfseries
  12} (2016) 007}, \href{http://arxiv.org/abs/1609.02908}{{\ttfamily
  arXiv:1609.02908 [astro-ph.CO]}}.

\bibitem{Desjacques:2016bnm}
V.~Desjacques, D.~Jeong, and F.~Schmidt, ``{Large-Scale Galaxy Bias},''
  \href{http://dx.doi.org/10.1016/j.physrep.2017.12.002}{{\em Phys. Rept.}
  {\bfseries 733} (2018) 1--193},
\href{http://arxiv.org/abs/1611.09787}{{\ttfamily arXiv:1611.09787
  [astro-ph.CO]}}.

\bibitem{Perko:2016puo}
A.~Perko, L.~Senatore, E.~Jennings, and R.~H. Wechsler, ``{Biased Tracers in
  Redshift Space in the EFT of Large-Scale Structure},''
\href{http://arxiv.org/abs/1610.09321}{{\ttfamily arXiv:1610.09321
  [astro-ph.CO]}}.

\bibitem{Chen:2020zjt}
S.-F. Chen, Z.~Vlah, E.~Castorina, and M.~White, ``{Redshift-Space Distortions
  in Lagrangian Perturbation Theory},''
  \href{http://dx.doi.org/10.1088/1475-7516/2021/03/100}{{\em JCAP} {\bfseries
  03} (2021) 100}, \href{http://arxiv.org/abs/2012.04636}{{\ttfamily
  arXiv:2012.04636 [astro-ph.CO]}}.

\bibitem{Cabass:2022avo}
G.~Cabass, M.~M. Ivanov, M.~Lewandowski, M.~Mirbabayi, and M.~Simonovi\'c,
  ``{Snowmass white paper: Effective field theories in cosmology},''
  \href{http://dx.doi.org/10.1016/j.dark.2023.101193}{{\em Phys. Dark Univ.}
  {\bfseries 40} (2023) 101193},
  \href{http://arxiv.org/abs/2203.08232}{{\ttfamily arXiv:2203.08232
  [astro-ph.CO]}}.

\bibitem{Baldauf:2015tla}
T.~Baldauf, E.~Schaan, and M.~Zaldarriaga, ``{On the reach of perturbative
  descriptions for dark matter displacement fields},''
  \href{http://dx.doi.org/10.1088/1475-7516/2016/03/017}{{\em JCAP} {\bfseries
  03} (2016) 017}, \href{http://arxiv.org/abs/1505.07098}{{\ttfamily
  arXiv:1505.07098 [astro-ph.CO]}}.

\bibitem{Baldauf:2015zga}
T.~Baldauf, E.~Schaan, and M.~Zaldarriaga, ``{On the reach of perturbative
  methods for dark matter density fields},''
  \href{http://dx.doi.org/10.1088/1475-7516/2016/03/007}{{\em JCAP} {\bfseries
  03} (2016) 007}, \href{http://arxiv.org/abs/1507.02255}{{\ttfamily
  arXiv:1507.02255 [astro-ph.CO]}}.

\bibitem{Schmittfull:2018yuk}
M.~Schmittfull, M.~Simonovi{\'c}, V.~Assassi, and M.~Zaldarriaga, ``{Modeling
  Biased Tracers at the Field Level},''
  \href{http://dx.doi.org/10.1103/PhysRevD.100.043514}{{\em Phys. Rev.}
  {\bfseries D100} no.~4, (2019) 043514},
\href{http://arxiv.org/abs/1811.10640}{{\ttfamily arXiv:1811.10640
  [astro-ph.CO]}}.

\bibitem{Taruya:2018jtk}
A.~Taruya, T.~Nishimichi, and D.~Jeong, ``{Grid-based calculation for
  perturbation theory of large-scale structure},''
  \href{http://dx.doi.org/10.1103/PhysRevD.98.103532}{{\em Phys. Rev. D}
  {\bfseries 98} no.~10, (2018) 103532},
  \href{http://arxiv.org/abs/1807.04215}{{\ttfamily arXiv:1807.04215
  [astro-ph.CO]}}.

\bibitem{McQuinn:2018zwa}
M.~McQuinn and A.~D'Aloisio, ``{The observable 21cm signal from reionization
  may be perturbative},''
\href{http://arxiv.org/abs/1806.08372}{{\ttfamily arXiv:1806.08372
  [astro-ph.CO]}}.

\bibitem{Schmittfull:2020trd}
M.~Schmittfull, M.~Simonovi\'c, M.~M. Ivanov, O.~H.~E. Philcox, and
  M.~Zaldarriaga, ``{Modeling Galaxies in Redshift Space at the Field Level},''
  \href{http://dx.doi.org/10.1088/1475-7516/2021/05/059}{{\em JCAP} {\bfseries
  05} (2021) 059}, \href{http://arxiv.org/abs/2012.03334}{{\ttfamily
  arXiv:2012.03334 [astro-ph.CO]}}.

\bibitem{Modi:2019hnu}
C.~Modi, M.~White, A.~Slosar, and E.~Castorina, ``{Reconstructing large-scale
  structure with neutral hydrogen surveys},''
\href{http://arxiv.org/abs/1907.02330}{{\ttfamily arXiv:1907.02330
  [astro-ph.CO]}}.

\bibitem{Schmidt:2020ovm}
F.~Schmidt, ``{An $n$-th order Lagrangian Forward Model for Large-Scale
  Structure},'' \href{http://dx.doi.org/10.1088/1475-7516/2021/04/033}{{\em
  JCAP} {\bfseries 04} (2021) 033},
  \href{http://arxiv.org/abs/2012.09837}{{\ttfamily arXiv:2012.09837
  [astro-ph.CO]}}.

\bibitem{Kokron:2021xgh}
N.~Kokron, J.~DeRose, S.-F. Chen, M.~White, and R.~H. Wechsler, ``{The
  cosmology dependence of galaxy clustering and lensing from a hybrid
  N-body\textendash{}perturbation theory model},''
  \href{http://dx.doi.org/10.1093/mnras/stab1358}{{\em Mon. Not. Roy. Astron.
  Soc.} {\bfseries 505} no.~1, (2021) 1422--1440},
  \href{http://arxiv.org/abs/2101.11014}{{\ttfamily arXiv:2101.11014
  [astro-ph.CO]}}.

\bibitem{Taruya:2021ftd}
A.~Taruya, T.~Nishimichi, and D.~Jeong, ``{Grid-based calculations of
  redshift-space matter fluctuations from perturbation theory: UV sensitivity
  and convergence at the field level},''
  \href{http://dx.doi.org/10.1103/PhysRevD.105.103507}{{\em Phys. Rev. D}
  {\bfseries 105} no.~10, (2022) 103507},
  \href{http://arxiv.org/abs/2109.06734}{{\ttfamily arXiv:2109.06734
  [astro-ph.CO]}}.

\bibitem{Obuljen:2022cjo}
A.~Obuljen, M.~Simonovi\'c, A.~Schneider, and R.~Feldmann, ``{Modeling HI at
  the field level},'' \href{http://arxiv.org/abs/2207.12398}{{\ttfamily
  arXiv:2207.12398 [astro-ph.CO]}}.

\bibitem{Schmidt:2018bkr}
F.~Schmidt, F.~Elsner, J.~Jasche, N.~M. Nguyen, and G.~Lavaux, ``{A rigorous
  EFT-based forward model for large-scale structure},''
  \href{http://dx.doi.org/10.1088/1475-7516/2019/01/042}{{\em JCAP} {\bfseries
  1901} (2019) 042},
\href{http://arxiv.org/abs/1808.02002}{{\ttfamily arXiv:1808.02002
  [astro-ph.CO]}}.

\bibitem{Elsner:2019rql}
F.~Elsner, F.~Schmidt, J.~Jasche, G.~Lavaux, and N.-M. Nguyen, ``{Cosmology
  inference from a biased density field using the EFT-based likelihood},''
  \href{http://dx.doi.org/10.1088/1475-7516/2020/01/029}{{\em JCAP} {\bfseries
  01} (2020) 029}, \href{http://arxiv.org/abs/1906.07143}{{\ttfamily
  arXiv:1906.07143 [astro-ph.CO]}}.

\bibitem{Cabass:2019lqx}
G.~Cabass and F.~Schmidt, ``{The EFT Likelihood for Large-Scale Structure},''
  \href{http://dx.doi.org/10.1088/1475-7516/2020/04/042}{{\em JCAP} {\bfseries
  04} (2020) 042}, \href{http://arxiv.org/abs/1909.04022}{{\ttfamily
  arXiv:1909.04022 [astro-ph.CO]}}.

\bibitem{Schmidt:2020tao}
F.~Schmidt, ``{Sigma-Eight at the Percent Level: The EFT Likelihood in Real
  Space},'' \href{http://dx.doi.org/10.1088/1475-7516/2021/04/032}{{\em JCAP}
  {\bfseries 04} (2021) 032}, \href{http://arxiv.org/abs/2009.14176}{{\ttfamily
  arXiv:2009.14176 [astro-ph.CO]}}.

\bibitem{Cabass:2020nwf}
G.~Cabass and F.~Schmidt, ``{The Likelihood for LSS: Stochasticity of Bias
  Coefficients at All Orders},''
  \href{http://dx.doi.org/10.1088/1475-7516/2020/07/051}{{\em JCAP} {\bfseries
  07} (2020) 051}, \href{http://arxiv.org/abs/2004.00617}{{\ttfamily
  arXiv:2004.00617 [astro-ph.CO]}}.

\bibitem{Schmidt:2020viy}
F.~Schmidt, G.~Cabass, J.~Jasche, and G.~Lavaux, ``{Unbiased Cosmology
  Inference from Biased Tracers using the EFT Likelihood},''
  \href{http://dx.doi.org/10.1088/1475-7516/2020/11/008}{{\em JCAP} {\bfseries
  11} (2020) 008}, \href{http://arxiv.org/abs/2004.06707}{{\ttfamily
  arXiv:2004.06707 [astro-ph.CO]}}.

\bibitem{Cabass:2020jqo}
G.~Cabass, ``{The EFT Likelihood for Large-Scale Structure in Redshift
  Space},'' \href{http://dx.doi.org/10.1088/1475-7516/2021/01/067}{{\em JCAP}
  {\bfseries 01} (2021) 067}, \href{http://arxiv.org/abs/2007.14988}{{\ttfamily
  arXiv:2007.14988 [astro-ph.CO]}}.

\bibitem{Barreira:2021ukk}
A.~Barreira, T.~Lazeyras, and F.~Schmidt, ``{Galaxy bias from forward models:
  linear and second-order bias of IllustrisTNG galaxies},''
  \href{http://dx.doi.org/10.1088/1475-7516/2021/08/029}{{\em JCAP} {\bfseries
  08} (2021) 029}, \href{http://arxiv.org/abs/2105.02876}{{\ttfamily
  arXiv:2105.02876 [astro-ph.CO]}}.

\bibitem{Lazeyras:2021dar}
T.~Lazeyras, A.~Barreira, and F.~Schmidt, ``{Assembly bias in quadratic bias
  parameters of dark matter halos from forward modeling},''
  \href{http://dx.doi.org/10.1088/1475-7516/2021/10/063}{{\em JCAP} {\bfseries
  10} (2021) 063}, \href{http://arxiv.org/abs/2106.14713}{{\ttfamily
  arXiv:2106.14713 [astro-ph.CO]}}.

\bibitem{Babic:2022dws}
I.~Babi\'c, F.~Schmidt, and B.~Tucci, ``{BAO scale inference from biased
  tracers using the EFT likelihood},''
  \href{http://dx.doi.org/10.1088/1475-7516/2022/08/007}{{\em JCAP} {\bfseries
  08} no.~08, (2022) 007}, \href{http://arxiv.org/abs/2203.06177}{{\ttfamily
  arXiv:2203.06177 [astro-ph.CO]}}.

\bibitem{Andrews:2022nvv}
A.~Andrews, J.~Jasche, G.~Lavaux, and F.~Schmidt, ``{Bayesian field-level
  inference of primordial non-Gaussianity using next-generation galaxy
  surveys},'' \href{http://dx.doi.org/10.1093/mnras/stad432}{{\em Mon. Not.
  Roy. Astron. Soc.} {\bfseries 520} no.~4, (2023) 5746--5763},
  \href{http://arxiv.org/abs/2203.08838}{{\ttfamily arXiv:2203.08838
  [astro-ph.CO]}}.

\bibitem{Kostic:2022vok}
A.~Kosti\'c, N.-M. Nguyen, F.~Schmidt, and M.~Reinecke, ``{Consistency tests of
  field level inference with the EFT likelihood},''
  \href{http://arxiv.org/abs/2212.07875}{{\ttfamily arXiv:2212.07875
  [astro-ph.CO]}}.

\bibitem{Stadler:2023hea}
J.~Stadler, F.~Schmidt, and M.~Reinecke, ``{Cosmology inference at the field
  level from biased tracers in redshift-space},''
  \href{http://arxiv.org/abs/2303.09876}{{\ttfamily arXiv:2303.09876
  [astro-ph.CO]}}.

\bibitem{Crocce:2007dt}
M.~Crocce and R.~Scoccimarro, ``{Nonlinear Evolution of Baryon Acoustic
  Oscillations},'' \href{http://dx.doi.org/10.1103/PhysRevD.77.023533}{{\em
  Phys. Rev. D} {\bfseries 77} (2008) 023533},
  \href{http://arxiv.org/abs/0704.2783}{{\ttfamily arXiv:0704.2783
  [astro-ph]}}.

\bibitem{Padmanabhan:2008dd}
N.~Padmanabhan, M.~White, and J.~D. Cohn, ``{Reconstructing Baryon
  Oscillations: A Lagrangian Theory Perspective},''
  \href{http://dx.doi.org/10.1103/PhysRevD.79.063523}{{\em Phys. Rev. D}
  {\bfseries 79} (2009) 063523},
  \href{http://arxiv.org/abs/0812.2905}{{\ttfamily arXiv:0812.2905
  [astro-ph]}}.

\bibitem{Sugiyama:2013gza}
N.~S. Sugiyama and D.~N. Spergel, ``{How does non-linear dynamics affect the
  baryon acoustic oscillation?},''
  \href{http://dx.doi.org/10.1088/1475-7516/2014/02/042}{{\em JCAP} {\bfseries
  02} (2014) 042}, \href{http://arxiv.org/abs/1306.6660}{{\ttfamily
  arXiv:1306.6660 [astro-ph.CO]}}.

\bibitem{Eisenstein:2006nk}
D.~J. Eisenstein, H.-j. Seo, E.~Sirko, and D.~Spergel, ``{Improving
  Cosmological Distance Measurements by Reconstruction of the Baryon Acoustic
  Peak},'' \href{http://dx.doi.org/10.1086/518712}{{\em Astrophys. J.}
  {\bfseries 664} (2007) 675--679},
  \href{http://arxiv.org/abs/astro-ph/0604362}{{\ttfamily
  arXiv:astro-ph/0604362}}.

\bibitem{Schmittfull:2017uhh}
M.~Schmittfull, T.~Baldauf, and M.~Zaldarriaga, ``{Iterative initial condition
  reconstruction},'' \href{http://dx.doi.org/10.1103/PhysRevD.96.023505}{{\em
  Phys. Rev.} {\bfseries D96} no.~2, (2017) 023505},
\href{http://arxiv.org/abs/1704.06634}{{\ttfamily arXiv:1704.06634
  [astro-ph.CO]}}.

\bibitem{Rimes:2005xs}
C.~D. Rimes and A.~J.~S. Hamilton, ``{Information content of the non-linear
  matter power spectrum},''
  \href{http://dx.doi.org/10.1111/j.1745-3933.2005.00051.x}{{\em Mon. Not. Roy.
  Astron. Soc.} {\bfseries 360} (2005) L82--L86},
  \href{http://arxiv.org/abs/astro-ph/0502081}{{\ttfamily
  arXiv:astro-ph/0502081}}.

\bibitem{Hamilton:2005dx}
A.~J.~S. Hamilton, C.~D. Rimes, and R.~Scoccimarro, ``{On measuring the
  covariance matrix of the nonlinear power spectrum from simulations},''
  \href{http://dx.doi.org/10.1111/j.1365-2966.2006.10709.x}{{\em Mon. Not. Roy.
  Astron. Soc.} {\bfseries 371} (2006) 1188--1204},
  \href{http://arxiv.org/abs/astro-ph/0511416}{{\ttfamily
  arXiv:astro-ph/0511416}}.

\bibitem{Mohammed:2016sre}
I.~Mohammed, U.~Seljak, and Z.~Vlah, ``{Perturbative approach to covariance
  matrix of the matter power spectrum},''
  \href{http://dx.doi.org/10.1093/mnras/stw3196}{{\em Mon. Not. Roy. Astron.
  Soc.} {\bfseries 466} no.~1, (2017) 780--797},
  \href{http://arxiv.org/abs/1607.00043}{{\ttfamily arXiv:1607.00043
  [astro-ph.CO]}}.

\bibitem{Barreira:2017kxd}
A.~Barreira and F.~Schmidt, ``{Response Approach to the Matter Power Spectrum
  Covariance},'' \href{http://dx.doi.org/10.1088/1475-7516/2017/11/051}{{\em
  JCAP} {\bfseries 11} (2017) 051},
  \href{http://arxiv.org/abs/1705.01092}{{\ttfamily arXiv:1705.01092
  [astro-ph.CO]}}.

\bibitem{Creminelli:2006gc}
P.~Creminelli, L.~Senatore, and M.~Zaldarriaga, ``{Estimators for local
  non-Gaussianities},''
  \href{http://dx.doi.org/10.1088/1475-7516/2007/03/019}{{\em JCAP} {\bfseries
  03} (2007) 019}, \href{http://arxiv.org/abs/astro-ph/0606001}{{\ttfamily
  arXiv:astro-ph/0606001}}.

\bibitem{McQuinn:2020yes}
M.~McQuinn, ``{On the primordial information available to galaxy redshift
  surveys},'' \href{http://dx.doi.org/10.1088/1475-7516/2021/06/024}{{\em JCAP}
  {\bfseries 06} (2021) 024}, \href{http://arxiv.org/abs/2008.12312}{{\ttfamily
  arXiv:2008.12312 [astro-ph.CO]}}.

\bibitem{Feng:2018for}
Y.~Feng, U.~Seljak, and M.~Zaldarriaga, ``{Exploring the posterior surface of
  the large scale structure reconstruction},''
  \href{http://dx.doi.org/10.1088/1475-7516/2018/07/043}{{\em JCAP} {\bfseries
  07} (2018) 043}, \href{http://arxiv.org/abs/1804.09687}{{\ttfamily
  arXiv:1804.09687 [astro-ph.CO]}}.

\bibitem{Peloso:2013zw}
M.~Peloso and M.~Pietroni, ``{Galilean invariance and the consistency relation
  for the nonlinear squeezed bispectrum of large scale structure},''
  \href{http://dx.doi.org/10.1088/1475-7516/2013/05/031}{{\em JCAP} {\bfseries
  05} (2013) 031}, \href{http://arxiv.org/abs/1302.0223}{{\ttfamily
  arXiv:1302.0223 [astro-ph.CO]}}.

\bibitem{Kehagias:2013yd}
A.~Kehagias and A.~Riotto, ``{Symmetries and Consistency Relations in the Large
  Scale Structure of the Universe},''
  \href{http://dx.doi.org/10.1016/j.nuclphysb.2013.05.009}{{\em Nucl. Phys. B}
  {\bfseries 873} (2013) 514--529},
  \href{http://arxiv.org/abs/1302.0130}{{\ttfamily arXiv:1302.0130
  [astro-ph.CO]}}.

\bibitem{Creminelli:2013mca}
P.~Creminelli, J.~Nore{\~n}a, M.~Simonovi{\'c}, and F.~Vernizzi,
  ``{Single-Field Consistency Relations of Large Scale Structure},''
  \href{http://dx.doi.org/10.1088/1475-7516/2013/12/025}{{\em JCAP} {\bfseries
  1312} (2013) 025},
\href{http://arxiv.org/abs/1309.3557}{{\ttfamily arXiv:1309.3557
  [astro-ph.CO]}}.

\bibitem{Creminelli:2013poa}
P.~Creminelli, J.~Gleyzes, M.~Simonovi\'c, and F.~Vernizzi, ``{Single-Field
  Consistency Relations of Large Scale Structure. Part II: Resummation and
  Redshift Space},''
  \href{http://dx.doi.org/10.1088/1475-7516/2014/02/051}{{\em JCAP} {\bfseries
  02} (2014) 051}, \href{http://arxiv.org/abs/1311.0290}{{\ttfamily
  arXiv:1311.0290 [astro-ph.CO]}}.

\bibitem{Creminelli:2013nua}
P.~Creminelli, J.~Gleyzes, L.~Hui, M.~Simonovi\'c, and F.~Vernizzi,
  ``{Single-Field Consistency Relations of Large Scale Structure. Part III:
  Test of the Equivalence Principle},''
  \href{http://dx.doi.org/10.1088/1475-7516/2014/06/009}{{\em JCAP} {\bfseries
  06} (2014) 009}, \href{http://arxiv.org/abs/1312.6074}{{\ttfamily
  arXiv:1312.6074 [astro-ph.CO]}}.

\bibitem{Mirbabayi:2014gda}
M.~Mirbabayi, M.~Simonovi\'c, and M.~Zaldarriaga, ``{Baryon Acoustic Peak and
  the Squeezed Limit Bispectrum},''
  \href{http://arxiv.org/abs/1412.3796}{{\ttfamily arXiv:1412.3796
  [astro-ph.CO]}}.

\bibitem{Baldauf:2015xfa}
T.~Baldauf, M.~Mirbabayi, M.~Simonovi\'c, and M.~Zaldarriaga, ``{Equivalence
  Principle and the Baryon Acoustic Peak},''
  \href{http://dx.doi.org/10.1103/PhysRevD.92.043514}{{\em Phys. Rev. D}
  {\bfseries 92} no.~4, (2015) 043514},
  \href{http://arxiv.org/abs/1504.04366}{{\ttfamily arXiv:1504.04366
  [astro-ph.CO]}}.

\bibitem{Blas:2016sfa}
D.~Blas, M.~Garny, M.~M. Ivanov, and S.~Sibiryakov, ``{Time-Sliced Perturbation
  Theory II: Baryon Acoustic Oscillations and Infrared Resummation},''
  \href{http://dx.doi.org/10.1088/1475-7516/2016/07/028}{{\em JCAP} {\bfseries
  1607} no.~07, (2016) 028},
\href{http://arxiv.org/abs/1605.02149}{{\ttfamily arXiv:1605.02149
  [astro-ph.CO]}}.

\bibitem{Carlson:2012bu}
J.~Carlson, B.~Reid, and M.~White, ``{Convolution Lagrangian perturbation
  theory for biased tracers},''
  \href{http://dx.doi.org/10.1093/mnras/sts457}{{\em Mon. Not. Roy. Astron.
  Soc.} {\bfseries 429} (2013) 1674},
  \href{http://arxiv.org/abs/1209.0780}{{\ttfamily arXiv:1209.0780
  [astro-ph.CO]}}.

\bibitem{Senatore:2014via}
L.~Senatore and M.~Zaldarriaga, ``{The IR-resummed Effective Field Theory of
  Large Scale Structures},''
  \href{http://dx.doi.org/10.1088/1475-7516/2015/02/013}{{\em JCAP} {\bfseries
  02} (2015) 013}, \href{http://arxiv.org/abs/1404.5954}{{\ttfamily
  arXiv:1404.5954 [astro-ph.CO]}}.

\bibitem{Vlah:2015zda}
Z.~Vlah, U.~Seljak, M.~Y. Chu, and Y.~Feng, ``{Perturbation theory, effective
  field theory, and oscillations in the power spectrum},''
  \href{http://dx.doi.org/10.1088/1475-7516/2016/03/057}{{\em JCAP} {\bfseries
  03} (2016) 057}, \href{http://arxiv.org/abs/1509.02120}{{\ttfamily
  arXiv:1509.02120 [astro-ph.CO]}}.

\bibitem{Senatore:2017pbn}
L.~Senatore and G.~Trevisan, ``{On the IR-Resummation in the EFTofLSS},''
  \href{http://dx.doi.org/10.1088/1475-7516/2018/05/019}{{\em JCAP} {\bfseries
  05} (2018) 019}, \href{http://arxiv.org/abs/1710.02178}{{\ttfamily
  arXiv:1710.02178 [astro-ph.CO]}}.

\bibitem{Ivanov:2018gjr}
M.~M. Ivanov and S.~Sibiryakov, ``{Infrared Resummation for Biased Tracers in
  Redshift Space},''
  \href{http://dx.doi.org/10.1088/1475-7516/2018/07/053}{{\em JCAP} {\bfseries
  07} (2018) 053}, \href{http://arxiv.org/abs/1804.05080}{{\ttfamily
  arXiv:1804.05080 [astro-ph.CO]}}.

\bibitem{Barreira:2017sqa}
A.~Barreira and F.~Schmidt, ``{Responses in Large-Scale Structure},''
  \href{http://dx.doi.org/10.1088/1475-7516/2017/06/053}{{\em JCAP} {\bfseries
  06} (2017) 053}, \href{http://arxiv.org/abs/1703.09212}{{\ttfamily
  arXiv:1703.09212 [astro-ph.CO]}}.

\bibitem{Wadekar:2020hax}
D.~Wadekar, M.~M. Ivanov, and R.~Scoccimarro, ``{Cosmological constraints from
  BOSS with analytic covariance matrices},''
  \href{http://dx.doi.org/10.1103/PhysRevD.102.123521}{{\em Phys. Rev. D}
  {\bfseries 102} (2020) 123521},
  \href{http://arxiv.org/abs/2009.00622}{{\ttfamily arXiv:2009.00622
  [astro-ph.CO]}}.

\bibitem{Modi:2016dah}
C.~Modi, E.~Castorina, and U.~Seljak, ``{Halo bias in Lagrangian Space:
  Estimators and theoretical predictions},''
  \href{http://dx.doi.org/10.1093/mnras/stx2148}{{\em Mon. Not. Roy. Astron.
  Soc.} {\bfseries 472} no.~4, (2017) 3959--3970},
  \href{http://arxiv.org/abs/1612.01621}{{\ttfamily arXiv:1612.01621
  [astro-ph.CO]}}.

\bibitem{Villaescusa-Navarro:2018vsg}
F.~Villaescusa-Navarro {\em et~al.}, ``{Ingredients for 21 cm Intensity
  Mapping},'' \href{http://dx.doi.org/10.3847/1538-4357/aadba0}{{\em Astrophys.
  J.} {\bfseries 866} no.~2, (2018) 135},
  \href{http://arxiv.org/abs/1804.09180}{{\ttfamily arXiv:1804.09180
  [astro-ph.CO]}}.

\bibitem{Taruya:2021jhg}
A.~Taruya and K.~Akitsu, ``{Lagrangian approach to super-sample effects on
  biased tracers at field level: galaxy density fields and intrinsic
  alignments},'' \href{http://dx.doi.org/10.1088/1475-7516/2021/11/061}{{\em
  JCAP} {\bfseries 11} no.~11, (2021) 061},
  \href{http://arxiv.org/abs/2106.04789}{{\ttfamily arXiv:2106.04789
  [astro-ph.CO]}}.

\bibitem{Baldauf:2016sjb}
T.~Baldauf, M.~Mirbabayi, M.~Simonovi\'{c}, and M.~Zaldarriaga, ``{Large-scale
  structure con{\-}straints with con{\-}trol{\-}led theo{\-}re{\-}ti{\-}cal
  un{\-}cer{\-}tain{\-}ties},''
\href{http://arxiv.org/abs/1602.00674}{{\ttfamily arXiv:1602.00674
  [astro-ph.CO]}}.

\bibitem{Chudaykin:2020hbf}
A.~Chudaykin, M.~M. Ivanov, and M.~Simonovi\'c, ``{Optimizing large-scale
  structure data analysis with the theoretical error likelihood},''
  \href{http://dx.doi.org/10.1103/PhysRevD.103.043525}{{\em Phys. Rev. D}
  {\bfseries 103} no.~4, (2021) 043525},
  \href{http://arxiv.org/abs/2009.10724}{{\ttfamily arXiv:2009.10724
  [astro-ph.CO]}}.

\bibitem{Baumann:2021ykm}
D.~Baumann and D.~Green, ``{The Power of Locality: Primordial Non-Gaussianity
  at the Map Level},'' \href{http://arxiv.org/abs/2112.14645}{{\ttfamily
  arXiv:2112.14645 [astro-ph.CO]}}.

\bibitem{Cabass:2022wjy}
G.~Cabass, M.~M. Ivanov, O.~H.~E. Philcox, M.~Simonovi\'c, and M.~Zaldarriaga,
  ``{Constraints on Single-Field Inflation from the BOSS Galaxy Survey},''
  \href{http://dx.doi.org/10.1103/PhysRevLett.129.021301}{{\em Phys. Rev.
  Lett.} {\bfseries 129} no.~2, (2022) 021301},
  \href{http://arxiv.org/abs/2201.07238}{{\ttfamily arXiv:2201.07238
  [astro-ph.CO]}}.

\bibitem{DAmico:2022gki}
G.~D'Amico, M.~Lewandowski, L.~Senatore, and P.~Zhang, ``{Limits on primordial
  non-Gaussianities from BOSS galaxy-clustering data},''
  \href{http://arxiv.org/abs/2201.11518}{{\ttfamily arXiv:2201.11518
  [astro-ph.CO]}}.

\bibitem{Cabass:2022ymb}
G.~Cabass, M.~M. Ivanov, O.~H.~E. Philcox, M.~Simonovi\'c, and M.~Zaldarriaga,
  ``{Constraints on multifield inflation from the BOSS galaxy survey},''
  \href{http://dx.doi.org/10.1103/PhysRevD.106.043506}{{\em Phys. Rev. D}
  {\bfseries 106} no.~4, (2022) 043506},
  \href{http://arxiv.org/abs/2204.01781}{{\ttfamily arXiv:2204.01781
  [astro-ph.CO]}}.

\bibitem{Heavens:2009nx}
A.~Heavens, ``{Statistical techniques in cosmology},''
  \href{http://arxiv.org/abs/0906.0664}{{\ttfamily arXiv:0906.0664
  [astro-ph.CO]}}.

\end{thebibliography}\endgroup


\end{document}